\documentclass{aa}
\usepackage{graphicx}
\usepackage{txfonts}
\usepackage{natbib}
\usepackage{array}
\usepackage{booktabs} 
\newcounter{rownum}

\usepackage{xspace}
\usepackage{color}
\usepackage{hyperref}
\usepackage{longtable}
\usepackage{booktabs}
\usepackage{graphicx}
\usepackage{url}
\usepackage{comment}
\usepackage{soul}
\usepackage{xcolor}
\graphicspath{{./figures/}{./figures/figures_appendix/}}

\newcommand{\nustar} {\textit{NuSTAR}\xspace}

\newcommand{\vzero}{V\,0332+53}
\newcommand{\Ecyc}{\ensuremath{E_{\mathrm{Cyc}}}}

\def \ergcmsec{\hbox{\ensuremath{\mathrm{erg\,cm}^{-2}\,\mathrm{s}^{-1}}}}

\def \arcmin {\hbox{\ensuremath{^\prime}}}

\def \redchisq {\ensuremath{\chi^2_{\mathrm{red}}}}
\def \S/Npem {\ensuremath{S/N_{\textrm{PEM}}}}
\def \esplit {\ensuremath{E_{\textrm{split}}}}

\def \pspin {\ensuremath{P_{\textrm{spin}}}}

\begin{document}
\title{Energy-resolved pulse profile changes in V\,0332+53:\\ 
Indications of wings in the cyclotron absorption line profile }
\author{Antonino D'Aì\inst{1}
\and Dimitrios K. Maniadakis\inst{1,2}
\and Carlo Ferrigno\inst{3,4}
\and Elena Ambrosi\inst{1}
\and Ekaterina Sokolova-Lapa\inst{5}
\and Giancarlo Cusumano\inst{1}
\and Peter A. Becker\inst{6}
\and Luciano Burderi\inst{1,7}
\and Melania Del Santo\inst{1}
\and Tiziana Di Salvo\inst{2}
\and Felix Fürst\inst{8}
\and Rosario Iaria\inst{2} 
\and Peter Kretschmar\inst{8}
\and Valentina La Parola\inst{1}
\and Christian Malacaria\inst{9,10}
\and Ciro Pinto\inst{1}
\and Fabio Pintore\inst{1}
\and Guillermo A. Rodr\'iguez-Castillo\inst{1}
}

\institute{INAF - IASF-Palermo, via Ugo La Malfa 153, 90146 Palermo, Italy  \email{antonino.dai@inaf.it}
\and
Dipartimento di Fisica e Chimica Emilio Segrè, Università di Palermo, via Archirafi 36, 90123 Palermo, Italy
\and
Department of Astronomy, University of Geneva, Chemin d’Écogia 16, 1290 Versoix, Switzerland 
\and
INAF, Osservatorio Astronomico di Brera, Via E. Bianchi 46, 23807 Merate, Italy
\and
Dr. Karl Remeis-Observatory and Erlangen Centre for Astroparticle Physics, Friedrich-Alexander Universität Erlangen-Nürnberg,
Sternwartstr. 7, 96049 Bamberg, Germany
\and
Department of Physics and Astronomy, George Mason University, Fairfax, VA 22030, USA
\and
Dipartimento di Fisica, Università degli Studi di Cagliari, SP Monserrato-Sestu KM 0.7,
09042 Monserrato,  Italy 
\and
European Space Agency (ESA), European Space Astronomy Centre (ESAC), Camino Bajo del Castillo s/n, 28692 Villanueva de la Cañada, Madrid, Spain
\and
International Space Science Institute, Hallerstrasse 6, 3012 Bern, Switzerland
\and
INAF-Osservatorio Astronomico di Roma, Via Frascati 33, 00078 Monteporzio Catone (RM), Italy
}
\date{ }
  \abstract
   {}
   {
   We aim to investigate the energy-resolved pulse profile changes   
   of the accreting X-ray pulsar \vzero\, focusing in the cyclotron line energy 
   range, using the full set of available \nustar\ observations. 
   }
   {We applied a tailored pipeline to study the 
   energy dependence of the pulse profiles and 
   to build the pulsed fraction spectra (PFS)
    for the different observations. 
    We also studied the profile changes using cross-correlation and lag spectra. 
    We re-analysed the energy spectra to search for links between the local
    features observed in the PFS and spectral emission components
    associated with the shape of the fundamental cyclotron line.
   }
   {
   In the PFS data, with sufficiently high statistics, we observe a consistent behaviour around the cyclotron line energy. Specifically, two Gaussian-shaped features appear symmetrically on either side of the putative cyclotron line. These features exhibit minimal variation with source luminosity, and their peak positions consistently remain on the left and right of the cyclotron line energy. Associated with the cyclotron line-forming region, we interpret them as evidence for the resonant cyclotron absorption line wings, as predicted by theoretical models of how the cyclotron line profile should appear along the observer's line of sight. A phase-resolved analysis of the pulse in the energy bands surrounding these features enables us to determine both the spectral shape and the intensity of the photons responsible for these peaks in the PFS. Assuming these features correspond to a spectral component, we used their shapes as priors for the corresponding emission components, finding a statistically satisfactory description of the spectra. To explain these results, we propose that our line of sight is close to the direction of the spin axis, while the magnetic axis is likely orthogonal to it.}
   {}
   \titlerunning{Energy-resolved pulse profile changes in V\,0332+53}
\authorrunning{D'Aì et al.}
   \maketitle

\section{Introduction} \label{sect:intro}

X-ray binary pulsars (XBPs) host a neutron star (NS) with 
a strong magnetic field ($B\approx10^{10-13}$ G), 
efficiently guiding accreting material from a companion star 
to the magnetic caps of the NS through the formation 
of an accretion column \citep{Basko1976} in most of the cases. 
Various physical processes, such as inverse-Compton, bremsstrahlung, 
synchrotron, and synchrotron-Compton emission, convert accretion power 
into radiation \citep[see e.g.][]{Becker2022}. At high mass-accretion rates, where the luminosity exceeds a critical threshold \citep{Basko1976}, the radiation pressure in the channel significantly affects the dynamics of the infalling material. In this scenario, radiation primarily escapes through the walls of the accretion column, which extends above the surface of the NS, and its angular redistribution is typically referred to as a fan beam. At lower accretion rates, several mechanisms for decelerating matter in the accretion channel have been proposed, such as the formation of a collisionless shock \citep{Shapiro1975, Langer1982} and Coulomb braking in the accreted plasma at the magnetic pole \citep{Zel'dovich1969, Staubert2007}. The radiation emitted from the pole is expected to escape as a pencil beam along the magnetic axis.
To a distant observer the 
radiation appears as a spin-modulated emission pattern due to the non-alignment of the 
rotational and magnetic axes, similar to classical radio pulsars.

The pulse profile (PP), computed by folding the XBP light curve over the spin period, \pspin, is found being one of their distinctive signatures. 
The PP is influenced by intrinsic factors such as the system and the accretion flow geometry, and the NS properties. 
Transient effects, such as the instantaneous mass accretion rate and local feedback
mechanisms, tend to further complicate the profiles. Disentangling all the physical
mechanisms behind the PP shapes is a challenging task due to the interplay of those
different effects and the importance of photon propagation in the relativistic regime
\citep{Beloborodov2002, Falkner2018}.
With the ultimate aim of understanding XBPs pulse profiles, 
efforts have been made to characterise them phenomenologically.
It has long been known that profile shapes depend on energy and luminosity \citep{Wang1981}. 
In general, the harmonic content is higher below $\sim$10 keV, with pulse amplitudes increasing with energy. 
This is likely due to the relative proportion of emitting spectral components in the pulsed and unpulsed part 
of the spectrum and/or, to the visibility of the pulsed and unpulsed emitting regions for the distant observer. 
A conceptual model includes a variable contribution from pencil and fan beams, with the former dominating at low luminosity and the latter at higher luminosity. 
This model extends to radiation from a thermal mound plus an extended atmosphere and direct radiation from the column. 
Gravitational lensing can significantly enhance the fan beam when the emitting column is behind the NS, 
causing substantial changes in the observed pulse for suitable geometries \citep{Falkner2018}.

Recently, \citep{ferrigno2023} demonstrated how the PP changes in
certain well-known accreting pulsars can be used  
to probe spectroscopic signatures. 
Specifically, the energy spectra of XBPs  often show 
emission and/or absorption lines, such as the iron fluorescence line 
at low energies and the cyclotron resonant scattering features (CRSFs) 
at higher energies \citep[see][for a review]{Mushtukov2022}.
The energy-resolved Pulsed Fraction (PF) Spectra (PFS) can display 
similar local features at the same energies. 
Considering a small sample of \nustar\ observations of well-known bright XRPs, 
\citet{ferrigno2023}  empirically fitted their PFS with a polynomial continuum  
and a small set of Gaussian lines. The positions 
and widths of such local features well matched the 
analogous spectral features at the same energies. 
PFS with sufficiently high statistics achieve the
same energy resolution of the corresponding energy spectra; 
in this case the relative uncertainties on fitted parameters 
are at the same level in both cases.
Based on these results, the analysis of spectral 
variation of the PFS can be seen as a complementary tool to detect and 
characterise spectral features. 
This technique can be particularly useful for the study of CRSFs, 
whose spectral detection and shape determination 
suffers from model degeneracy and/or strong correlation 
of the spectral parameters as well as from technical 
issues such as lack of sufficient statistics and/or energy resolution.
 
\subsection{Cyclotron line shapes} \label{sect:lineshapes}

In the presence of a strong magnetic field such as those found within the accretion 
column of accreting neutron stars (B $\sim 10^{12}$ Gauss), transversal electron
energies are confined to the Landau levels, i.e. they are quantized, with the energy 
of the fundamental state depending on the magnetic field, following the so-called "12-B-12 rule": 

\begin{equation}
    E_{\textrm{cyc}} \approx 11.57\,  \frac{B_{12}}{1+z}\,\mathrm{keV} ,
\end{equation}

\cite{Canuto1977}, where $B_{12}$ is the magnetic field in units of $10^{12}$ Gauss. A gravitational 
redshift, $z$, has the effect of lowering the $E_{\textrm{cyc}}$ estimate by a factor (1+$z$).

Absorption and resonant scattering of photons on these electrons will result 
in absorption-like features in the X-ray spectrum, the so-called cyclotron 
resonance scattering features  \citep[CRSFs, see][for a detailed review]{Staubert2019}. 
\citet{Harding1991} demonstrated that there is no significant difference in the 
cross section for both processes, as long as the magnetic field is 
below $\sim$4\,$\times$10$^{12}$ G (at least for the first two harmonics), 
though the details of the cyclotron line
formation are still expected to differ, as complex (partial)
redistribution takes place in case of the line formation due to resonant
Compton scattering.

To calculate the line profile of CRSFs, \citet{Nishimura1994} computed 
the multi-angle radiative transfer of the first and second harmonics of a slab 
illuminated from below, taking into account the effects of non-coherent scattering. 
In his calculations, the cyclotron line profile of the fundamental is complex, consisting of an absorption line core and line wings on one or both sides of the line. 
The wings (also called shoulders by various authors; 
we use these terms synonymously) would appear as local emission components on the side of the deep cyclotron absorption feature in the spectra.
The shape of such wings depends strongly on the viewing angle formed between 
the direction of the magnetic field (which is assumed normal to the slab's surface) 
and the line of sight. A greater intensity is predicted when the slab is viewed at larger angles. 
The formation of the wings was explained as the result of non-coherent scattering of photons 
out from the line core to the wings, in combination with photon spawning from higher harmonics.   

\citet{Isenberg1998a} performed generalized Monte-Carlo (MC) 
simulations and found some results at odds with those of \citet{Nishimura1994}. 
They tested different geometries: a slab with the illuminating photon source below, 
a slab with the illuminating source in its midplane, and a slab of cylindrical shape 
and the illuminating source along its axis. Their results showed that the emission 
wings of the fundamental were actually stronger when the slab is observed from above, 
therefore for a small angle between the line of sight and the magnetic field, 
in some contradiction with the results of \citet{Nishimura1994}. 

\citet{Araya1996, Araya1999} presented a set of MC simulations for very 
hard spectra of X-ray pulsars with near-critical fields ($B_{\textrm{crit}} \sim 4 \times 10^{13}$ G). 
They studied 
propagation in the slab and cylinder geometry. For both cases the emission 
wings were present, however weaker in the latter case. \citet{Schonherr2007} 
further refined the same code of  \citet{Araya1996, Araya1999} by following a Green's 
functions approach for the calculation of the CRSF shapes.
The emission wings were also apparent in the results of the their MC simulations. 
Based on this they developed the \texttt{xspec} model \texttt{cyclomc} with 
which they attempted to fit \textit{RXTE} and \textit{INTEGRAL} data from the 
January 2005 outburst of \vzero.
However they did not find evidence for emission wings from these data.

The most recent developments in this area have been contributed 
by \citet{Schwarm2017a} and \citet{Kumar2022}. 
\citet{Schwarm2017a} also employed the Green's functions approach, 
expanded the set of possible geometries by making it possible to 
combine together cylinders of arbitrary dimensions.
Their results were consistent with the \citet{Isenberg1998a} work, 
with some slight deviations in the line wings shapes, additionally
they showed that, adopting the cylindrical instead of the slab
geometry, wings would appear rather at large angles to the magnetic fields.
They developed a new fitting \texttt{xspec} model called \texttt{cyclofs} 
and successfully fitted \nustar\ data from the June 2014 outburst of Cep X-4,
in this case not showing any evidence for the existence of emission 
wings around the absorption core of the fundamental cyclotron line. 
\citet{Kumar2022} followed the approach of \citet{Araya1999} and added 
the relativistic effects of light bending and gravitational redshift. 
He cautioned that by using a correct GR approach, emission line 
wings could completely disappear.\\
In addition to this, \citet{Mushtukov2021a}
and \citet{Sokolova-Lapa2021} 
demonstrated the creation of broad wings due to Compton up-scattering of
continuum photons and resonant redistribution in the context of
low-luminosity accretion. \citet{Loudas2024} noted a similar
phenomenon in the case of the formation of the CRSF in the radiative
shock, where thermal and bulk Comptonization result in a pronounced blue
wing. In these cases, the broadened wings are often difficult to
disentangle from the continuum.\\
In all the aforementioned studies in the case of a slab geometry, 
the fundamental cyclotron line exhibits two emission wings 
which are more prominent under specific conditions. 
Firstly, when the angle between the viewing direction and the slab's normal is small, 
due to the lower optical depths. 
Secondly, in the case where the X-ray continuum is hard, 
since the number of transition photons which tend to fill in the 
fundamental feature is larger. 
In addition, they also depend on the optical thickness of the slab 
and the geometry of the illuminating source of the photons. 
Nonetheless, despite their predicted appearance in theoretical works,  CRSF profiles are found to be usually well fitted either by simple Gaussian, or Lorentzian, profiles and no spectral residuals 
around these lines clearly indicated the presence of such wings 
\citep[see e.g.][for a spectral decomposition attempt with wings
emission for the source GX\,301--2]{Zalot2024}.

\subsection{The X-ray pulsar V\,0332+53} \label{sect:v0332}
\vzero\ is a recurrent outbursting young X-ray pulsar \citep{Terrell1984}. 
With the O8-9 Ve star BQ Cam \citep{Honeycutt1985, Negueruela1999}, 
they form a Be/X-ray binary system which has exhibited 
several giant X-ray outbursts since its discovery. 
The most recent X-ray timing analysis study constrained the 
orbital period, the projected major semi-axis and the eccentricity of the
system to 33.85 days, 77.8 light seconds, and 0.371, respectively \citep{Doroshenko2016}. 
The most updated distance estimate from the third Gaia data release (DR) is 5.6$_{+0.7}^{-0.5}$ kpc
\citep[consistent with  previous published estimate of 5.1$_{+0.8}^{-1.0}$ kpc from Gaia DR2][]{Arnason2021}. 

During the 1989 outburst a CRSF  at $\sim$\,28.5 keV was discovered \citep{Makishima1990},  
from which a NS magnetic field of 3 $\times$ 10$^{12}$ G was estimated. 
The following giant outburst took place between November of 2004 and February of 2005, 
and was monitored by \textit{RXTE} and \textit{INTEGRAL} 
\citep{Mowlavi2006, Tsygankov2006, Tsygankov2010a}. 
Three CRSFs were  detected \citep{Pottschmidt2005} 
at $\sim$\,27 keV, $\sim$\,51 keV, and $\sim$\,74 keV. 
The energy of the fundamental cyclotron line showed a complex evolution in time. 
An anti-correlation between the cyclotron centroid energy and the source luminosity was detected \citep{Tsygankov2010a}, 
which has been explained by either the changing height of the accretion column \citep{Tsygankov2016, Mushtukov2015a}, 
or by the changing of the reflective area of the accretion column emission 
from the surface of the pulsar \citep{Poutanen2013, Lutovinov2015, Mushtukov2018},
though the latter interpretation has been recently challenged by Monte-Carlo 
simulations of reflected spectra, suggesting
that overlapping spectra from the different illuminate rings on the NS surface would
result in washed-out features \citep{Kylafis2021}.

In 2015, another giant outburst occurred, followed by a period of relative X-ray soft, 
low-luminosity state \citep{Wijnands2016}, and a failed mini-outburst in 2016 \citep{Baum2017, Doroshenko2017}. 
\citet{Cusumano2016a} using \textit{Swift}/BAT data  during the 2015 outburst, 
found that the energy of the fundamental cyclotron line decreased with luminosity. 
Additionally, it was observed that, in the declining phase of the outburst, for similar levels of luminosity,
there was a systematically lower value of the cyclotron line energy 
compared to the rising phase. 
This resulted in a drop of approximately $1.7\times 10^{11}$ G 
of the observed magnetic field strength between the onset and the end of the outburst.
This result was further confirmed by the analysis of \textit{INTEGRAL} \citep{Ferrigno2016} 
and \nustar\ observations \citep{Doroshenko2017, Vybornov2018}.
The observed shift in energy was attributed either to the 
screening of the pulsar's dipole magnetic field by the accreting matter 
on to the polar cap \citep{Cusumano2016a} or to a change of the observed emitting regions \citep{Mushtukov2018}. 
Since 2016, \vzero\ has been in quiescence. 
Search for possible pulsations in radio did not yield any detection \citep{vandenEijnden2024}. 

In this work we mainly focus on the analysis of the 
energy-resolved pulse profiles, particularly on the  
PFS properties of \vzero\ observed with 
\nustar\ during its latest major outburst in 2015 and the failed outburst of 2016.

\section{Observations and data analysis} \label{sect:observations}

\begin{figure}
\centering
\includegraphics[width=\columnwidth]{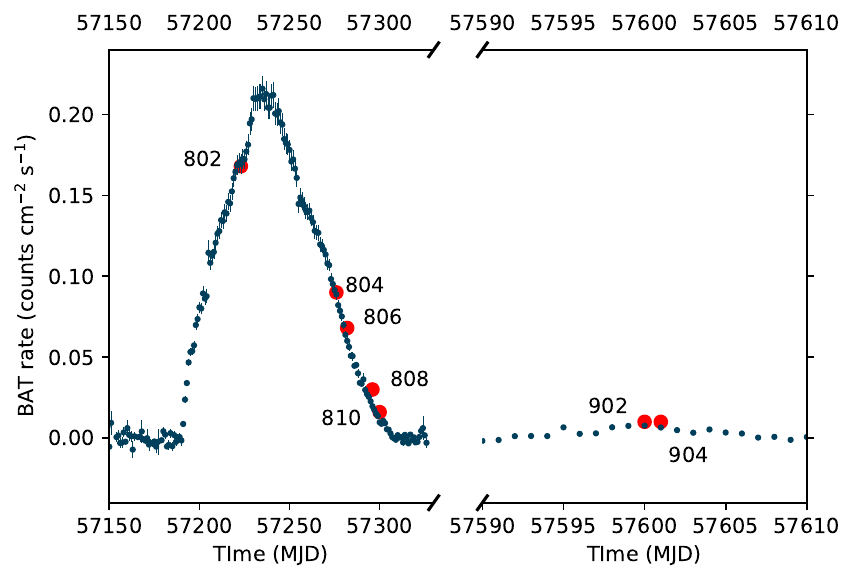} 
\caption{$Swift$/BAT light curve of \vzero\ during its 2015 outburst and 
at the time of the 2016 failed outburst. Times of \nustar\ observations are
marked by the red points.}
\label{fig:batlc}
\end{figure}

We retrieved the \nustar\ \citep{Harrison2013} observations from the public HEASARC archive.
Each observation is uniquely identified by an Observation Identification number (ObsID), 
and we will use the ObsID shortcuts as declared in Table \ref{tab:observing_log} to 
refer to the single ObsIDs. We show in Fig. \ref{fig:batlc} a light curve of the
2015 outburst and a snapshot of the 2016 failed outburst as observed by 
the $Swift$/BAT instrument in the 15--50 keV range. $NuSTAR$  observation times are marked by the red points. 
We processed the data using the \nustar\ Data Analysis
Software \textsc{nustardas} v.\,1.9.7
available in \textsc{HEASoft} v.\,6.31 along with the latest \nustar\
calibration files (CALDB v20240104), using the custom-developed 
\textsc{nustarpipeline}\footnote{\url{https://gitlab.astro.unige.ch/ferrigno/nustar-pipeline}} wrapper.
First, we obtained calibrated level 2 event files of the Focal Plane Module A
(FPMA) and Focal Plane Module B (FPMB).
We defined a circular source region (centred on the best-known source X-ray coordinates 
and having a radius of 2\arcmin, thus encompassing $\sim$\,95\% of the source signal) 
and a background region of similar size chosen in a detector region free of contaminating sources. 
We used the \textsc{SAO ds9} software for a visual inspection 
and verified that none of the examined ObsIDs was affected by stray light issues.

The events extracted from these source and background regions were then used 
for spectral and timing analysis. To this aim, we used our developed analysis toolkit, 
detailed in \citep{ferrigno2023}. 

Here, we give a brief summary of  the pipeline steps:
\begin{itemize}
\item We filter out time intervals when the source shows extreme spectral 
or flux variability (e.g. dip, eclipse or flaring episodes).
\item We barycentre the arrival time of each event in the Solar System frame
and  performed a Lomb-Scargle (LS) search for coherent signals in the 3--7 keV 
range (this is usually already sufficient to find pulsations in case of bright pulsars).
We take the highest peak in the LS spectrum as the preliminary 
spin frequency ($P_\mathrm{spin}$), checking its consistency with the literature value.
\item We refine $P_\mathrm{spin}$ using an epoch folding search in a frequency
interval around $\pm5\%$\,$\times P_\mathrm{spin}$, after correcting for the binary
motion, by using the  binary orbital parameters taken from the \textit{Fermi Gamma-ray Burst Monitor} (GBM) online
database\footnote{\url{https://gammaray.nsstc.nasa.gov/gbm/science/pulsars.html}} \citep{Malacaria2020}.
\item We extract folded pulse profiles with 128 phase-bins, in the 3--30 keV range,
in intervals of 1500\,s for the whole duration of the observation;
we store these values in a time-phase matrix. 
We impose a minimum signal-to-noise ratio (S/N) of 15 for each time-selected
folded profile and, when this threshold is not met, we sum adjacent profiles 
until obtaining the desired S/N. 
\item We determine through a fast Fourier transform (FFT) 
the phase of the first harmonic in each 
time-selected pulse profile as a function of time. 
Subsequently, we examine the presence of a linear trend in the
phase values and calculate a spin derivative term when the 
linear trend is significant at  a confidence 
level greater than 95\% . In this case, we adopt the new refined timing solution for further analysis.
\item We derive for the whole observation under exam 
an energy-phase matrix 
(i.e. a pulse profile for each energy bin) by folding data with the best
spin and spin derivative values obtained from the previous step. 
There is a matrix associated to a specific $N_{\textrm {bins}}$, 
where $N_{\textrm{bins}}$ is the number of the phase bins 
in the pulse profile (e.g. 8, 10, 16, 32).
The default energy spacing of the bins matches 
the intrinsic energy resolution of the FPMs.
We compute separate source and background matrices 
for the two FPMA and FPMB detectors, and then sum 
the resulting matrices.
For each energy bin, we compute the S/N of the pulse 
taking the background events into account. We then sum over adjacent bins to reach 
a minimum S/N over the whole \nustar\ energy band.
\item We compute the PF value in each re-binned energy bin, 
adopting the FFT method \citep[see][]{ferrigno2023} and derive the harmonic decomposition 
simultaneously, which we will also use in the following 
analysis:
\begin{equation}\label{eq:fft}
\mathrm{PF}_{\mathrm{FFT}} = \frac {\sqrt { \sum _{k = 1} ^{N_\mathrm{harm}}\mid A_{k} \mid ^2 } } {\mid A_{0}\mid }.
\end{equation}
Here $\mid A_{0}\mid$ is the average value of the pulse profile, 
which is the zero term of the FFT transform, and the terms $A_{1...k}$ 
are the $k$-th terms of the discrete Fourier transform, 
so that each $ \mid A_{k} \mid $ represents the amplitude of the $k$-th harmonic.
\item We truncate the Fourier spectral decomposition, 
using a number of harmonics that describe the pulse 
with a statistical acceptance level of at least 10\% (usually 
3 to 5 harmonics). 
\item We compute the uncertainty on the PF value using a 
bootstrap method. For each energy-resolved profile, we 
simulated 1000 faked profiles, assuming Poisson statistics in
each phase bin; we then derive the average and 
the standard deviation of this sample. 
After verifying that the average is compatible with the value 
computed from data, we  adopt the sample standard deviation as an 
estimate of the uncertainty of the PF at a 1$\sigma$ confidence level.
\item We compute the correlation and the lag spectra for each 
observation according to the methods described in \citet{ferrigno2023}, 
but slightly modifying the procedure to improve the error estimation (see Sect.\ref{sect:crossandlag}). 
We use the 3--70 keV band, with the exclusion of  
the energy bin which we want to cross-correlate,
as the band to extract the PP of reference. 
\end{itemize}

For the spectral analysis, we used \texttt{numkarf} and \texttt{numkrmf}
to obtain the ancillary response and the redistribution
matrix files, respectively. We rebin the source spectra using 
the \texttt{ftgrouppha} tool\footnote{\url{https://heasarc.gsfc.nasa.gov/lheasoft/help/ftgrouppha.html}} 
optimization algorithm of \citet{Kaastra2016} and with the additional
requirement of a S/N at least of three for each rebinned energy bin.

\subsection{Energy and luminosity dependence of pulse profiles} \label{sect:pp_perobsid}
\begin{figure*}
\centering
\begin{tabular}{cc}
\includegraphics[width=0.8\columnwidth]{figures2/colmap802.pdf}  & 
\includegraphics[width=0.8\columnwidth]{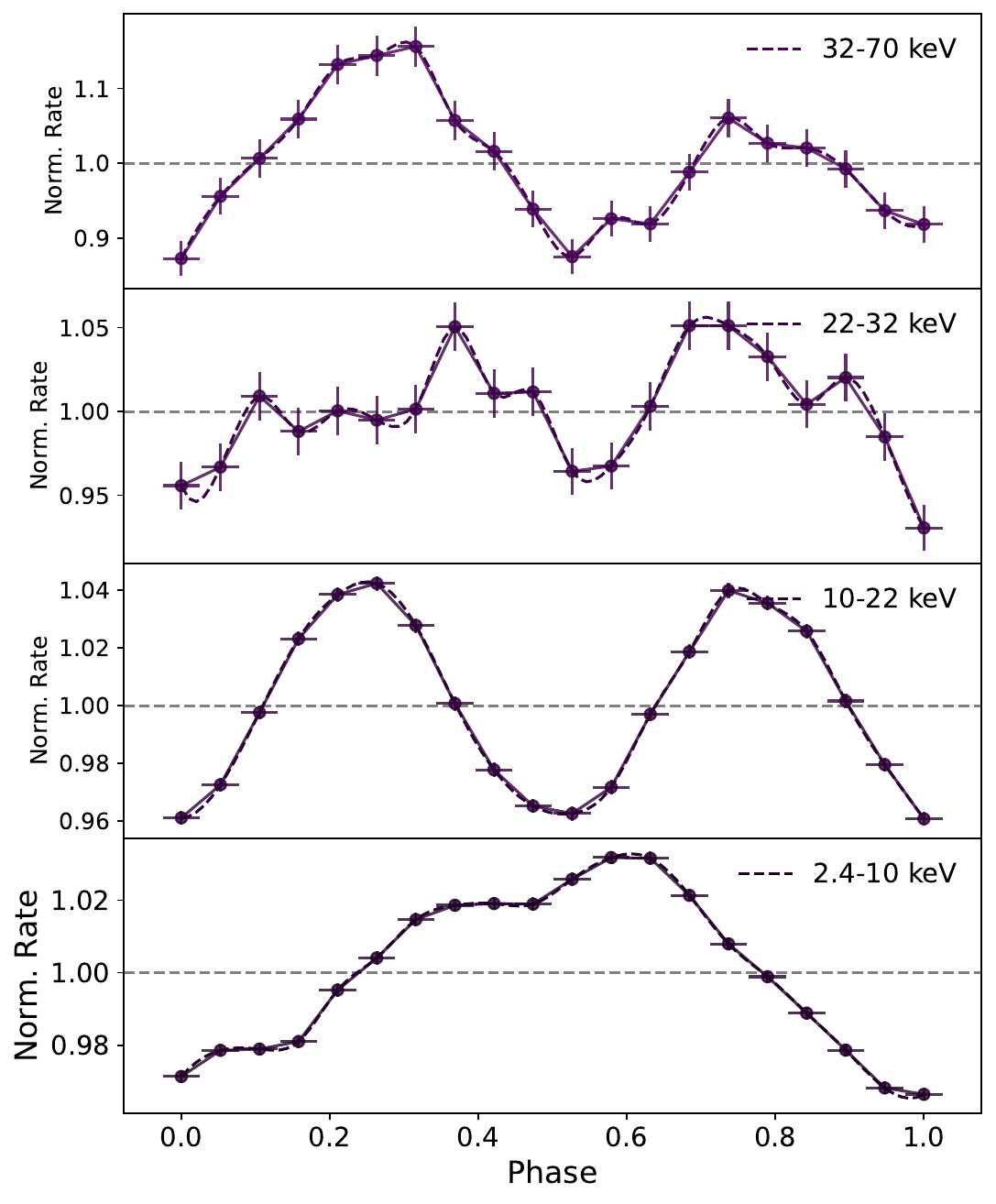} \\
\includegraphics[width=0.8\columnwidth]{figures2/colmap804.pdf}  & 
\includegraphics[width=0.8\columnwidth]{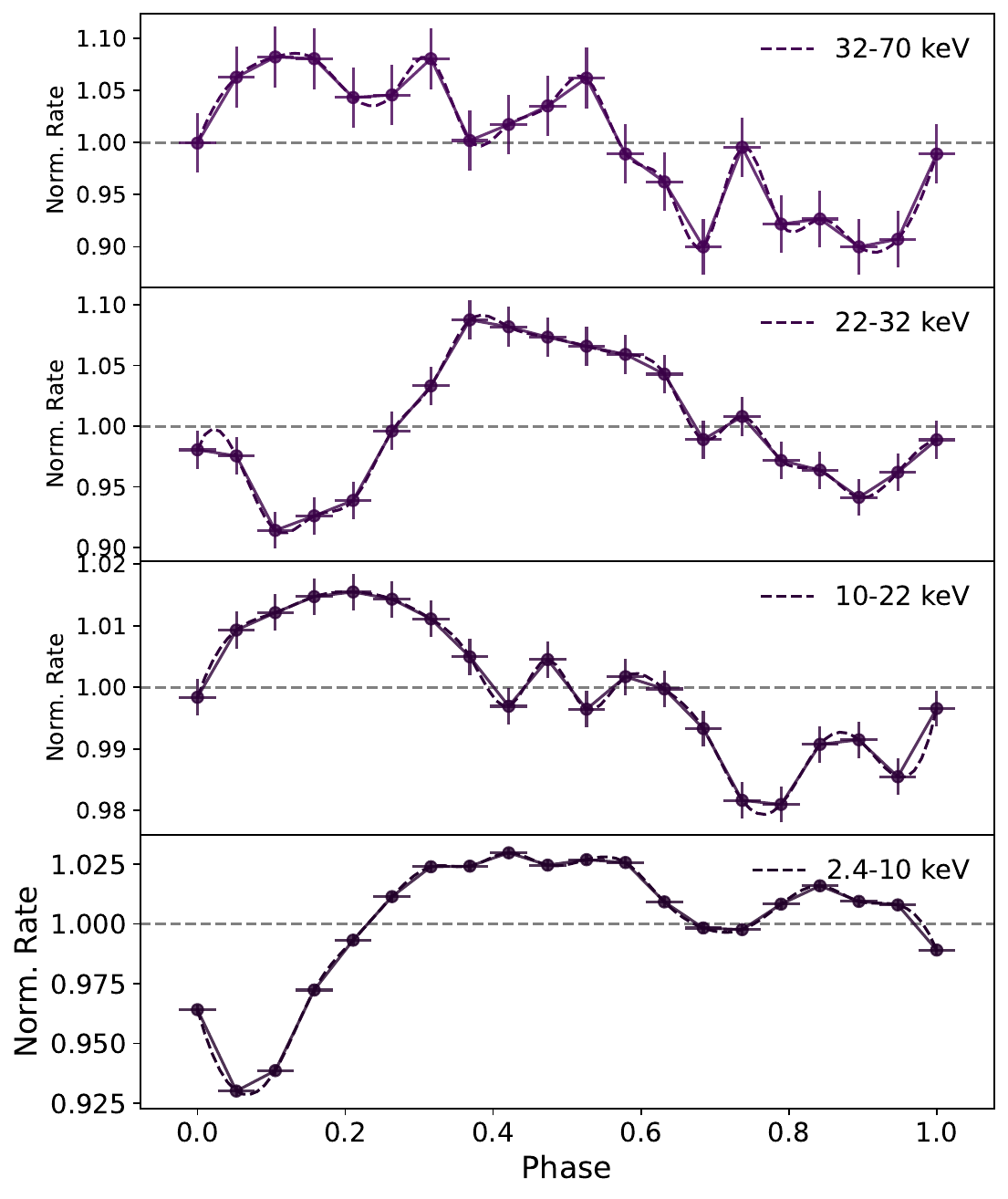} \\
\end{tabular}
\caption{Energy-phase colour maps and energy-selected pulse profiles 
for ObsIDs 802 and 804.}
\label{fig:pp802_pp804}
\end{figure*}

An energy-phase map provides a quick visual overview of the energy-phase matrix. For each array of phase bin values, we calculate the average and standard deviation. A normalized pulse profile is then created by subtracting this average and dividing by the standard deviation for each phase bin value. These maps are also useful for identifying broad energy bands in which the pulse profiles exhibit the most pronounced variations.
We show for example in the right panels of Fig. \ref{fig:pp802_pp804} two maps 
taken from the first two ObsID 802 and 804, 
which at once give an idea of the complex and strongly energy dependent
changes in the PP, especially around the cyclotron line energy, at \Ecyc $\sim$28 keV,
\citep[see e.g.][for a comparison where similar maps were 
extracted from $INTEGRAL$ data]{Tsygankov2006}. The magenta lines
show the energy ranges (2.4--10 keV, 10--22 keV, 22--32 keV, and 32--70 keV) from which some reference pulse profiles are extracted (right panels of Fig. \ref{fig:pp802_pp804}). Similar plots for the other observations examined
in this work are shown in the appendix \ref{appendix:pprofiles}. 
It is evident that the profiles vary significantly in their harmonic
content both as a function of energy and accretion state. Although the
range of luminosity encompasses likely accretion states above and below
the critical threshold, it is difficult to clearly identify a single component or pattern that might indicate a switch of the accretion mode.

\subsection{Modelling of the pulsed fraction spectra} \label{sect:fitting_method}

From a qualitative description of the pulse-energy dependence provided by the
energy phase maps, we moved on to a quantitative description of the PP changes
using the spectral decomposition of each profile.
To this end, the study of the PFS provided the 
most robust way of tracking such changes, although 
their description is still phenomenological.
Because we are mostly interested in accounting and 
describing the characteristics of the local features, we rely on a 
polynomial function to model the PF continuum as described in \citet{ferrigno2023}. 
To prevent the polynomial order to be too high and, thus,  
to avoid describing the shapes of local features with the polynomial continuum,
we split the fit into two energy bands: a low- and a high-energy one. 
The energy where we split the modelling (\esplit) is chosen 
in the 10--15 keV range (i.e. sufficiently distant from 
typical cyclotron line energies).
We use a \textit{spline} function to interpolate
the data points in this range and search for a zero in the period derivative. 
If such a value is found,  \esplit\ is set to the corresponding energy 
bin, otherwise \esplit\ is set to the energy value corresponding 
to the minimum of the first derivative of the interpolating function.
The polynomial order is chosen so that the fitted model has 
a corresponding $p$-value higher than 0.05.

In the low-energy band (typically between 3 and 12 keV), 
we do not find significant hints to features corresponding to the iron emission line
in any of the observations (see details in Sect.\ref{sect:pflowenergy}), so that a 
simple, $n$-order polynomial is sufficient to reach a statistically 
acceptable description of the PF spectrum.

In the high-energy part of the PFS, when  
local features are clearly spotted, we model them
using Gaussian profiles, where the line centroid, the 
width and amplitude are free fitting parameters, after the fit
seed  values are set by eye. We caution the reader to not implicitly 
assume that Gaussian components with positive, or negative, amplitudes are direct 
analogues of \textit{emission}, or \textit{absorption}, lines as in the energy spectra. 
Our description of the PFS remains purely phenomenological and 
any tempting matching at a specific energy close
to a spectral feature needs to be carefully examined case by case. 

We use the \texttt{lmfit} python package \citep{lmfit} for obtaining 
the best-fitting parameters and calculating the associated uncertainties
using Markov chain Monte Carlo (MCMC) algorithms from the \texttt{emcee} package \citep{emcee}.
We use the Goodman \& Weare algorithm \citep{Goodman2010}
with 50 walkers, a burning phase of 600 steps, and a length of 6000, 
to prevent auto-correlation effects.
From the sample, we extract the 50th, the 16th and the 84th percentile 
of the marginalised distributions as the best-fitting value 
and associated uncertainty. 
We report for each fit the reduced
chi-squared value (\redchisq) and the corresponding degrees of freedom (dof).

As shown in \citet{ferrigno2023}, the PF spectrum slightly depends on the method by which the PF 
is calculated, on the number of phase-bins for the folded profiles and on the minimum S/N adopted to  rebin the original PF spectrum (which 
is produced on a first instance at the instrumental spectral resolution) when needed. 
We found no general recipe which could be systematically applied, as much 
depends on the average value of the PF, on the source count rate and on the 
background level. We produced different PFS by varying all the possible 
criteria. For this source and the set of the observations that we analysed, 
we generally found that a S/N\,=\,4 and a number of phase-bins of 16 provided 
the best compromise between S/N and spectral resolution. In a few cases, 
for better visual clarity, we excluded from the fit a $N_{\textrm{excl}}$ number
of last bins, affected by larger uncertainties.
We present a complete log of the observations and the 
parameters of the PFS used in this work  in Table \ref{tab:observing_log}.

The same pipeline (model fitting and error evaluations) and criteria were 
applied to the  description of the energy-dependence of the amplitudes of the first two
harmonics (see details in Appendix \ref{appendix:harmonics}, best-fitting 
results and uncertainties in the $1^{st}$ and $2^{nd}$ labelled columns of Tables
~\ref{tab:fit_parameters1} and \ref{tab:fit_parameters2}). 

\begin{table*}
    \scriptsize
    \caption{Log of \nustar\ observations used in this work and 
    adopted parameters for the PF spectra. 
    }
    \label{tab:observing_log}
    \renewcommand{\arraystretch}{1.3}
    \centering
    \begin{tabular}{lccccr@{}lc}
        \hline
        \hline
       ObsID (shortcut) & Exposure time & Total counts (FPMA$+$FPMB)   & Flux\tablefootmark{a} & Luminosity\tablefootmark{b} & \multicolumn{2}{c}{$P_\mathrm{spin}$} & $S/N_{\textrm min}$/$N_{\textrm bins}$/$N_{\textrm excl.}$\tablefootmark{c}\\
        \hline
        & ks & 10$^{6}$ counts &  $10^{-9}\ergcmsec$ & 10$^{37}$ erg s$^{-1}$ & \multicolumn{2}{c}{s}& \\
       80102002002 (802) & 10.5 & 8.88   &  29.6  & 11 & $4.375737$ & $\pm{0.000016}$  & 4/10/0\\
        80102002004 (804) & 14.9 & 6.00  &  14.7  & 5.5 & $4.375775$ & $\pm{0.000015}$ & 4/16/0 \\
        80102002006 (806) & 17.0 & 4.45  &  9.96  & 3.7 & $4.374639$ & $\pm{0.000016}$  & 4/16/0 \\
        80102002008 (808) & 18.1 & 1.45  &  3.12  & 1.2 & $4.377008$ & $\pm{0.000008}$ & 4/16/0\\
        80102002010 (810) & 20.8 & 0.88  &  1.65  & 0.6 & $4.377090$ & $\pm{0.000014}$ & 3/16/4\\
        90202031002 (902) & 25.2 & 0.65  &  1.05  & 0.4 & $4.377252$ & $\pm{0.000014}$  & 3/16/8  \\
        90202031004 (904) & 25.0 & 0.52  &  0.85  & 0.3 & $4.377448$ & $\pm{0.000014}$  & 3/16/3 \\
        \hline
        \hline
    \end{tabular}
    \tablefoot{
    \tablefoottext{a}{Absorbed flux in the 3--70 keV.}
    \tablefoottext{b}{Assuming isotropic flux emission at a distance of 5.6 kpc. The bolometric correction is of the order of few percent and it is ignored.}
    \tablefoottext{c}{Minimum signal-to-noise ratio, number of phase bins in the pulse profile and number of excluded highest-energy bins in the PFS.}
   }
\end{table*}

\section{PFS fit results} 

\begin{figure*}
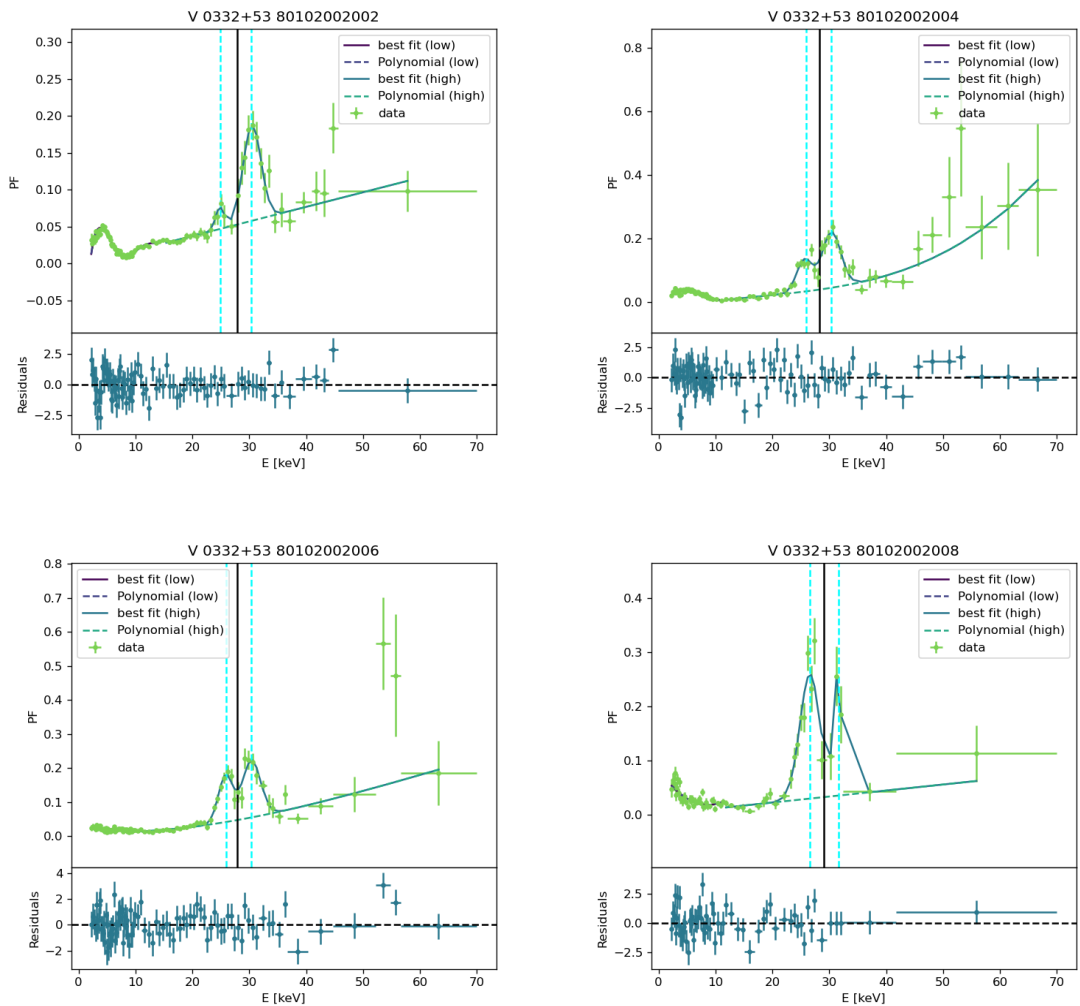

\centering
\begin{tabular}{cc}
\includegraphics[width=0.8\columnwidth]{figures2/V_0332+53_80102002002pulsed_fitted.pdf}  & 
\includegraphics[width=0.8\columnwidth]{figures2/V_0332+53_80102002004pulsed_fitted.pdf} \\
\includegraphics[width=0.8\columnwidth]{figures2/V_0332+53_80102002006pulsed_fitted.pdf} & 
\includegraphics[width=0.8\columnwidth]{figures2/V_0332+53_80102002008pulsed_fitted.pdf} \\
\end{tabular}
\caption{PFS, best-fitting models and residuals for ObsID 802, 804, 806 and 808.
Cyan dashed lines mark energy positions of the Gaussian peaks. The black line marks the energy of the fundamental cyclotron line \citep[from][]{Doroshenko2017}.}
\label{fig:pfspectra}
\end{figure*}

\begin{figure*}
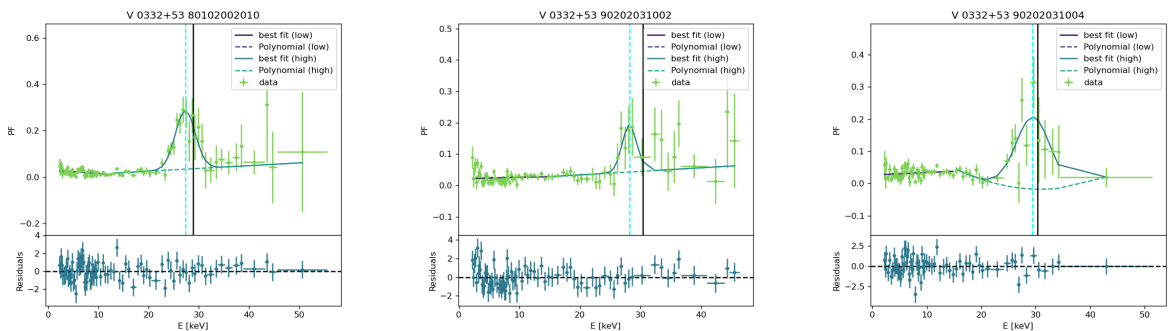

\centering
\begin{tabular}{ccc}
\includegraphics[width=5cm]{figures2/V_0332+53_80102002010pulsed_fitted.pdf}  & 
\includegraphics[width=5cm]{figures2/V_0332+53_90202031002pulsed_fitted.pdf}  &
\includegraphics[width=5cm]{figures2/V_0332+53_90202031004pulsed_fitted.pdf} \\
\end{tabular}
\caption{PFS, best-fitting models and residuals for ObsID 810, 902, and 904.
Cyan dashed lines mark best-fitting energy positions of the Gaussian peaks. 
The black line marks the energy of the fundamental cyclotron line \citep[from][]{Doroshenko2017}.}
\label{fig:pfspectra2}
\end{figure*}

We are able to satisfactorily describe, in statistical terms,  
the PFS using a very limited number of local components (either one or two) 
and polynomial functions of low order.
A general picture of all the PFS, the best-fitting models 
and the residuals in units of standard deviations are shown in Fig.~\ref{fig:pfspectra}
and in Fig.~\ref{fig:pfspectra2}. The first four observations
of Table~\ref{tab:observing_log} have sufficient 
statistics to allow a detailed view of the PF behaviour 
around the cyclotron line range (25--35 keV), 
while the remaining three observations have less statistics and 
do not allow a finer characterisation of the features. 
We report the best-fitting parameters and errors of the PFS,
in the PF column of
Tables~\ref{tab:fit_parameters1} and \ref{tab:fit_parameters2}.
\begin{table}
\caption{PFS best-fit parameters of the \vzero\ for observations 802, 804, 
806, and 808 (PF column). The columns $1^\mathrm{st}$ and $2^\mathrm{st}$ are the 
best-fit parameters and uncertainties for the energy-dependent amplitudes of the 
first and second harmonic. The CRSF parameters in the \textit{Spectral} column
are taken as reference from \citet{Doroshenko2017}.}
\label{tab:fit_parameters1}
\scalebox{0.75}{
\begin{minipage}{\textwidth}
\renewcommand{\arraystretch}{1.15}
\begin{tabular}{ lr@{}lr@{}lr@{}lr@{}ll}
\hline 
\hline
 & PF & & $1^\mathrm{st}$ & & $2^\mathrm{st}$ & & \multicolumn{2}{c}{Spectral}\\
\toprule
\multicolumn{9}{c}{802} \\
\midrule
$\chi^2_\mathrm{red,lo}$/dof &  1.4 &/48 &   1.4 &/45 &   1.2 &/50 & --&--  \\
$\chi^2_\mathrm{red,hi}$/dof &  0.8 &/28 &   0.8 &/31 &   1.4 &/31 & --&--  \\
$n_\mathrm{pol}^\mathrm{(hi)}$ &1 & & 1 & & 1 & & -- & --  \\
$n_\mathrm{pol}^\mathrm{(lo)}$ &5 & & 5 & & 3 & & -- & --  \\
$E_\mathrm{split}$(\text{keV}) &14.97 & & 13.47 & & 14.97 & & -- & -- \\
$A_\mathrm{1}$ &0.049 &$\pm$0.014 & 0.092 &$\pm$0.007 & -0.049 &$\pm$0.008 &--&-- \\
$E_\mathrm{1}$(\text{keV})  &25.1 &$_{-0.2}^{+0.4}$ & 24.88 &$\pm$0.07 & 24.81 &$_{-0.10}^{+0.14}$ &--&-- \\
$\sigma_\mathrm{1}$(\text{keV}) &0.94 &$\pm$0.18 & 0.90 &$\pm$0.06 & 0.87 &$_{-0.19}^{+0.14}$ &--&-- \\
$A_\mathrm{2}$ &0.52 &$\pm$0.04 & 0.58 &$\pm$0.02 & --&--&--&-- \\
$E_\mathrm{2}$(\text{keV})  &30.49 &$\pm$0.11 & 30.66 &$\pm$0.08 & --&--&--&--\\
$\sigma_\mathrm{2}$(\text{keV})  &1.63 &$_{-0.10}^{+0.12}$ & 2.11 &$\pm$0.08 & --&--&--&--\\
$D_\mathrm{cyc}$ & --  & -- & -- &--  & -- & -- &  0.863&$\pm$0.002 \\
$E_\mathrm{cyc}$ (\text{keV})  & --  & -- & -- &--  & -- & -- &  27.93&$\pm$0.02 \\
$\sigma_\mathrm{cyc}$ (\text{keV})  & --  & -- & -- &--  & -- & -- &  8.47&$\pm$0.08 \\
\midrule
\multicolumn{9}{c}{804} \\
\midrule
$\chi^2_\mathrm{red,lo}$/dof & 1.2   &/42 & 1.3   &/42 & 1.3   &/45 & -- & --  \\
$\chi^2_\mathrm{red,hi}$/dof & 1.7   &/36 & 1.7   &/36 & 1.4   &/38 & -- & --  \\
$n_\mathrm{pol}^\mathrm{(hi)}$  & 3     &    & 3     &    & 3     &    & -- & -- \\
$n_\mathrm{pol}^\mathrm{(lo)}$  & 3     &    & 3     &    & 1     &    & -- & -- \\
$E_\mathrm{split}$(\text{keV})  & 11.33 &    & 11.59 &    & 12.43 &    & -- & -- \\
$A_\mathrm{1}$ &0.36 &$\pm$0.04 & 0.172 &$\pm$0.012 & 0.138 &$\pm$0.014  & -- & --\\
$E_\mathrm{1}$(\text{keV})  &25.74 &$_{-0.15}^{+0.18}$ & 25.30 &$\pm$0.08 & 26.91 &$\pm$0.14  & -- & --\\
$\sigma_\mathrm{1}$(\text{keV}) &1.48 &$_{-0.13}^{+0.15}$ & 1.14 &$\pm$0.09 & 1.27 &$\pm$0.16 & -- & --\\
$A_\mathrm{2}$ &0.70 &$\pm$0.05 & 0.460 &$\pm$0.019 & -- & -- & -- & --\\
$E_\mathrm{2}$(\text{keV}) &30.44 &$\pm$0.10 & 30.33 &$\pm$0.06 & -- & -- & -- & -- \\
$\sigma_\mathrm{2}$  &1.63 &$\pm$0.12 & 1.41 &$\pm$0.06 & -- & --  & -- & --\\
$D_\mathrm{cyc}$ & --  & -- & -- &--  & -- & --        &0.861&$\pm$0.002 \\
$E_\mathrm{cyc}$ (\text{keV})  & --  & -- & -- &--  & -- & --       &28.25&$\pm$0.02 \\
$\sigma_\mathrm{cyc}$ (\text{keV})  & --  & -- & -- &--  & -- & --  &7.59&$\pm$0.08 \\
\midrule
\multicolumn{9}{c}{806} \\
\midrule
$\chi^2_\mathrm{red,lo}$/dof &  1.0 &/43 &   1.2 &/43 &   0.9 &/45 & -- & --  \\
$\chi^2_\mathrm{red,hi}$/dof &  1.2 &/36 &   1.9 &/36 &   1.3 &/33 & -- & --  \\
$n_\mathrm{pol}^\mathrm{(hi)}$ &3 & & 3 & & 3 & & -- & --  \\
$n_\mathrm{pol}^\mathrm{(lo)}$ &1 & & 1 & & 2 & & -- & --  \\
$E_\mathrm{split}$ (keV)&10.01 & & 10.01 & & 12.08 & & -- & -- \\
$A_\mathrm{1}$ &0.38 &$\pm$0.02 & 0.228 &$\pm$0.011 & 0.28 &$_{-0.10}^{+0.06}$ & 0.86 &  \\
$E_\mathrm{1}$ (keV) &25.78 &$\pm$0.08 & 25.81 &$\pm$0.06 & 27.4 &$_{-0.5}^{+0.3}$ & 27.93 & \\
$\sigma_\mathrm{1}$ (keV) &1.20 &$\pm$0.06 & 1.12 &$\pm$0.05 & 1.96 &$_{-0.22}^{+0.18}$ & 8.47 & \\
$A_\mathrm{2}$ &0.66 &$\pm$0.05 & 0.340 &$\pm$0.019 & 0.17 &$_{-0.05}^{+0.10}$ & -- & --  \\
$E_\mathrm{2}$ (keV) &30.28 &$\pm$0.11 & 30.56 &$\pm$0.07 & 31 &$_{-2}^{+2}$ & -- & --  \\
$\sigma_\mathrm{2}$ (keV) &1.68 &$\pm$0.14 & 1.16 &$\pm$0.06 & 2.0 &$_{-0.7}^{+0.5}$ & -- & -- \\
$D_\mathrm{cyc}$  & --  & -- & -- &--  & -- & -- &  0.862 &$\pm$0.002 \\
$E_\mathrm{cyc}$ (\text{keV})  & --  & -- & -- &--  & -- & -- & 28.40 &$\pm$0.02 \\
$\sigma_\mathrm{cyc}$ (\text{keV}) & --  & -- & -- &--  & -- & -- &  7.59 &$\pm$0.08 \\
\midrule
\multicolumn{9}{c}{808} \\
\midrule
$\chi^2_\mathrm{red,lo}$/dof &  1.5 &/36 &   0.7 &/38 &   2.0&/36 & -- & -- \\
$\chi^2_\mathrm{red,hi}$/dof &  1.6 &/17 &   1.3 &/16 &   1.5 &/15 & -- & -- \\
$n_\mathrm{pol}^\mathrm{(hi)}$ &1 & & 1 & & 1 & & -- & -- \\
$n_\mathrm{pol}^\mathrm{(lo)}$ &2 & & 1 & & 3 & & -- & -- \\
$E_\mathrm{split}$(\text{keV}) & 11.81 &  & 12.13 & & 12.40 & &  -- & --  \\
$A_\mathrm{1}$ &1.02 &$\pm$0.05 & 0.80 &$_{-0.15}^{+0.30}$&0.22 &$\pm$0.02 &  -- & -- \\
$E_\mathrm{1}$ (\text{keV})&26.62 &$\pm$0.09 & 27.2 &$_{-0.4}^{+0.7}$ & 25.95 &$\pm$0.20 & -- & -- \\
$\sigma_\mathrm{1}$ (\text{keV})&1.79 &$\pm$0.09 & 2.0 &$_{-0.3}^{+0.2}$ & 1.78 &$_{-0.11}^{+0.09}$ &  -- & -- \\
$A_\mathrm{2}$ &0.45 &$_{-0.07}^{+0.09}$ & -0.24 &$_{-0.15}^{+0.31}$ & 0.64 &$\pm$0.13 & -- & -- \\
$E_\mathrm{2}$(\text{keV}) &31.70 &$_{-0.18}^{+0.30}$ & 29.40 &$_{-0.13}^{+1.75}$ & 32.8 &$_{-0.4}^{+0.3}$& -- & -- \\
$\sigma_\mathrm{2}$(\text{keV}) &0.98 &$\pm$0.17 & 0.76 &$\pm$0.19 1.46 &$\pm$0.16 & -- & -- \\
$D_\mathrm{cyc}$  & --  & -- & -- &--  & -- & -- &  0.899 &$\pm$0.003 \\
$E_\mathrm{cyc}$ (\text{keV})  & --  & -- & -- &--  & -- & -- &  29.03 &$\pm$0.04 \\
$\sigma_\mathrm{cyc}$ (\text{keV})  & --  & -- & -- &--  & -- & -- & 6.73 &$\pm$0.08 \\
\bottomrule
\hline
\end{tabular}
\end{minipage}
}
\end{table}
\begin{table}  \label{tab:fit_parameters2}
\caption{PFS best-fit parameters of the \vzero\ for observations 810, 902, 
and 904 (PF column). The columns $1^\mathrm{st}$ and $2^\mathrm{st}$ are the 
best-fit parameters and uncertainties for the energy-dependent amplitudes of the 
first and second harmonic. The CRSF parameters in the \textit{Spectral} column
are taken as reference from \citet{Doroshenko2017}.}
\scalebox{0.75}{
\begin{minipage}{\textwidth}
\renewcommand{\arraystretch}{1.15}
\begin{tabular}{ lr@{}lr@{}lr@{}lr@{}ll}
\hline 
\hline
 & PF & & $1^\mathrm{st}$ & & $2^\mathrm{st}$ & & \multicolumn{2}{c}{Spectral}\\
\toprule
\multicolumn{9}{c}{810} \\
\midrule
$\chi^2_\mathrm{red,lo}$/dof &  1.0 &/48 &   0.8 &/47 &   0.7 &/47 & -- & -- \\
$\chi^2_\mathrm{red,hi}$/dof &  0.9 &/32 &   0.9 &/33 &   1.0 &/33 & -- & -- \\
$n_\mathrm{pol}^\mathrm{(hi)}$ &2 & & 2 & & 2 & & -- & -- \\
$n_\mathrm{pol}^\mathrm{(lo)}$ &1 & & 1 & & 1 & & -- & -- \\
$E_\mathrm{split}$(\text{keV}) &12.84 & & 12.26 & & 12.31 & & -- & --  \\
$A_\mathrm{1}$ (\text{keV})&1.31 &$\pm$0.09 & 0.81 &$\pm$0.05 & 0.61 &$\pm$0.05 & -- & --\\
$E_\mathrm{1}$ (\text{keV})&27.30 &$\pm$0.15 & 27.07 &$\pm$0.14 & 28.9 &$\pm$0.2 &--  & -- \\
$\sigma_\mathrm{1}$ (\text{keV})&2.11 &$\pm$0.12 & 2.04 &$\pm$0.11 & 2.69 &$\pm$0.19 & -- & -- \\

$D_\mathrm{cyc}$ (\text{keV})  & --  & -- & -- &--  & -- & -- &  0.895 &$\pm$0.004 \\
$E_\mathrm{cyc}$ (\text{keV})  & --  & -- & -- &--  & -- & -- &  28.90 &$\pm$0.05 \\
$\sigma_\mathrm{cyc}$ (\text{keV}) & --  & -- & -- &--  & -- & -- & 6.30 &$\pm$0.01 \\
\midrule
\multicolumn{9}{c}{902} \\
\midrule
$\chi^2_\mathrm{red,lo}$/dof &  1.5 &/26 &   1.0 &/28 &   1.5 &/23 & -- & -- \\
$\chi^2_\mathrm{red,hi}$/dof &  0.6 &/9 &   1.7 &/7 &   1.5 &/12 & -- & -- \\
$n_\mathrm{pol}^\mathrm{(hi)}$ &2 & & 2 & & 2 & & -- & -- \\
$n_\mathrm{pol}^\mathrm{(lo)}$ &1 & & 1 & & 1 & & -- & -- \\
$E_\mathrm{split}$ (\text{keV} ) &13.42 & & 14.99 & & 11.24 & & -- & -- \\
$A_\mathrm{1}$ (\text{keV}) &0.50 &$\pm$0.11 & 0.51 &$_{-0.07}^{+0.09}$ & 0.14 &$_{-0.14}^{+0.25}$ & --&-- \\
$E_\mathrm{1}$ (\text{keV})& 28.0 &$_{-0.3}^{+0.2}$ & 28.53 &$\pm$0.07 & 32 &$_{-4}^{+3}$ & -- & --\\
$\sigma_\mathrm{1}$ (\text{keV}) &1.13 &$_{-0.22}^{+0.18}$ & 1.07 &$\pm$0.15 & 0.6 &$\pm$0.3 & -- & -- \\
$D_\mathrm{cyc}$ (\text{keV})  & --  & -- & -- &--  & -- & -- &  0.925 &$\pm$0.004 \\
$E_\mathrm{cyc}$ (\text{keV})  & --  & -- & -- &--  & -- & -- & 30.42 &$\pm$0.05 \\
$\sigma_\mathrm{cyc}$ (\text{keV}) & --  & -- & -- &--  & -- & -- &  6.80 &$\pm$0.01 \\
\midrule
\multicolumn{9}{c}{904} \\
\midrule
$\chi^2_\mathrm{red,lo}$/dof &  1.3 &/48 &   1.2 &/42 &   1.4 &/42 & -- & -- \\
$\chi^2_\mathrm{red,hi}$/dof &  1.0 &/17 &   1.3 &/23 &   1.6 &/23 & -- & -- \\
$n_\mathrm{pol}^\mathrm{(hi)}$ &2 & & 2 & & 2 & & -- & -- \\
$n_\mathrm{pol}^\mathrm{(lo)}$ &1 & & 1 & & 1 & & -- & -- \\
$E_\mathrm{split}$ (\text{keV})&14.96 & & 11.74 & & 11.78 & & -- & --  \\
$A_\mathrm{1}$ (\text{keV})&3.0 &$\pm$1.1 & 0.53 &$\pm$0.05 & 1.1 &$_{-0.2}^{+0.3}$ & -- & --\\
$E_\mathrm{1}$ (\text{keV})&30.7 &$\pm$0.7 & 29.03 &$\pm$0.19 & 33 &$\pm$1 & -- & -- \\
$\sigma_\mathrm{1}$ (\text{keV})&4.2 &$\pm$0.5 & 1.8 &$\pm$0.2 & 4.9&$\pm$0.5 &-- &--  \\

$D_\mathrm{cyc}$ (\text{keV}) & --  & -- & -- &--  & -- & -- & 0.921&$\pm$0.004 \\
$E_\mathrm{cyc}$ (\text{keV})  & --  & -- & -- &--  & -- & -- &  30.28&$\pm$0.06 \\
$\sigma_\mathrm{cyc}$ (\text{keV})  & --  & -- & -- &--  & -- & -- &  6.90&$\pm$0.01 \\
\bottomrule
\hline
\end{tabular}
\end{minipage}
}
\tablefoot{
    $\chi^2_{\mathrm{red,lo}}$ and $\chi^2_{\mathrm{red},hi}$ are the reduced $\chi^2$ for the lower-
    and higher-energy sections, respectively, with
    $n_\mathrm{pol}^\mathrm{(lo)}$ and $n_\mathrm{pol}^\mathrm{(hi)}$ the corresponding polynomial orders. 
    $E_\mathrm{split}$ is the energy at which we separate the two regions.
	}
\end{table}

All the examined PFS share some common characteristics, 
which can be assumed to weakly depend on the accretion state: 
the PF values below \esplit\ are generally very low ($<$\,0.1), 
especially compared with other well-known accreting binaries 
\citep[see e.g.][]{ferrigno2023}, and tend to show a global minimum around 8--12 keV. 
After this minimum the PF starts to increase following a linear trend, 
which is then interrupted by the presence of a complex  structure that we decided 
to model as two Gaussians with positive amplitudes. 
Although the modelling of this complex feature is not unique 
(for instance, a similar good fit is  achieved by fitting with  
a broader Gaussian with positive amplitude and a narrower one with negative amplitude), 
the model we adopted appears to us better motivated both because PPs do significantly 
and rapidly change in this band (as we will show in Sect.\ref{sect:pprofiles})
and because this decomposition can be also
linked to a physical interpretation of the energy spectra 
(see Sect. \ref{sect:spectralanalysis}).

\begin{figure}
\centering
\includegraphics[width=0.9\columnwidth]{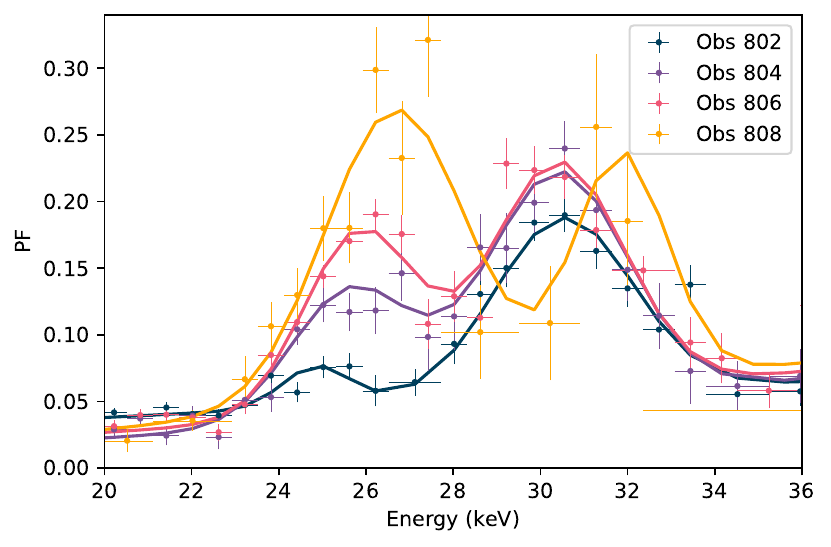} 
\caption{
PFS comparison for the first four observations in 
the 20--36 keV energy band. We over-impose the best-fitting models
from Table\ref{tab:fit_parameters1}}
\label{fig:peakcomparison}
\end{figure}

As shown in the four panels of Fig.~\ref{fig:pfspectra}, 
these two peaks show some variations 
in their relative amplitudes (see also Fig.~\ref{fig:peakcomparison}, 
where PF data and best-fitting models for the four ObsID are superimposed), 
though their positions and widths remain
always consistent within the uncertainties. 

\begin{figure}
\centering
\hspace*{-0.2cm}
\includegraphics[width=\columnwidth]{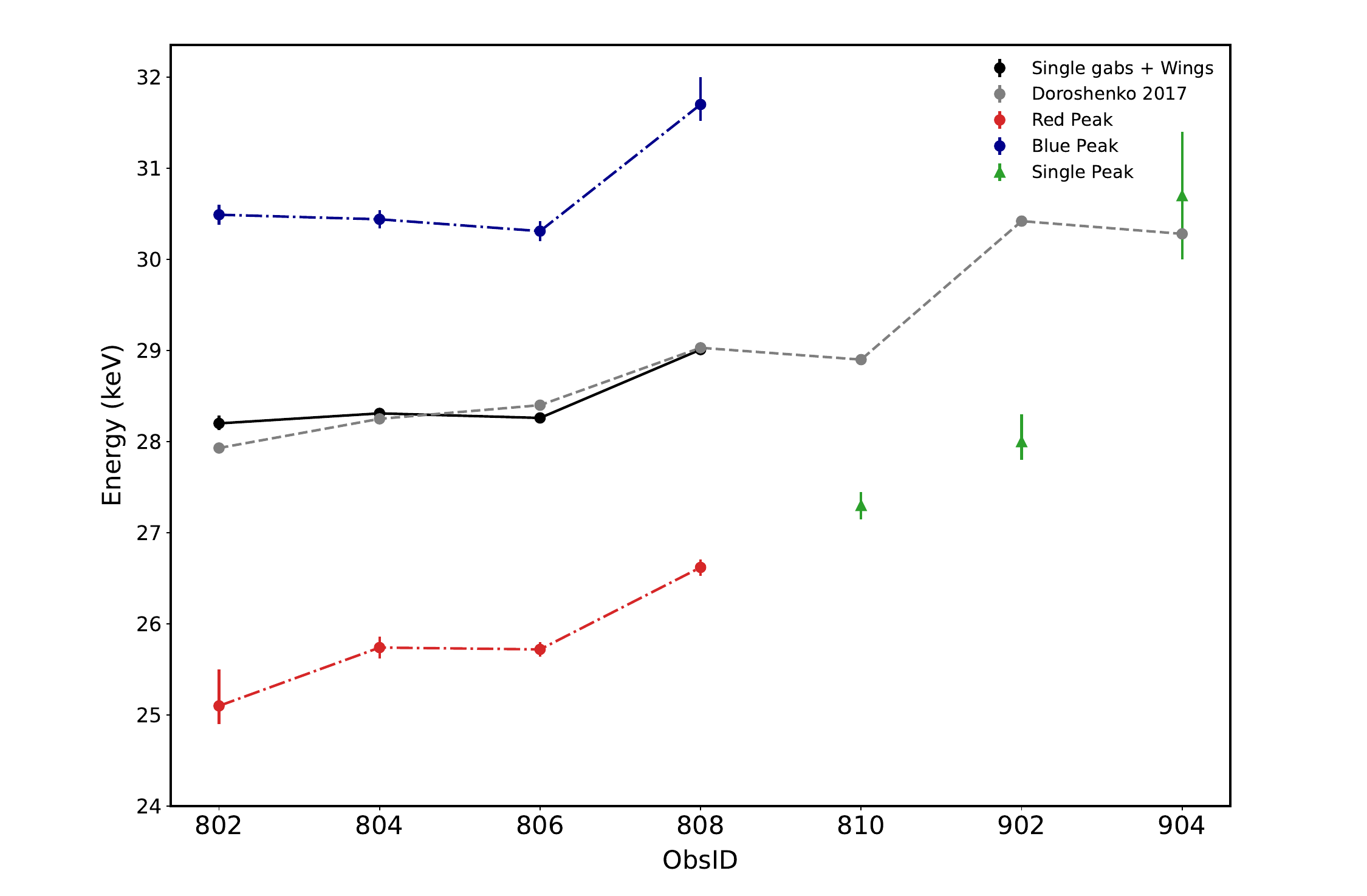}
\caption{
The best-fitting peak energies of the PFS features and the energies of the fundamental cyclotron line are shown. The values of the former are taken from the spectral analysis of  \citet{Doroshenko2017} (grey dots) and this work (black dots).  Red, blue and green points sign the best-fitting values from the PFS features, 
when two lines are clearly identified (red and blue dots) and when only one 
peak has been detected (green dots) (from Tables~\ref{tab:fit_parameters1} and \ref{tab:fit_parameters2}).}
\label{fig:obsid_energy}
\end{figure}

In the first four observations the positions of the two features  
do not display significant variations and their distance, 
relative to the spectral position of the fundamental cyclotron 
line \citep{Doroshenko2017}, remains constant as shown in Fig.~\ref{fig:obsid_energy}.
By reference to the peak position with respect to the cyclotron position,
we shall refer to these PFS features as the red and blue 
(also left and right) peaks.
 
After the blue peak,  the PF values clearly increase again
for the two brightest observations (802 and 804), while for all the 
remaining observations a lack of good statistics prevents us 
to determine any clear pattern. 

The three observations at lower statistics (810, 902, and 904) in Fig.~\ref{fig:pfspectra2} 
still confirm the significant increase in PF around the cyclotron energy range, though we can only 
constrain the presence of a single peak (though some hints for a still more complex, 
double peaked profile, is suggested by the fact that at the best-fitting Gaussian 
peak energy, one, or more, data points remain always below the best-fitting model).
The peak is clearly below the expected cyclotron line energy in the first
two observations, while it is compatible just in ObsID 904. 
In this latter case the peak position error is larger and we guess a blending of the two features, if their amplitudes become comparable, is occurring. 
To have a comprehensive picture of all these results, we report all the peak positions for each ObsID in comparison 
with the spectral cyclotron line values in Fig.~\ref{fig:obsid_energy}. 
\subsection{Changes in the pulse profile around $E_{\textrm cyc}$}  \label{sect:pprofiles}
\begin{figure*}
\centering
\begin{tabular}{cc}
\vspace{-0.5cm}
\hspace{-1.0cm}
\includegraphics[width=0.8\columnwidth]{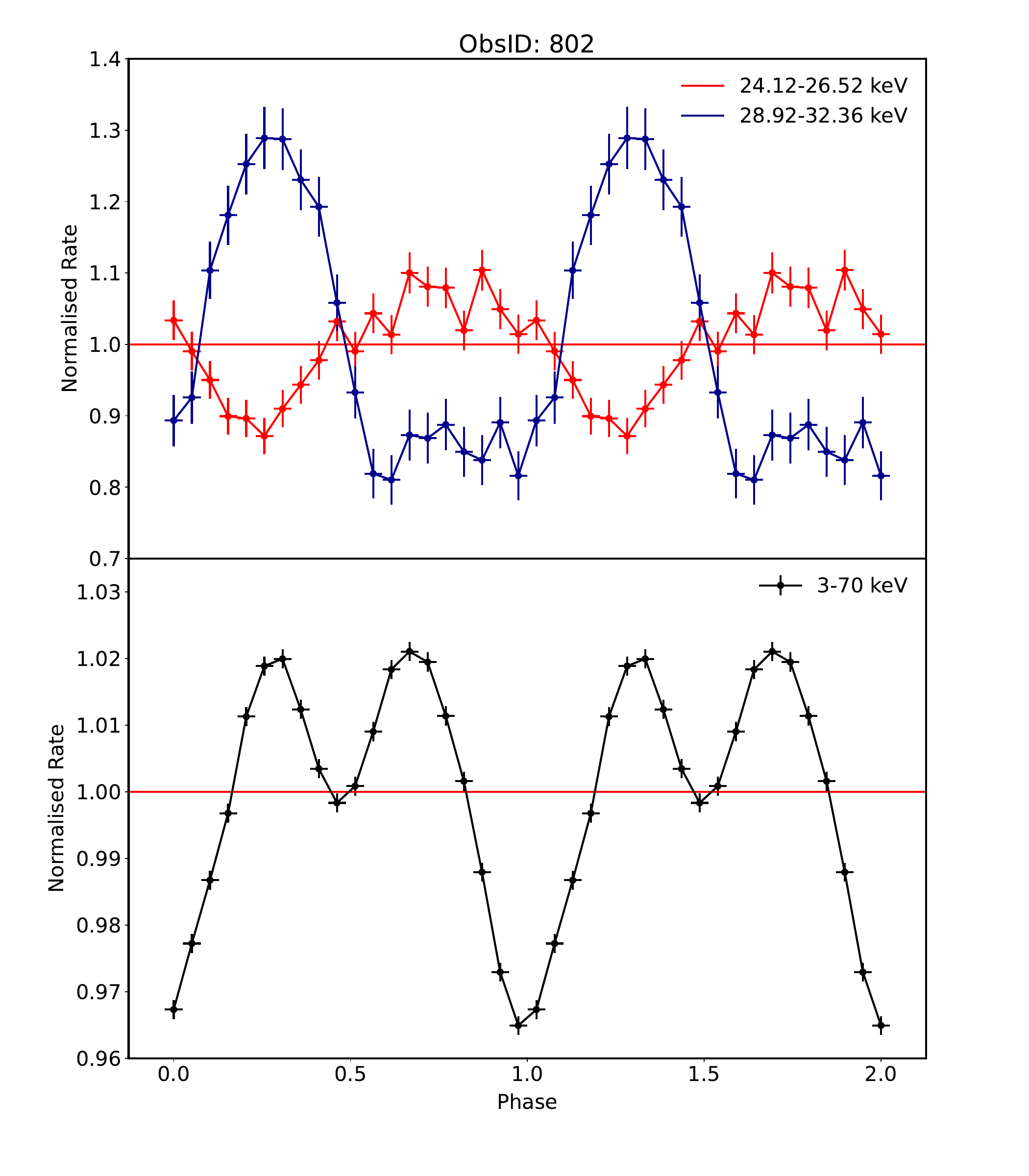}  & 
\hspace{-1.0cm}
\includegraphics[width=0.8\columnwidth]{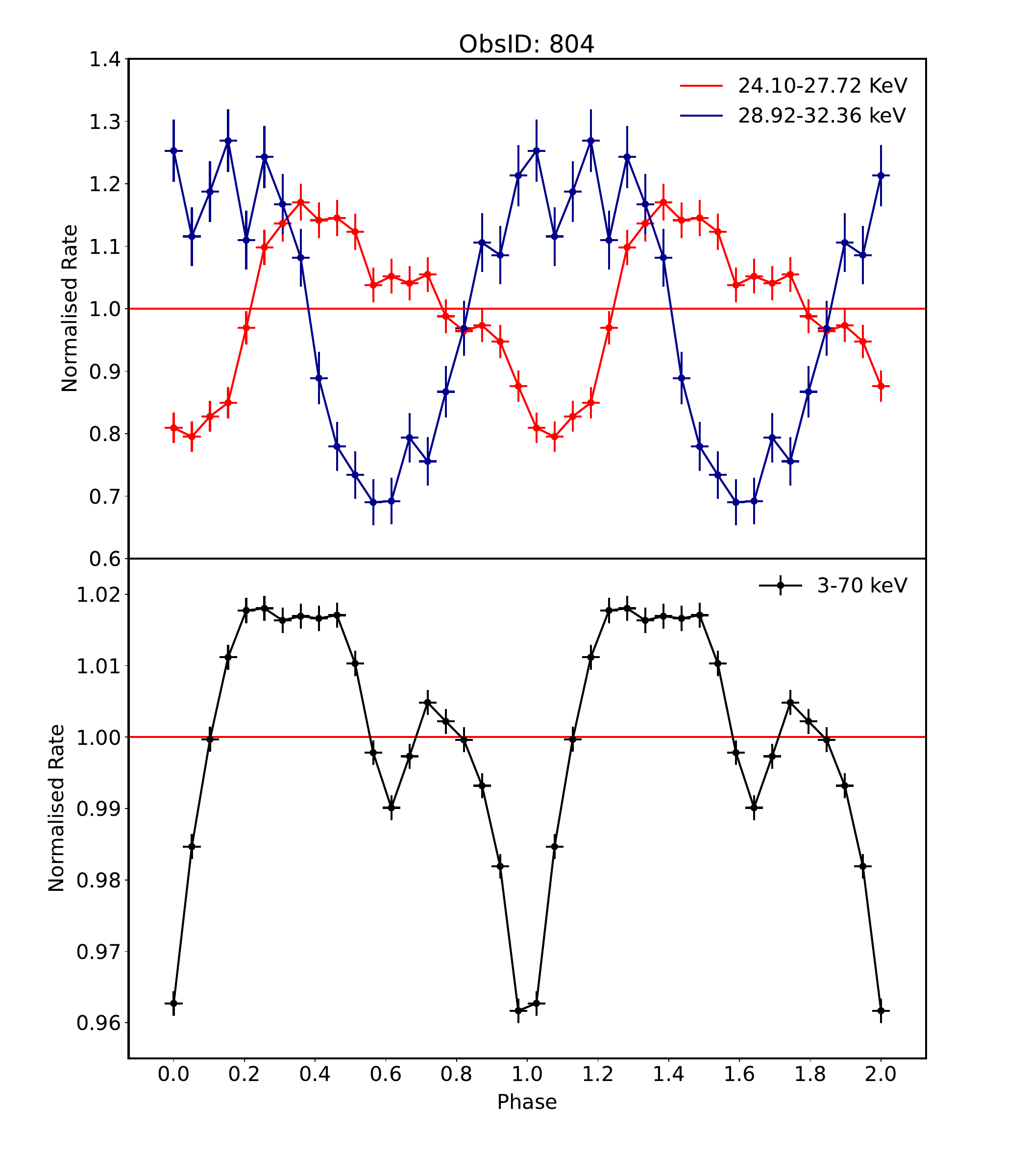} \\
\vspace{-0.5cm}
\hspace{-1.0cm}
\includegraphics[width=0.8\columnwidth]{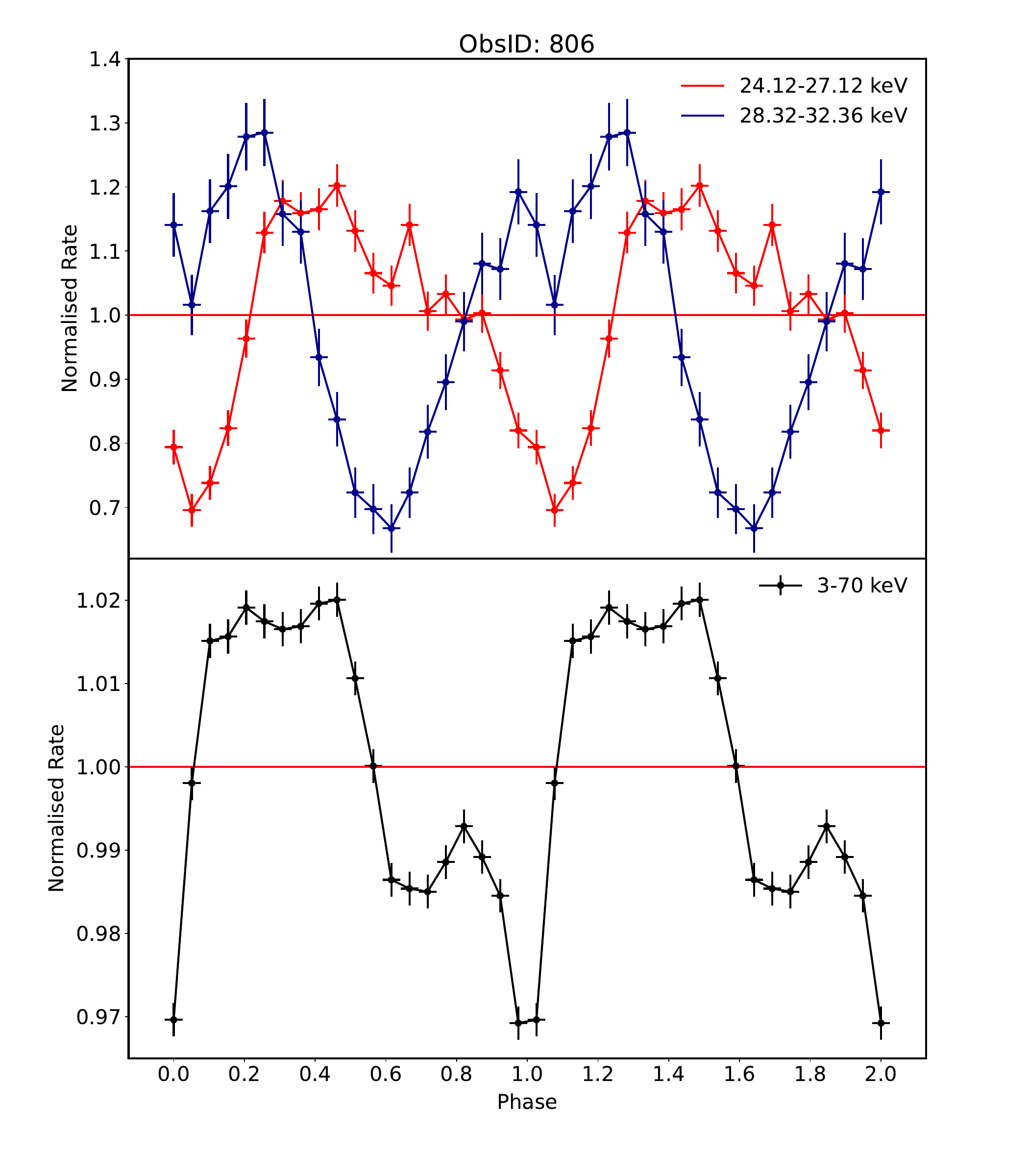} & 
\hspace{-1.0cm}
\includegraphics[width=0.8\columnwidth]{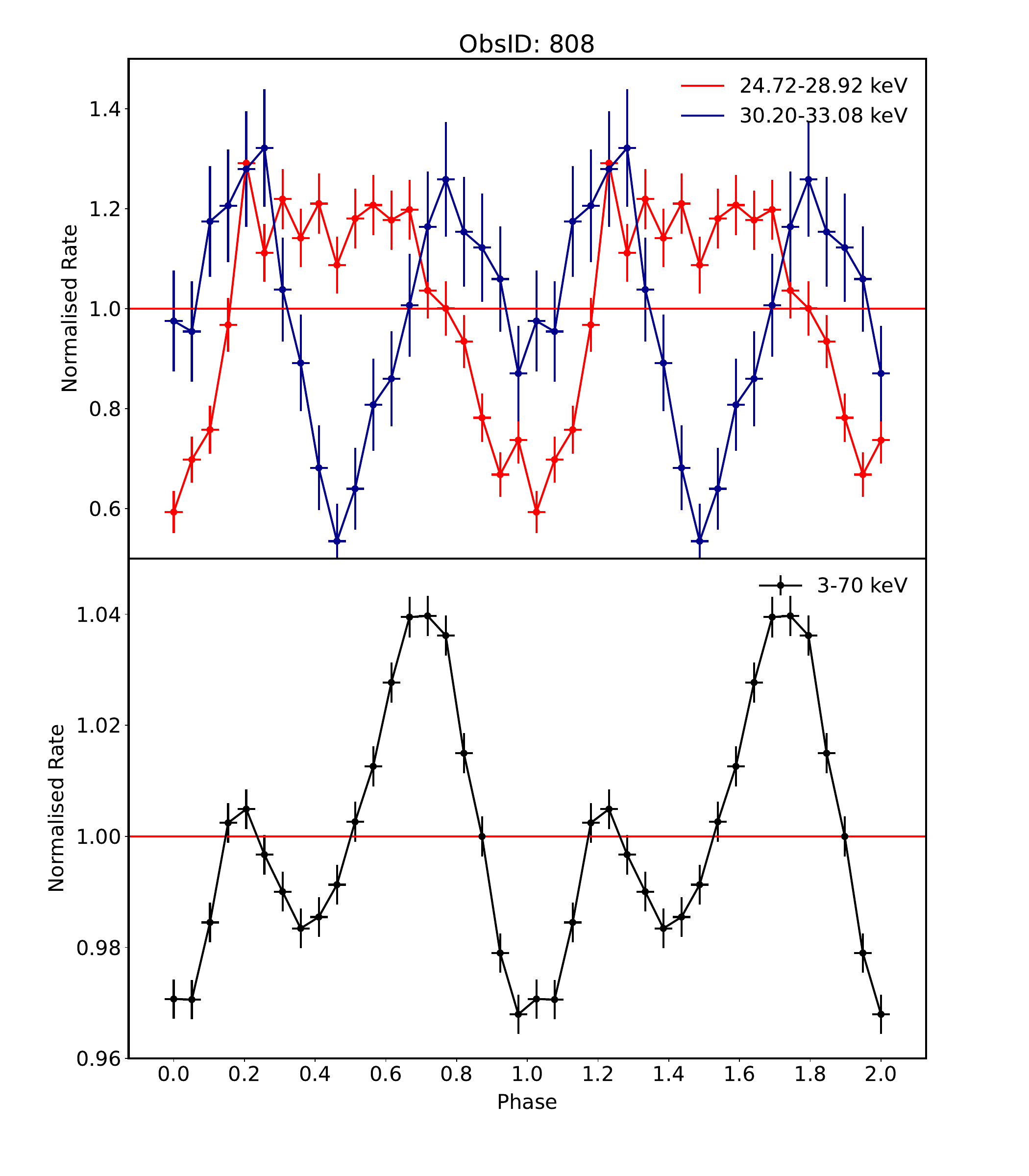} \\
\end{tabular}
\caption{Normalised energy-resolved pulse profiles for the ObsID 
802, 804, 806 and 808. Upper panels show energy-selected profiles 
around the two PF peaks, lower panels show the full 3--70 keV 
energy-averaged PP.}
\label{fig:enselectedpp}
\end{figure*}

We examine here the changes in PPs
for selected energy ranges as derived from the modelling of the PFS. 
We note that a detailed description of the pulse shape 
requires a good number of phase bins (typically more than 10), which, however, 
might result in noisy profiles if the profiles are selected from too narrow energy intervals.
To have enough statistics, but still preserving our objective to best describe 
the profile changes at the characteristic energies of the 
PF features, for each observation we extracted the PP
using the best-fitting parameters in Table \ref{tab:fit_parameters1}. 
For each clearly identified Gaussian line centred at $E_{\textrm{peak}}$, 
we selected the energy  range from $E_{\textrm{peak}}-\sigma$ and $E_{\textrm{peak}}+\sigma$.
We show the results of the four ObsIDs where two peaks are present 
in Fig.\ref{fig:enselectedpp}, where we superimposed in the upper panels
the two energy-selected profiles (red/blue colours show PP
extracted from the red/blue peak energy ranges).

The red peak profile is characterised by a variable broad peak at phase $\sim$\,0.5. 
The peak broadens and its amplitude significantly increases as the luminosity decreases. 
The PP of the blue peak is more variable with the source luminosity, with its maxima and minima generally in  phase opposition with those of the red peak. 
It is single peaked in the first two
observations, there is a hint of a double peak in ObsID 806 
and it becomes clearly double peaked in ObsID 808. 

\subsection{Cross and lag spectra around $E_{\textrm cyc}$}  \label{sect:crossandlag}

\begin{figure*}[h]
\centering
\begin{tabular}{cc}
\includegraphics[width=0.8\columnwidth]{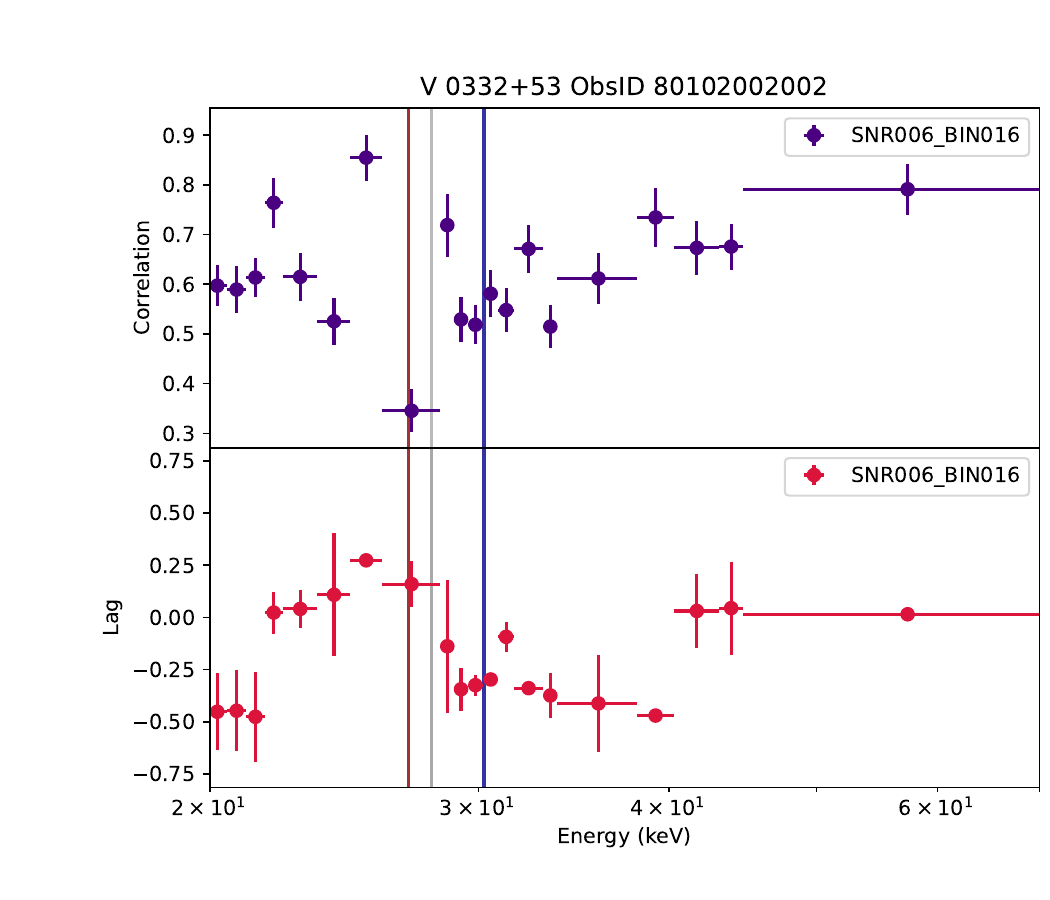}  & 
\includegraphics[width=0.8\columnwidth]{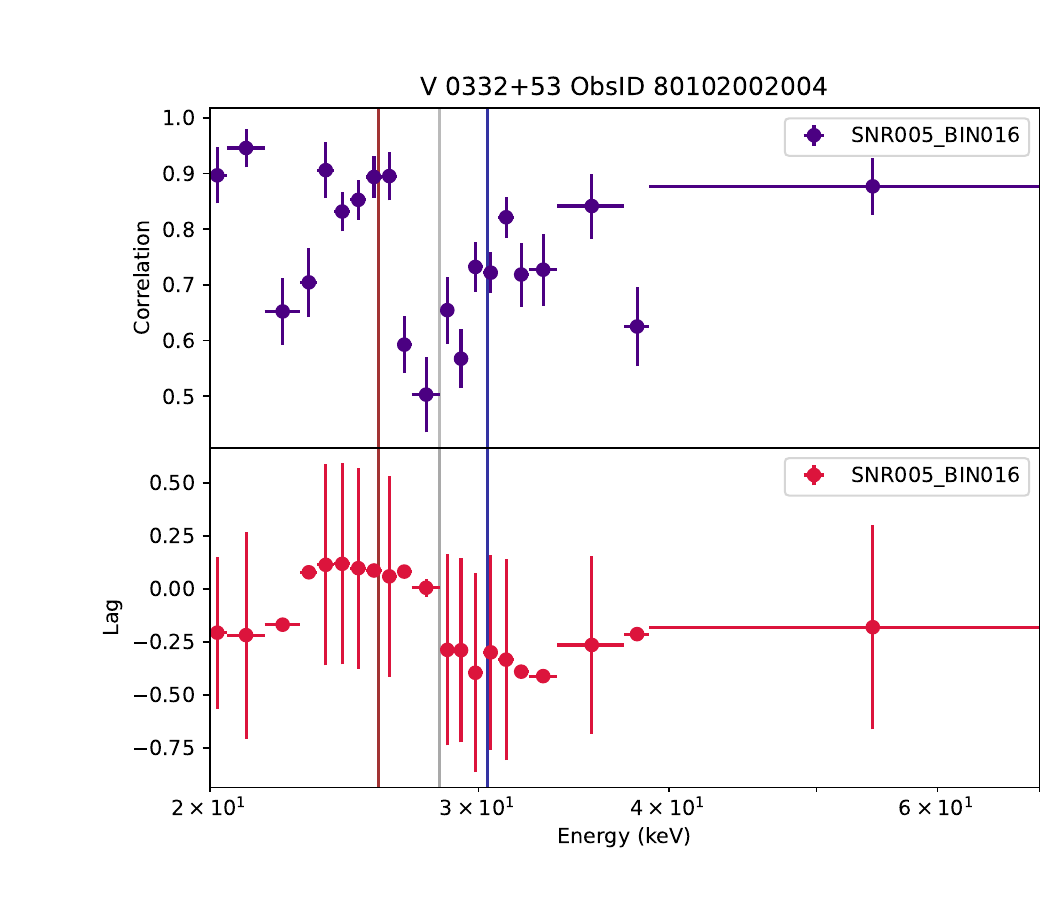} \\
\includegraphics[width=0.8\columnwidth]{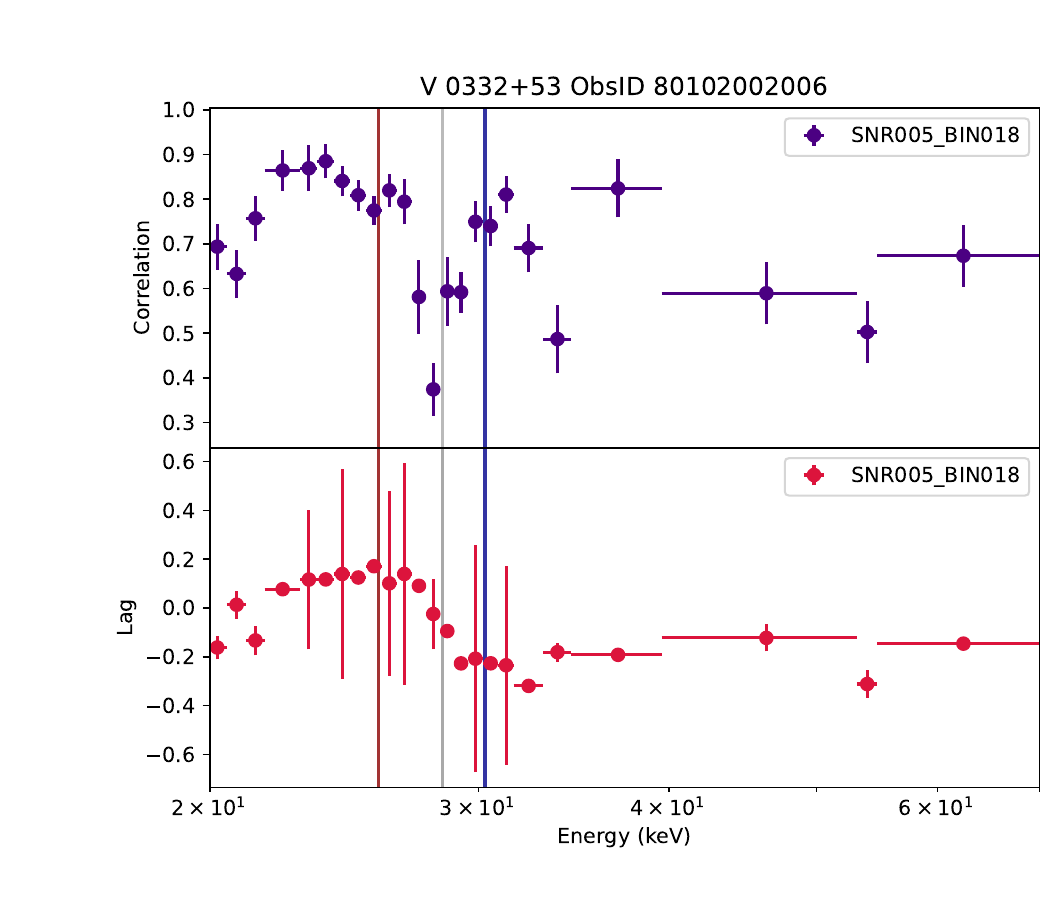} & 
\includegraphics[width=0.8\columnwidth]{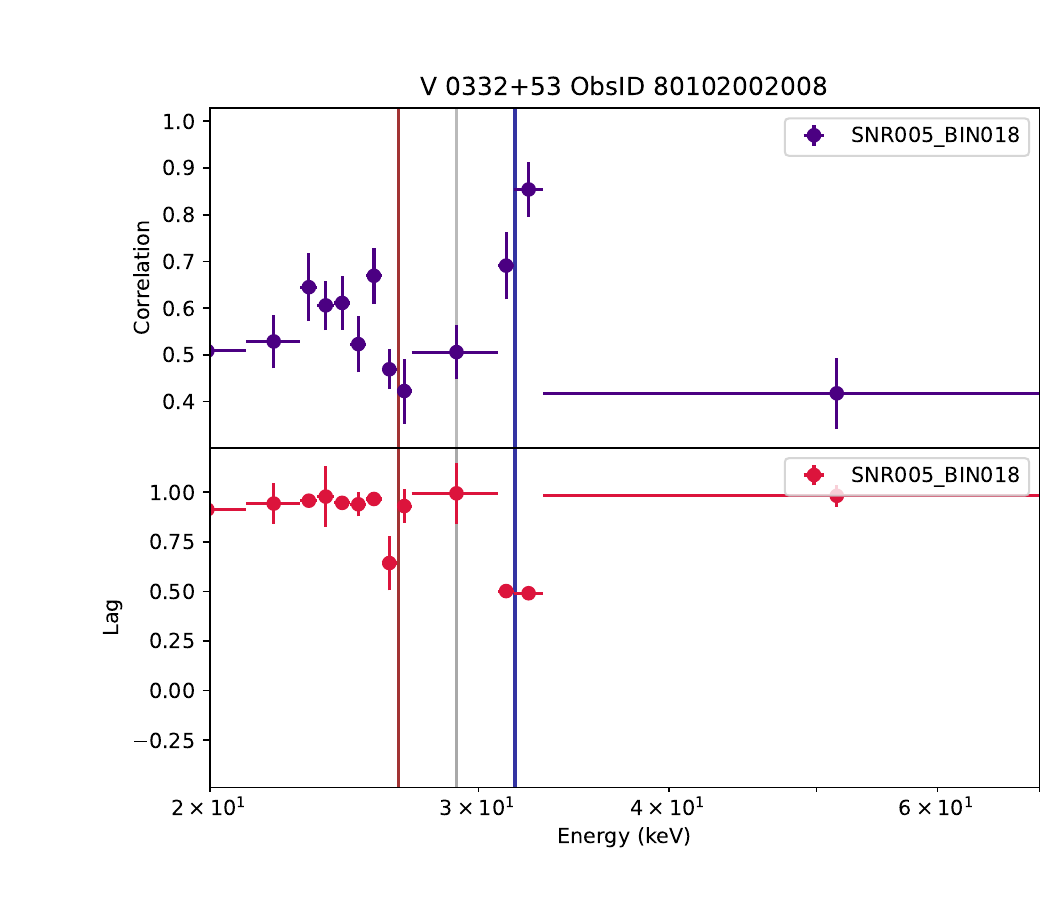} \\
\end{tabular}
\caption{Cross-correlation and lag spectra for ObsID 802, 804, 806 and 808.
The spectra are shown in the high-energy (20--70 keV) range for 
highlighting the behaviour around the cyclotron line energies. 
The brown and blue vertical lines mark the positions of the best-fit 
Gaussians of the PFS. The vertical grey line mark the spectral best-fit
position of the cyclotron absorption line \citep{Doroshenko2016}. Lags are given 
in phase units.}
\label{fig:corrlagspectra}
\end{figure*}

We computed the cross-correlation (CC, hereafter) and lag spectra for the 
first four observations (ObsID 802, 804, 806 and 808).  
To improve the quality of the spectra, we used PP
with 16 phase-bins and a S/N of 6. 
We comprehensively show them in Fig.\ref{fig:corrlagspectra}.
As explained in \citet{ferrigno2023}, we computed the error-bars on these points
by Monte-Carlo simulating $N$ faked PP with the same Poisson statistics
and taking the standard deviation from each set of simulations.
We noted that some points were affected by larger uncertainties with 
respect to adjacent bins, due to the presence of some outliers in the
distribution. To avoid this condition, we use here $N$\,=\,125 faked PP
and a trimmed standard deviation, where we rejected the left and right tails
of the MC distribution at a 5\% level. 
This choice drastically improved the 
quality of the spectra by significantly lowering the 
uncertainties (generally more than 50\% in CC spectra, and 
a factor 3-4 for the lag spectra) and their interpretation. 

\subsection{Pulse profile changes in the low-energy band and at the iron line} \label{sect:pflowenergy}

\begin{figure}[!ht]
\centering
\begin{tabular}{cc}
\includegraphics[width=0.9\columnwidth]{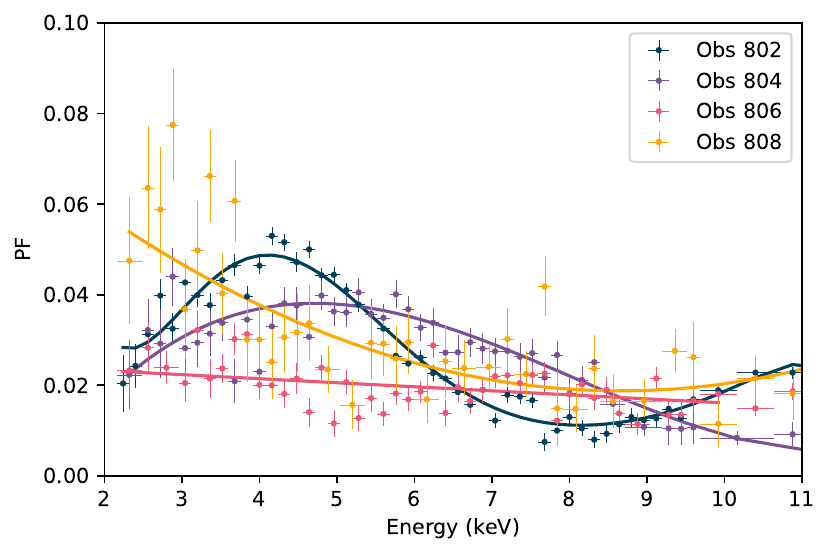} 
\end{tabular}
\caption{Comparative PF spectra in the 2---11 keV band for 
the first four observations of Table~\ref{tab:observing_log}.}
\label{fig:pflowenergy}
\end{figure}

\begin{figure}[!ht]
    \centering
    \includegraphics[width=0.8\columnwidth]{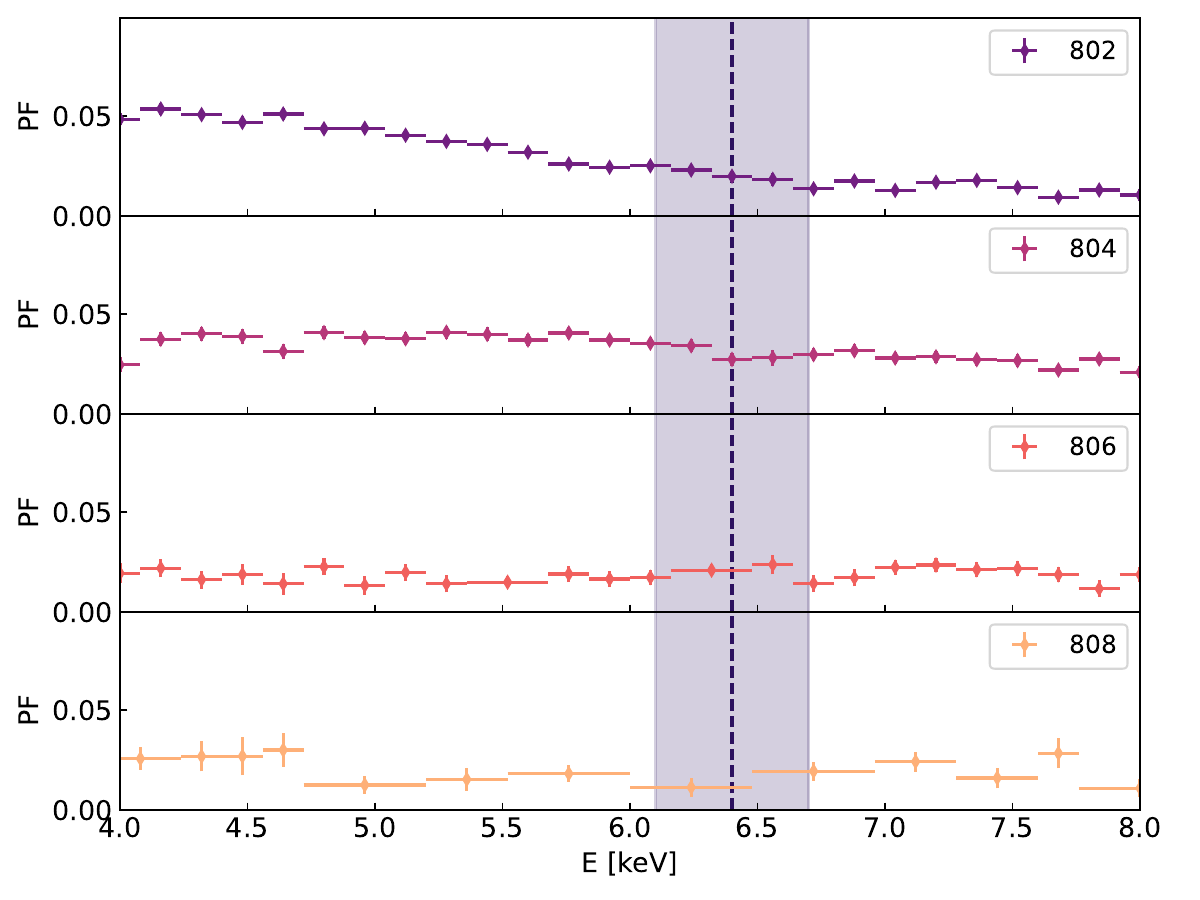}
    \includegraphics[width=0.8\columnwidth]{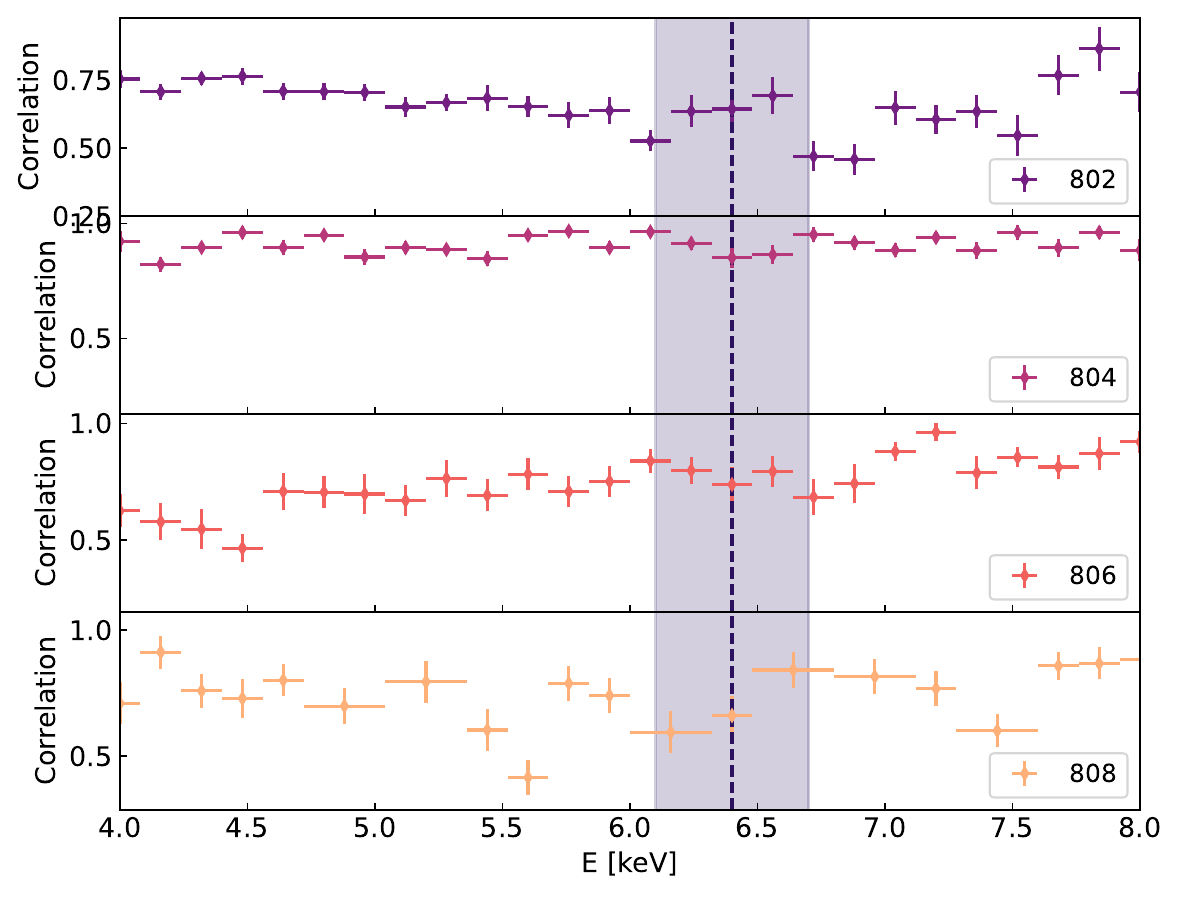}
    \caption{PF values (upper panel) and CC spectra (lower panel) for ObsID 802,804,806 and 808 in the
    4--9 keV energy range. The position and width of the spectral emission iron line is marked by 
    the dotted line and the purple-coloured area (1-sigma contour), respectively.} 
    \label{fig:pf_iron}
\end{figure}

Below 12 keV the PFS are quite peculiar
if compared to the sources analysed in \citet{ferrigno2023}.
In Fig.~\ref{fig:pflowenergy} we show the representative PFS for 
observations 802, 804, 806 and 808, which have the best statistics to search 
for local substructures. 
With the exception of 806, whose PF is flat in the entire low energy band,
the other three spectra show a local peak between 3 and 6 keV. 
We expected to find a drop in the PF at energies around $\sim$ 6.4 keV, 
similarly to what observed in PF spectra of the other sources
with relatively broad ($\sigma \sim$ 0.4 keV) iron emission lines \citep{ferrigno2023}.
However, we found no evidence of a drop in PF values 
as clearly shown from the PFS shown in upper panels of Fig.~\ref{fig:pf_iron}. 
Similarly, the CC spectra (bottom panel of Fig.~\ref{fig:pf_iron}) 
do not show any significant change in the pulse profiles. 

\citep{Bykov2021} reported that the iron emission line of \vzero\ pulsates, based on an analysis of its 2004-2005 outburst observed with Rossi XTE. This suggests that part of the emission likely originates in the neutron star's magnetosphere. Our findings align with this claim, as the iron line photons appear to be emitted coherently. The optical depth of the fluorescence process path is consistent with the main pulsation, implying that the reflector is either close to the primary hard emission source or that the pulsation is sufficiently long to maintain coherence or both. This may explain the lack of a significant dip in the PFS at these energies. However, a detailed analysis of the relationship between pulse profiles, shapes, and the intensity of the frequently observed iron fluorescence line in this and similar sources is beyond the scope of this paper and will be addressed in a future study.
\section{Spectral Analysis} \label{sect:spectralanalysis}
\citet{Doroshenko2016} and \citet{Vybornov2017}
have reported on the spectral analysis of these \nustar\ observations. 
The former performed phase-averaged spectroscopy, while the latter 
analysed pulse-resolved spectra. For this work, we do not aim to
perform a new detailed spectral analysis of these datasets; instead, we
investigate if, and how, the features we identified
in the PFS might have possible spectral counterparts. 
We will limit this analysis to the first four ObsID of Table
\ref{tab:observing_log} because of their higher statistics.
First, we use the tools of the phase-resolved analysis  
in Sect.\ref{sect:phaseresolved} to investigate 
the spectral distribution of the pulsed photons around
the peaks of the energy-selected PPs.
Because the PF is a measure of the ratio of the pulsed to the total
emission (at a particular energy bin), 
this step will provide evidence that the shape observed in the
PFS is linked to excess
photons at the pulsed peak and their energy distributions
are compatible with what is observed in the 
corresponding PFS. 
Moreover, this will also allow us to
estimate the number of counts responsible for the
observed peaks in the PFS.

In Sect. \ref{sect:spectralmodel} we will test a set of possible spectral 
decomposition of the phase-averaged spectrum using some  
constraints derived from the PFS modelling and the phase-resolved spectroscopy. 
The results of these steps will support a coherent interpretation of all our 
results linking the features in the PFS with spectral residuals around the 
cyclotron line energy. 

\subsection{Phase-resolved spectral characterisation 
of the peak excesses} \label{sect:phaseresolved}

To obtain a first-order estimate of the emission shape responsible 
of the pulse peak, we consider the PPs shown in Fig.~\ref{fig:enselectedpp}.
For each PP we visually identified the phase 
bin with the highest amplitude value, we selected a continuous phase interval expanding to the neighbour phase bins with largest amplitudes 
for a total of 4 phase bins. From these bin intervals 
we derived a phase range for extracting the \textit{peak pulse spectrum}. 
We then selected the bin with the lowest amplitude value and 
derived the phase interval using the same logic as for the 
\textit{peak pulse spectrum} to extract a corresponding 
\textit{background pulse spectrum}. 
We choose this fixed range of phase intervals 
because the peak phase and the 
bottom part of the PP are thus uniformly covered 
for all the PPs examined. 

\begin{table*}
    \begin{center}
    \caption{Comparison of best-fitting values for the red and 
    blue peaks of the PF spectra versus phase-resolved analysis of counts excesses.}
    \label{tab:timing_vs_resolved}
    \renewcommand{\arraystretch}{0.7}
    \resizebox{\textwidth}{!}{
\begin{tabular}{cccccccc|r}

        \toprule
    
         \multicolumn{1}{c}{} & \multicolumn{3}{c}{PF Analysis} & \multicolumn{4}{c}{Excess Counts Analysis} & Spectral Analysis \\
            
            \midrule
            
             ObsID & Energy (keV) & Sigma (keV) & \# of counts & Energy (keV) & Sigma (keV) & \# of counts  & $\chi^2 / \mathrm{dof}$  & \# of counts \\
            \toprule
            \multicolumn{9}{c}{Red Peak} \\
            \toprule
            
            802 & 25.10 $^{+0.40}_{-0.20}$ & $0.94_{-0.18}^{+0.18}$   & $710\pm210$& $24.64_{-0.25}^{+0.24}$ & $1.08_{-0.24}^{+0.19}$ & $1500\pm270$ & 11.15 / 15 & 3600$\pm$600\\
               
            804 &  $25.79_{-0.12}^{+0.12}$ & $1.43_{-0.11}^{+0.11}$ & $3170\pm350$& $25.13_{-0.25}^{+0.27}$ & $1.72_{-0.25}^{+0.27}$ & $2500_{-270}^{+360}$ & 20.72 / 15 & 8200$\pm$360 \\
           
            806 & $25.78_{-0.08}^{+0.08}$ & $1.20_{-0.06}^{+0.06}$ & $2410\pm190$& $25.04_{-0.22}^{+0.20}$ & $1.64_{-0.21}^{+0.27}$ &$2650_{-270}^{+330}$ & 10.08 / 15 & 7200$\pm$1000\\

            808 &  $26.62_{-0.09}^{+0.09}$ &  $1.79_{-0.09}^{+0.09}$ & $2050\pm100$ &$25.00_{-0.63}^{+0.43}$ & $2.22_{-0.47}^{+0.94}$ &$1350_{-270}^{+450}$ & 18.09 / 15 & $<$1000\\
            \bottomrule
            \toprule
            \multicolumn{9}{c}{Blue Peak} \\                          
            \toprule
            802 & $30.49^{+0.11}_{-0.11}$ & 1.63 $^{+0.12}_{-0.10}$ & $2000\pm160$ &$31.40_{-0.38}^{+0.36}$ & $2.40_{-0.38}^{+0.56}$ &$2160_{-270}^{+450}$ & 22.34 / 21 & 5000$\pm$800\\
               
            804 &  $30.43_{-0.09}^{+0.09}$ & $1.44_{-0.08}^{+0.08}$  &$2110\pm150$& $30.71_{-0.38}^{+0.33}$ & $2.05_{-0.35}^{+0.39}$ &$1530\pm270$ & 20.86 / 21 & 3600$\pm$270 \\
           
            806 & $30.28_{-0.11}^{+0.11}$ & $1.68_{-0.14}^{+0.14}$ & $960\pm60$ &$30.68_{-0.33}^{+0.29}$ & $1.92_{-0.33}^{+0.47}$ &$1440_{-180}^{+450}$ & 13.74 / 21 & 2000$\pm$270 \\

            808 &  $31.70^{+0.30}_{-0.18}$ & $0.98_{-0.17}^{+0.17}$ & $200\pm40$ &$32.90_{-0.43}^{+0.55}$ & $1.63_{-0.41}^{+0.60}$ & $400_{-100}^{+100} $ & 17.03 / 21 & 630$\pm$200\\
            \bottomrule
          
        \end{tabular}}
       
    \end{center} 
\end{table*}

We used \texttt{xspec} and $\chi^2$ statistics to fit 
the \textit{peak pulse spectrum}
using the \textit{background pulse spectrum} as background and
the \textit{addspec} tool\footnote{\url{https://heasarc.gsfc.nasa.gov/ftools/caldb/help/addspec.txt}} 
to sum the FPMA and FPMB spectra and corresponding 
background and response files. 

We noted positive excesses around the expected energies of the 
Gaussian peaks of the PF spectra. We fitted them using a 
simple Gaussian model in the restricted energy bands 22--28 keV and 
28--36 keV for the red and blue peaks, respectively.
In all cases, we found satisfactory reduced $\chi^2$ for all fits performed.

\begin{figure*}[!ht]
\centering
\begin{tabular}{cc}
\vspace{-0.5cm}
\includegraphics[width=0.8\columnwidth]{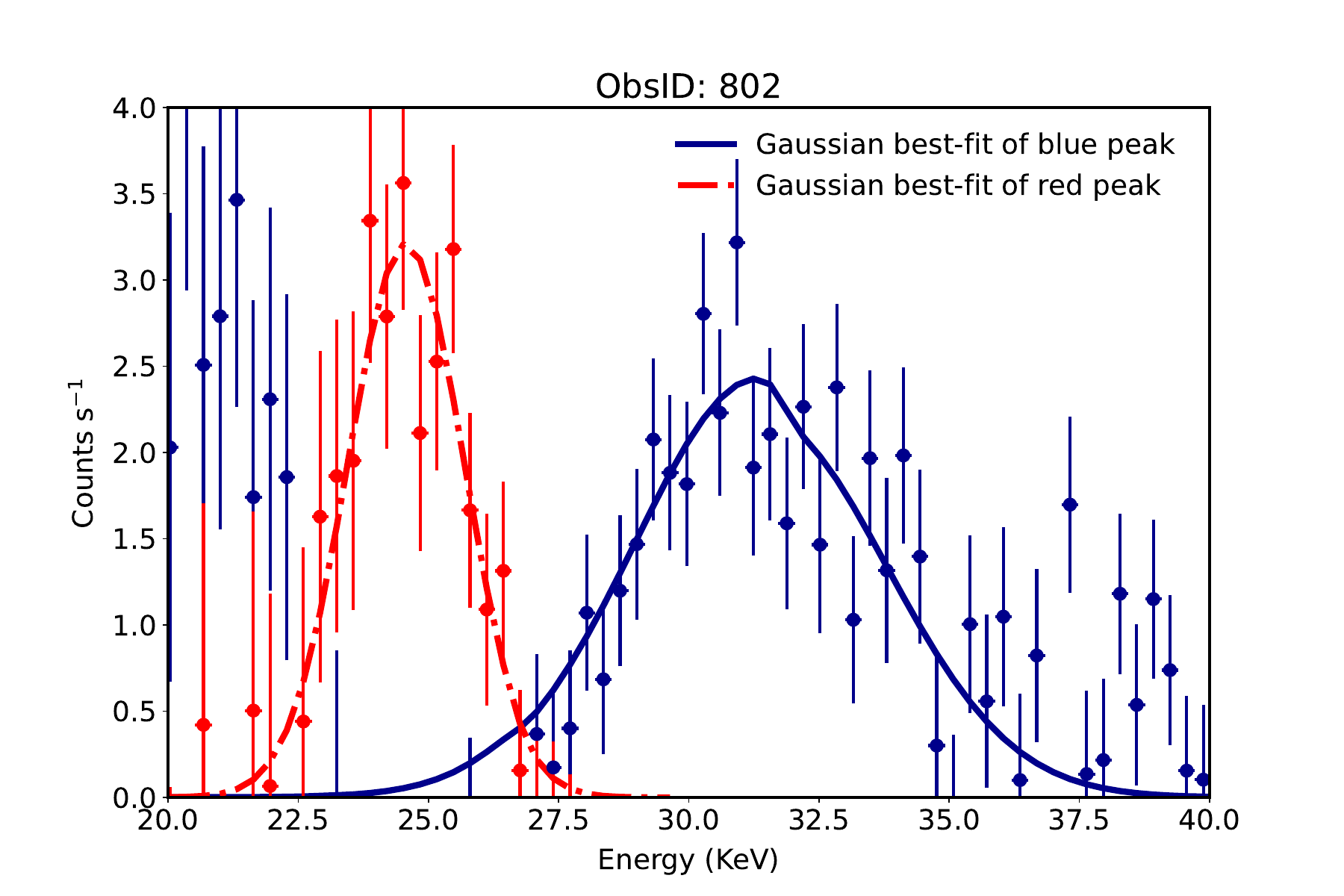}  & 
\includegraphics[width=0.8\columnwidth]{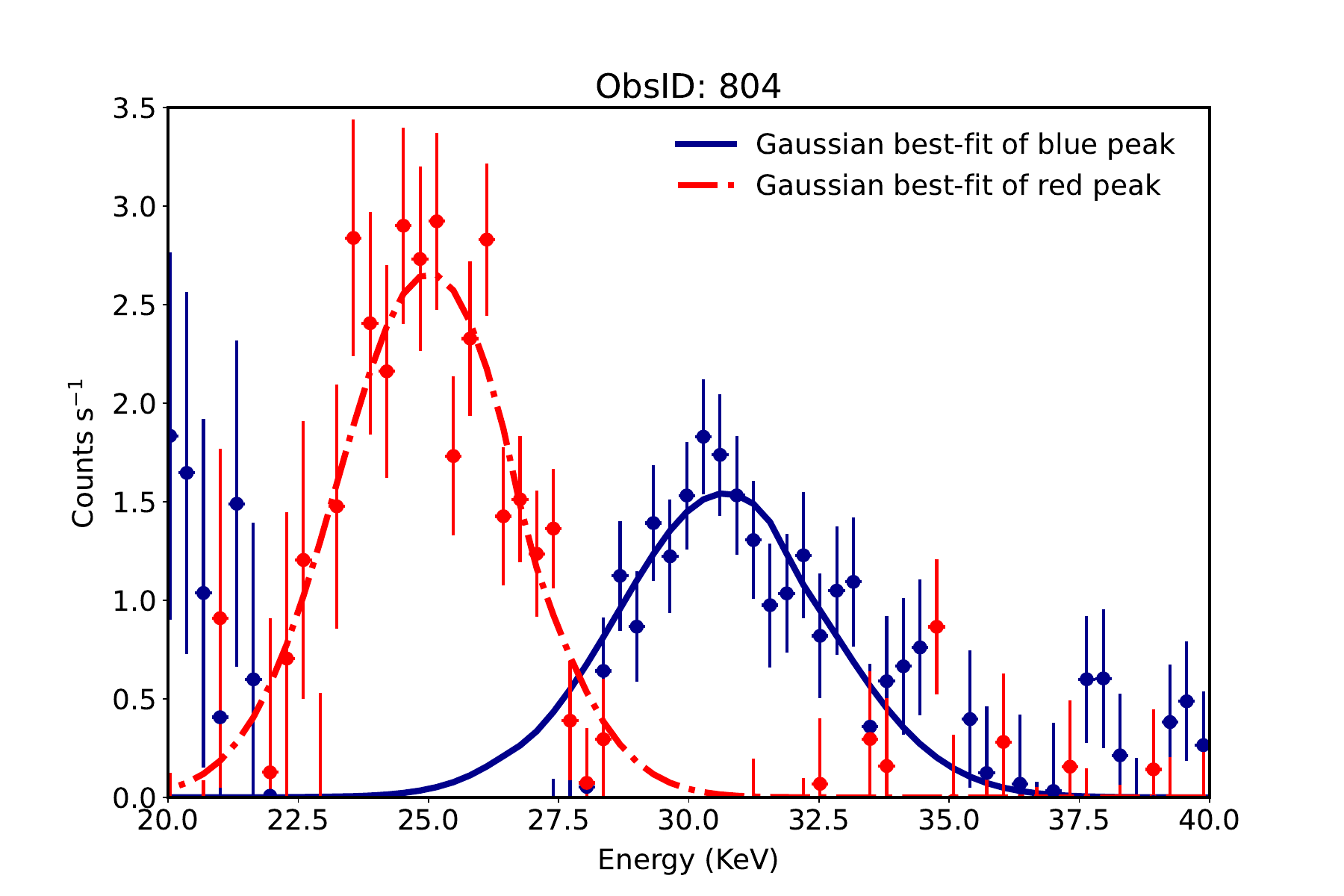} \\
\vspace{-0.5cm}
\includegraphics[width=0.8\columnwidth]{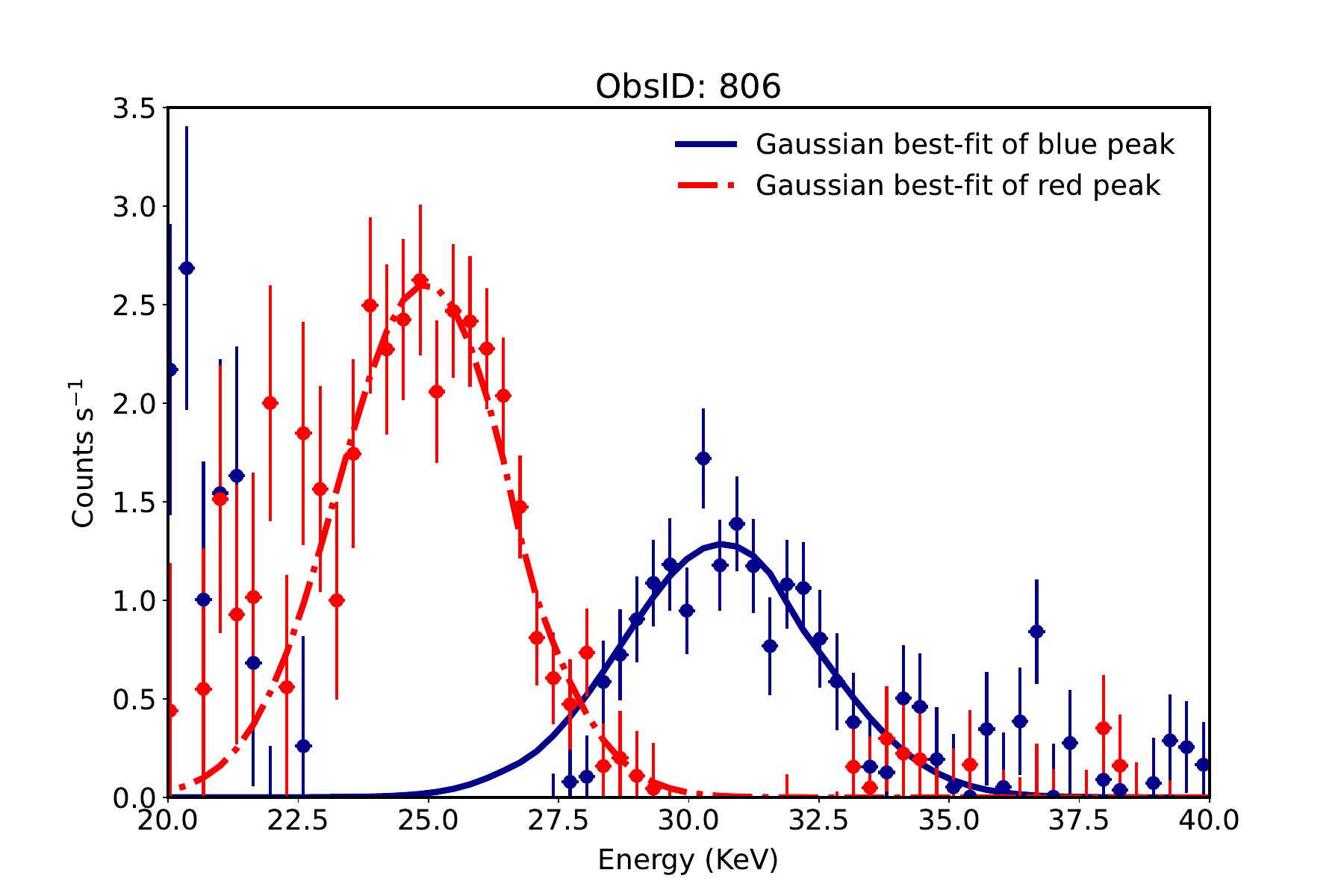} & 
\includegraphics[width=0.8\columnwidth]{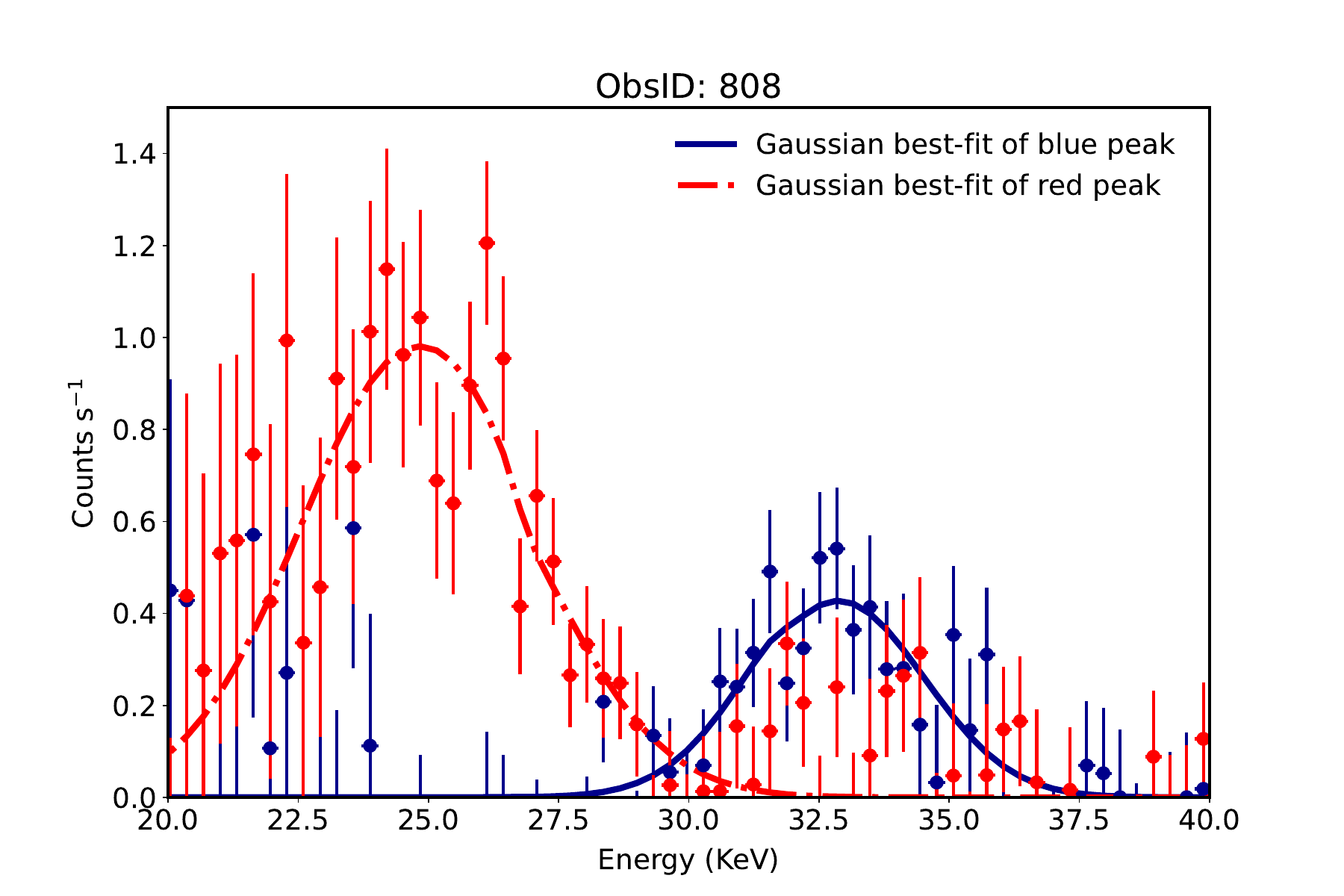} \\
\end{tabular}
\vspace{+0.5cm}
\caption{Spectral characterisation  of the phase-resolved peak excesses 
for ObsID 802, 804, 806, and 808. 
Each panel shows the photon excesses per energy bin both 
for the blue (in blue colour) and the red peak (in red colour). 
The Gaussian best-fitting model for the red (blue) data is marked by
a thick red (blue) line. 
}
\label{fig:phase_resolved}
\end{figure*}

We show in Fig.~\ref{fig:phase_resolved}  the data and the best-fit model 
for the four datasets. In Table~\ref{tab:timing_vs_resolved}, 
we compare the best-fitting values from the phase-resolved spectral fits and the 
best-fitting parameters of the PF spectra (same as Table~\ref{tab:fit_parameters1}). 
The parameters are generally well in agreement within the statistical uncertainties. 
As a consistency check, we evaluated if the number of counts 
in these excesses are  sufficient to account for the increase
in the blue/red peaks of the PFS. 
The number of excess photons in the spectra 
is simply the line normalisation multiplied by the exposure time
(see Table~\ref{tab:observing_log})
and the effective area (we take this value in correspondence with the
Gaussian peak energies, as the effective area is nearly constant for this 
narrow energy range). We, then, multiplied this number by a
corrective factor of 0.9 (which we derived through simple simulations 
to convert the number of photons into counts,
and therefore compensate for the effects of the redistribution matrix). 
The excess counts are reported in the seventh column of Table~\ref{tab:timing_vs_resolved}.

The number of counts in the PFS is obtained by 
element-wise multiplication of the best-fitting 
Gaussian model values with the total number of counts 
for each energy bin, and then summing over the values of the
resulting vector. 
Because PF values depend on the method how the PF is actually computed, 
we double checked these values adopting also the PF area definition \citep[see
PFS comparisons in the Appendix of][]{ferrigno2023}, as 
this method gives the most direct way to count the number of 
pulsed photons in the profile. We found that both PF definitions 
led to consistent estimates, and we report 
the values, obtained with the FFT method, in Table~\ref{tab:timing_vs_resolved}. 
The number of counts from this spectral excess analysis and the ones observed in the PF feature 
are of the same order of magnitude, and in most cases, even consistent
within the error bars. Given the rough approximations used for this comparative 
analysis, we retain that the local increase in the PF can be attributed to 
a certain increase in pulsed photons which have a clear and local energy range
distribution.

\subsection{Spectral analysis: modelling the phase-averaged spectra} \label{sect:spectralmodel}

Based on the results of the phase-resolved spectroscopy, which 
clearly showed that characteristic emission features can be 
associated to the pulsed part of the emission at the energies
of the PF blue and red peaks, we evaluate here 
the conditions under which such emission peaks at the two sides of 
the broad absorption CRSF can be statistically detected in 
the phase-averaged spectra. 
As our main interest lies mostly in the high-energy part of the spectrum,
we  fitted the data in the 5--60 keV energy range and 
analysed only the first four observations, which have good statistics.
We used the $W$ statistics, analogue to the Cash statistics (C-stat) when a background 
dataset is provided, assuming Poisson distributions for both source and background spectra \citep[see Sect. 7 in][]{Arnaud2011}.
We performed an initial fit with the usual minimization 
algorithm and  explored the neighbouring parameter space with a Monte Carlo Markov Chain MCMC, 
using the Goodman \& Weare algorithm \citet{Goodman2010} with 200 walkers, 
a burn-in phase of 20\,000 and a length of 200\,000 steps. 
We visually inspected the fit statistic distribution and the parameter 
corner plots to verify that the chains converged properly.

We model the continuum using a power-law with a 
high-energy cut-off (\texttt{highecut})
smoothed with a third-order polynomial 
at the cut-off to avoid the  
artificial discontinuity created by the model 
\citep[\texttt{newhcut} component, as defined in][]{Burderi2000}.
This phenomenological approach has been extensively 
used in literature for a long time, though it is 
well known to be an approximation to the true continuum shape.
We are aware that other phenomenological and even some physical 
continuum models are available. However, as we will discuss later, 
our aim is to evaluate the statistical significance 
of these possible detection and their correlation with the broad 
and deep CRSF.

We used the \texttt{tbabs} component for the photo-electric 
Galactic extinction. Because the value of the Galactic absorption column parameter, $N_{\textrm{H}}$, 
is poorly constrained, we fixed it to the  expected interstellar 
value\footnote{\url{https://heasarc.gsfc.nasa.gov/cgi-bin/Tools/w3nh/w3nh.pl}}  
of  7\,$\times$\,10$^{21}$ cm$^{-2}$ .

All spectra clearly showed the presence of two CRSF, at $\sim$\,28 keV and $\sim$\,51 keV.
The use of a single absorption Gaussian (\texttt{gabs} model in \texttt{Xspec})
to model these CRSFs resulted 
always in poor fits (see Table~\ref{tab:chi_reduced_models}) 
and evident residuals especially around the fundamental line, 
thus indicating a complex spectral shape of this feature. 
We shall refer to this model as the \textit{one Gabs}.
As already noted by \citet{Pottschmidt2005}, 
the cyclotron profile of \vzero\ is, at least phenomenologically, 
better described by two nested Gaussian absorption lines, i.e.
a superposition of a shallow/narrow core and 
a broader/deeper Gaussian wing. 
We noted no statistically significant shift between the energies
of these two absorption lines, and therefore, we tied these 
positions to be the same in the best-fitting model.
We shall refer to this model as the \textit{nested Gabs}.
In general, also for the \nustar\ spectra,
this model gives a significant statistical improvement with respect to 
the \textit{one Gabs} model. 
Finally, based on the hints from the PF spectrum modelling and 
the phase-resolved analysis, 
we modelled the CRSF using a broad absorption Gaussian 
with two, narrower, Gaussian emission lines: one above 
and one below the absorption line energy,
we shall refer to this as the \textit{wings} model. 
To avoid  over-fitting and model degeneracy, 
we impose Gaussian priors on the centroid energy 
and width of the emission lines from the phase-resolved 
spectroscopy as described in Sect.~\ref{sect:phaseresolved}.
For the other spectral parameters, we use non-informative priors 
in linear or logarithmic space, as appropriate. 
The posterior error distributions 
of all the spectral parameters of this model appear generally well constrained 
(corner plots of the posterior error distributions of the four fits are presented
in Appendix~\ref{appendix:cornerplots}).

Tables \ref{tab:spectralfits802},  \ref{tab:spectralfits804}
\ref{tab:spectralfits806}, and \ref{tab:spectralfits808}
reported in Appendix~\ref{sect:appendix2} 
show the best-fitting parameters and uncertainties of these
three models, while 
we give a quick comparative view 
of the best-fit results (C-stat values and dof)
in Table~\ref{tab:chi_reduced_models}.
We use the \textit{Akaike information criterion}  
\citep[AIC][]{Akaike1974}  to evaluate the 
statistical improvements of the  \textit{nested gabs} 
and \textit{wings}  model with respect to the \textit{one gabs} model. 
The AIC values are defined according to the formula:
\begin{align*}
\rm AIC=2\textit{k}-2ln(\mathcal{L}_{max})
\end{align*}
\noindent where $k$ is the number of fitted parameters of the model and $\mathcal{L}_{max}$ is the maximum likelihood value. The C-statistics value and the maximum likelihood are related as $\rm Cstat=-2ln(\mathcal{L}_{max})$ \citep{Cash1979}.  We calculate the AIC difference ($\rm \Delta AIC$) over all candidate models with respect to the simpler, \textit{one gabs} model. 
Models  with $\rm \Delta AIC>>10$ are strongly more favoured with respect to the baseline model. 
The \textit{wings} model is always statistically
more favoured  with respect to the \textit{one gabs} model, 
whereas the \textit{nested gabs} model is the preferred model 
in the first three observations, but it is the least statistically 
significant one for spectrum 806. In three, out of four spectra, 
the \textit{nested gabs} model 
is statistically favoured over the \textit{wings} model, although 
we noted that such improvement can be mostly attributed by a better
description of the overall continuum shape, whereas residuals at the cyclotron line 
energy are distributed similarly in both models.

\begin{table} [h]
\centering
\renewcommand{\arraystretch}{1.0}
\caption{Quick-look summary report 
for spectral best-fitting comparison of models of Tables \ref{tab:spectralfits802}, \ref{tab:spectralfits804},
\ref{tab:spectralfits806}, and \ref{tab:spectralfits808}. The $\Delta$AIC values 
are given as difference between the \textit{one gabs} AIC model value and the AIC
values of the tested, more complex, \textit{wings} and \textit{nested gabs} models.}
\label{tab:chi_reduced_models}
\begin{tabular}{l r|rr}
\hline    
          & \textit{one gabs} & \textit{wings} & \textit{nested gabs} \\ 
    ObsID & AIC   & $\Delta$AIC  &$\Delta$AIC \\
    \hline
    802 & 715 & 145 & 190 \\
    804 & 749 & 126 & 299 \\
    806 & 698 & 115 & 143 \\
    808 & 408 & 38  & -2   \\
    \hline
\end{tabular}
\end{table}

For the \textit{wings} model, the best-fit normalisations (or the upper limits, 
when lines are not statistically detected) of the emission CRSF lines 
give a total number of photons in the line that is a factor of 2-3 higher 
than the excess counts derived from the phase-resolved analysis (last column of Table \ref{tab:timing_vs_resolved}). 
This is not surprising, given the systematic large uncertainties which 
are involved in the choice of the phase intervals for the excess analysis 
and in subtracting, or disentangling, the effect of absorption from the CRSF. 
In conclusion, we retain these count estimates to be satisfactorily in agreement.

\subsection{Spectral analysis: phase-resolved spectroscopy of ObsID 806} \label{sect:ph_res_806}

As a final step in establishing a relationship between the spectral
blue and red line components and the features of the PFS we looked 
for phase-dependent changes of the line fluxes. If these spectral lines
are responsible for the local increase in the PF and their pulsed emission
is clearly peaked at some spin phase, we expect a corresponding modulation of the
fluxes. To test this conclusion, we took the ObsID 806 where both 
spectral lines are clearly detected and their line normalisations are the highest.
We phase-resolved the spectra using ten equally spaced phase bins and applied
to these spectra the \textit{wings} spectral model, assuming the same priors of the 
phase-averaged analysis and calculating the best-fitting parameters and errors. 

\begin{figure*}
  \centering
  \begin{tabular}{cc}
  \includegraphics[width=0.9\columnwidth]{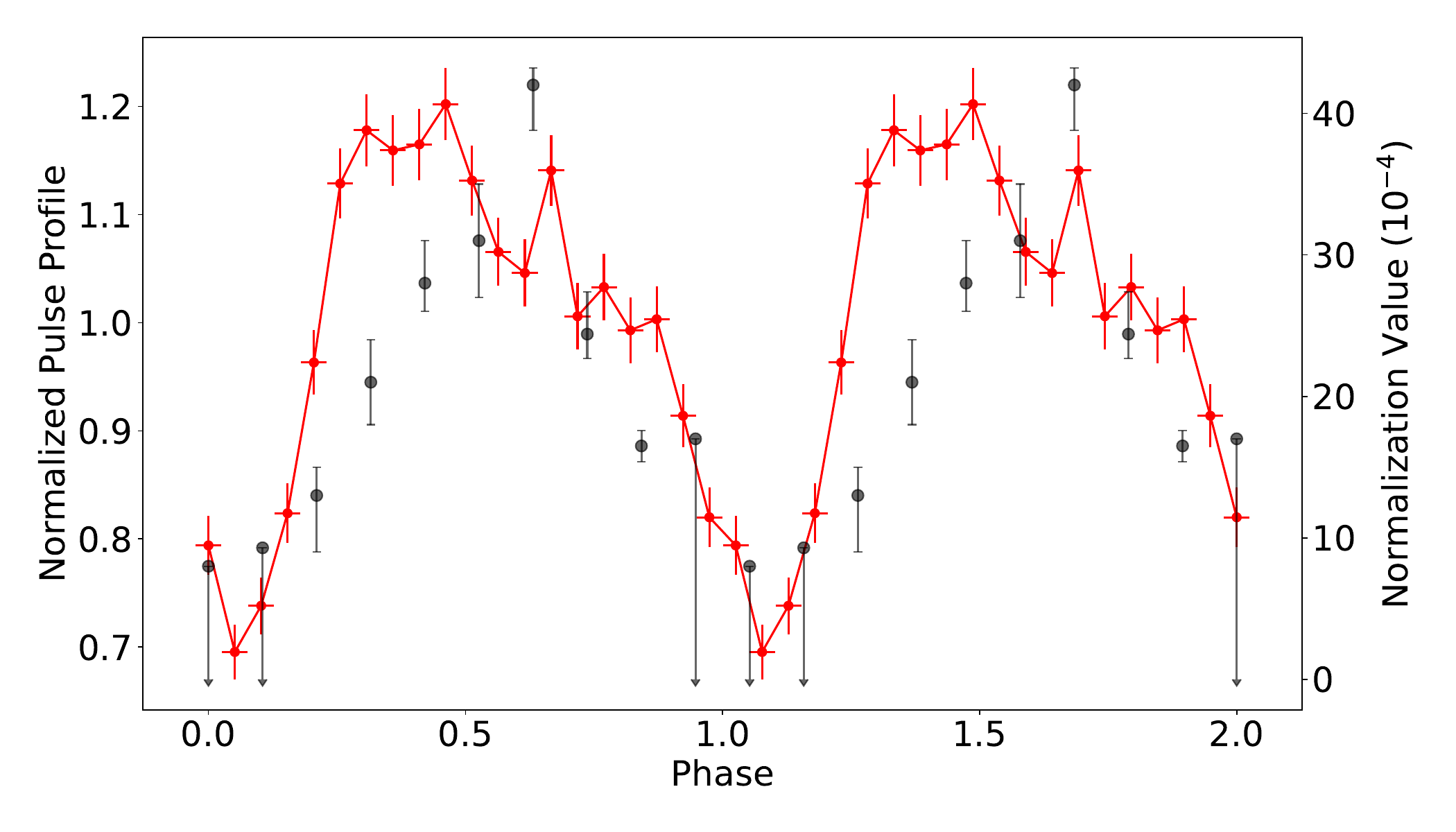} &
  \includegraphics[width=0.9\columnwidth]{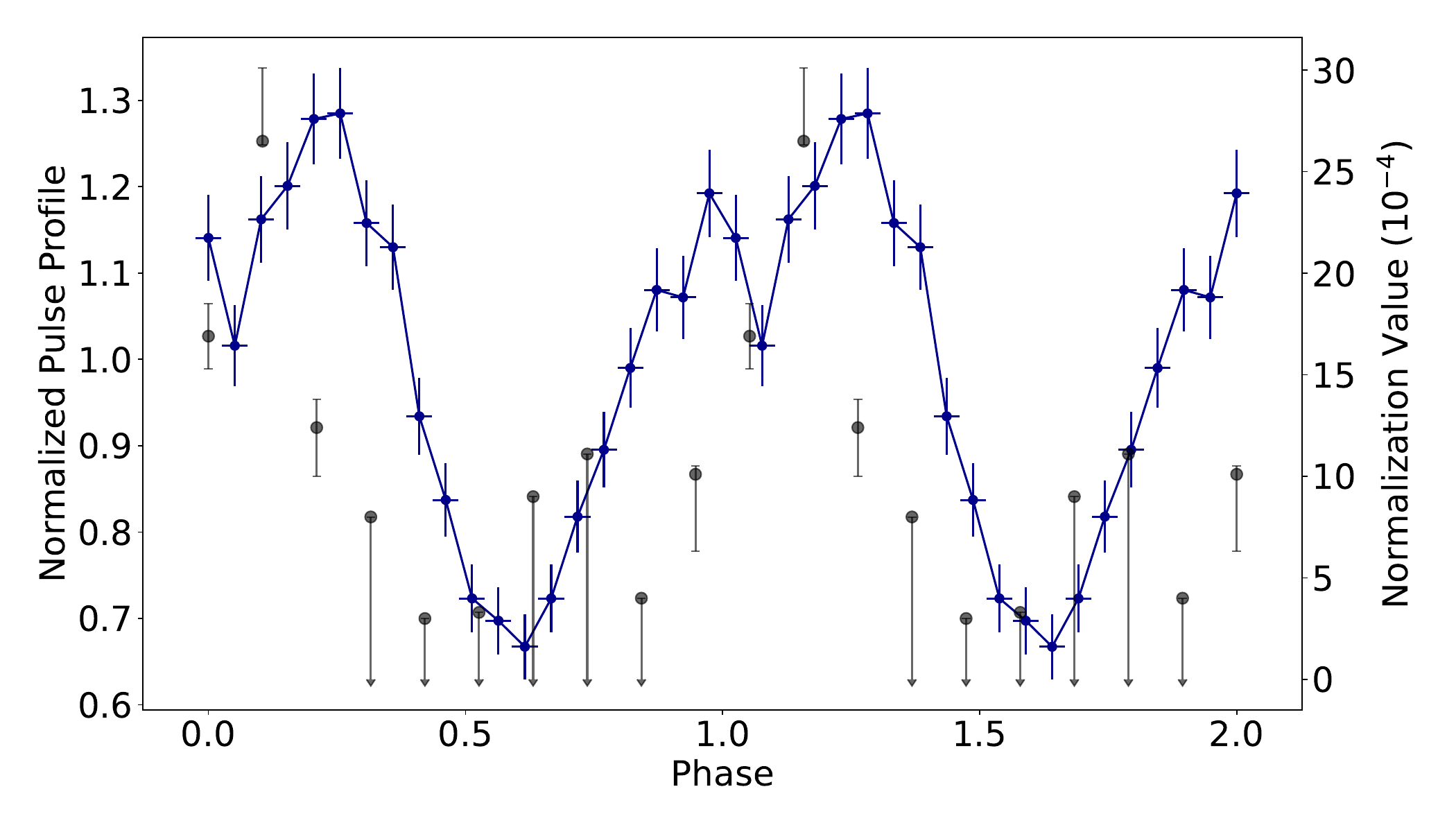} \\
  \end{tabular}
  \caption{Phase-resolved spectroscopy for ObsID 806. Left (Right) panel: phase-dependence of the red (blue) line normalisation (black dots in units of 10$^{-4}$ photons/cm$^{-2}$/s) superimposed with the energy-resolved pulse-profile (from Fig. \ref{fig:enselectedpp}).} \label{fig:phase_linenorm} 
\end{figure*}

In Fig.\ref{fig:phase_linenorm}, we illustrate the variation of line normalizations across the spin phase. For improved clarity, we overlay the energy-resolved pulse profiles from Fig.\ref{fig:enselectedpp}, specifically those extracted in the 24.1–-27.1 keV range for the red peak and the 28.3–32.4 keV range for the blue peak. 

The modulations are clearly visible and there is a remarkable 
good match with the corresponding pulse profiles. 
In conclusion, we find that there is a consistent and robust picture 
that from the features that appear in the PFS of \vzero\ 
lead to a complex but well-consistent modelling 
of the energy spectra around the fundamental cyclotron
line energy. We emphasise that such modelling is yet based
on a phenomenological approach and the true cyclotron line shape 
might be certainly more complex than what is envisaged here (a mix
of emission/absorption Gaussians at different energies). 
However, this modelling would have been difficult to do at all had 
it not been for the analysis of the energy-resolved pulse profile changes.

\section{Discussion} \label{sect:discussion}

We presented a comprehensive analysis of the 
variations of the pulse profile with energy of the accreting 
X-ray pulsar \vzero, using observations from \nustar, which is today's best
observatory to perform this kind of study in the 3--70 keV band at the 
highest energy resolution.
Compared to the majority of other well known X-ray pulsars, \vzero\ 
has a remarkable PF spectrum. The general PFS characteristics seem to
only mildly depend on the accretion rate. The examined dataset span a wide range of luminosity:
from something close to the Eddington limit ($L_{\textrm{Edd}}=2\times 10^{38}$ erg s$^{-1}$ 
for a classical 1.4 M$_{\textrm{sun}}$ NS, e.g. ObsID 802) to observations when the luminosity was
only a few percent of the Eddington limit (e.g. ObsID 904). 
In all cases, the PF values are generally low in the soft (3--10 keV) and in  the intermediate (10--22 keV) X-ray bands, 
followed by a significant increase only around the energy of the  
fundamental cyclotron line. 
Above the cyclotron energy range we note the only 
significant change with luminosity, as the PF trend values tend to 
flatten as the luminosity decreases 
(see Fig. \ref{fig:pfspectra} and \ref{fig:pfspectra2}).
Contrary to what observed in other 
sources, where the PFS dip around the cyclotron line energies
\citep[see][]{ferrigno2023}, \vzero\ shows a peculiar wiggle, which could
be described, in the most simple terms, as two close Gaussian lines with positive amplitudes.
The fact that this pattern is similarly reproduced in four different datasets
is evidence that the shape has a physical origin, and cannot be due to statistical fluctuations. 

\citet{Tsygankov2006} gave a detailed description
of the profile variations with energy of this source, noting  both the abrupt shape change below 
and above the cyclotron energy and the PF increase at these energies.
We were able to  resolve spectrally this increase by showing how it
appears complex overall when compared with PFS of other similar 
sources \citep[see e.g.][]{ferrigno2023}. 
One might, however, argue that a \textit{drop} in the PF is still present in-between 
the two lines, albeit the bottom value of the PF appears consistent 
with the expected trend of the underneath continuum. However, the fact that 
these two peaks are physically not connected is supported by the 
evident change of the pulse profile shapes (Fig.~\ref{fig:peakcomparison}) and by the clear trends in the 
CC and lag spectra (Fig.~\ref{fig:corrlagspectra}) that show how 
the discontinuity of the shape is produced around the cyclotron line energy: the 
lag profile drifts rapidly, 
yet continuously, in phase with the steepest derivative very close 
to the cyclotron line energy. This can also be appreciated by looking 
at the energy maps (see Fig.~\ref{fig:pp802_pp804} and Fig.~\ref{fig:pp806}), 
where the drift of the PP peak in the 22--32 keV range is clearly visible.
The spectral shape of the pulsed photons responsible 
of these local PF peaks adds another piece of evidence of their uncommon origin. 
The Gaussian shaped distribution of excess photons, as derived by
their respective energy-resolved peaks in pulse profiles, have spectral widths 
that well match the  red and blue PF Gaussian widths, 
with little leakage from adjacent energy ranges, apart from 
some red photons in the blue energy range for ObsID 808 (Fig.~\ref{fig:phase_resolved}).

The trend of the peak positions with 
source luminosity as shown in Fig.~\ref{fig:obsid_energy} trails the 
trend of the cyclotron line energy: the energy of the peaks remains always 
on the two sides of the cyclotron line, though there is no clear correlation
with source luminosity. The ratio of the peak amplitudes, on the contrary, 
varies: as the source luminosity decreases the red peak 
amplitude  increases, whereas the blue peak amplitude does not 
display significant variability, as shown in Fig.~\ref{fig:peakcomparison}, 
though a shift of about 1 keV is hinted in ObsID 808.

\subsection{Physical origin of the PF peaks}
After having established the general properties of these features, 
we discuss their possible physical origin. As noted for many other 
sources \citep[see e.g.][]{Ferrigno2011a}, 
a sharp pulse change at the cyclotron energy is expected because of the 
complex and sharp resonant cross-section energy-dependence of the 
cyclotron absorption for the extra-ordinary mode photons, while 
ordinary mode photons are not affected.
Because the relative fraction 
of ordinary vs extra-ordinary mode photons might be a function of 
energy and there are preferential angles for both modes according 
to the energy at which the scattered, or re-emitted, photon, is 
escaping, the pulse shape can be fundamentally altered. 
Thus, as the shape changes,  a change in the pulse amplitude
(and consequently in the PF value) is expected. 
In consideration of their peak energies, variability, and associated energy-dependent, 
pulse shapes, we retain, therefore, the two peaks in the PFS physically 
related with the fundamental cyclotron line. 
The next step is to ask why these features produce an increase the PF values, 
rather than to decrease it with respect to what is observed in other sources. 
\citet{ferrigno2023} showed the  consistency 
between the cyclotron energies as derived by the spectral analysis and 
the position of the  absorption features in the PFS for the four sources
analysed (e.g. Cen X--3, 4U 1626--67, Her X--1 and Cep X--4), while for \vzero\ 
the mismatch in the positions is evident: none of the two features can be 
reliably appointed to the spectral energy position \Ecyc. 
However, theoretical models of cyclotron line 
absorption have since long time pointed out that the absorption profile is far from 
being either a smooth Gaussian, or a Lorentzian, and many physical processes intervene to distort the 
profile \citep{Isenberg1998, Schonherr2007, Schwarm2017b}. One key prediction has been the presence of \textit{wings}. 
Such features would appear in emission with respect to a simpler Gaussian profile fit 
and MC simulations have confirmed the formation 
of these structures on both sides of cyclotron energy (see references in Sect.\ref{sect:lineshapes}) in certain conditions. 

As a next step, we attempted to find a spectral confirmation to this scenario.
To this aim, we first looked at the spectral shape responsible 
of the pulsed part of the profile for each PF peak. We found 
(as expected) a good agreement between the PF shape and the count excesses shape (Fig. \ref{fig:phase_resolved}). Interpreting these excesses as the source 
of the wings components (here simply modelled as Gaussian 
emission features), we re-examined the phase-averaged energy spectra of 
the first four \nustar\ observations.
We found that residuals around the fundamental CRSF
 are always present. We also made 
different spectral fits by changing both the continuum shapes and the line shapes 
and noted that residuals were always there. On the contrary, \citet{Doroshenko2017}
fitting the same data found no evidence for significant residuals beyond 
a simple Gaussian fit; we were able to obtain very similar best-fitting
parameters as the ones shown in \citet{Doroshenko2017}, though in all cases 
the reduced $\chi^2$ were much worse. We argue that this is due to how the 
spectra were re-binned. We used the optimal binning 
prescription from \citet{Kaastra2016} to avoid oversampling the spectra, 
whereas \citet{Doroshenko2017} used the simpler requirement of a minimum 25 counts 
per noticed channel, thus resulting in spectra with $\sim$ a factor 4 
more channels than the ones used by us. We note that over-sampled spectra 
have generally lower reduced $\chi^2$, which can lead to miss some 
residual patterns, especially if they are restricted to a narrow energy range.

We retain that these spectral residuals provide instead good evidence
for a complex cyclotron line shape, and we showed that the addition of 
these  emission lines as the ones found from the phase-resolved spectroscopy 
analysis gives generally a statistically  significant improvement with respect
to the null model (here the \textit{one Gabs} model). 
However, we also noticed that by allowing the absorption line shape to take
two additional free parameters (as shown with the \textit{nested Gabs} model), 
the statistical description of the data could be even better and 
the need to add additional components around the energy 
of the fundamental CRSF appears no longer motivated.
 
This fact is clearly comprehensible if we make some computations 
on the relative flux components involved in the spectral components of the fits. 
For instance, if we take the best-fitting model \texttt{wings} from 
Table \ref{tab:spectralfits804} for ObsID 804 (where there is good statistics and 
both emission lines) and compute the best-fitting flux model in the 
energy range $E_{\textrm cyc}$-2$\times \sigma_{cyc}$ and 
$E_{\textrm cyc}$+2$\times \sigma_{cyc}$ (21.1--35.5 keV), we see 
that the cyclotron feature absorbs about half of the total emitted flux
in this range, while the two emission lines contribute 
for only 2\% to the total flux. It is, therefore, no wonder that
by adding complexity to how the broad and deep absorption feature is modelled, 
any detection claim for such, relatively small, feature by means of 
pure spectral analysis becomes challenging (leaving aside that 
inherent complexity of finding a suitable spectral continuum). 
If this also applies  to past observations of \vzero,  we can understand how 
previous attempts to  find evidence of such emission wings in the
energy spectra of this source proved hard to succeed. 
The prominent peaks in the PFS raise questions 
about whether this behaviour is supported by theoretical radiative transfer models 
and what is the physical mechanism behind the significant increase in the pulsed fraction. 
Theoretical predictions for pulse profile formation rely on complex 
dependencies from multiple factors such as system geometry, light bending, and location of the regions of spectral formation \citep{Falkner2018}, 
but, to our knowledge, no predictive quantitative assessment 
has been made regarding the expected fractional pulsed energy dependency 
for complex CRSF shapes with wings.
Observationally we see that photons in the wings exhibit greater coherence than those of the underneath continuum. 
This suggests that the last scattering region of the wing photons, or their
emission cone, is distinct and likely more asymmetric than the region producing the pulsed continuum.
Some theoretical considerations play a role in this context.
Assuming photons appearing in wings are spawned from higher excited
Landau levels, then the escape beam pattern are different in case
the electron de-excites from the first to ground state, or from the third 
to the second \citep{Schwarm2014}. 

Assuming that the red wing is mainly populated by spawned photons from third to second excited state (the observed energy difference of the second and fundamental cyclotron line energies is close to the red peak energy), and the blue wing by the transition from the first to the ground state, then the escape paths of these photons might also result in some phase shift, depending on the overall angles that the escape cones make with the observer's line of sight and the assumed system geometry. Arguably 
we might intercept spawned photons populating the red wing with their emission cone only from one column, while we intercept the blue photons, having a different beam angle only from the opposite column, thus explaining
the large (almost 180$^\circ$) phase shift observed in the energy-resolved pulse profiles extracted from the two features energy ranges (Fig.\ref{fig:peakcomparison}).\\
It remains also to be addressed the expected variations in the PFS 
around the cyclotron line for different luminosities 
(e.g., sub-critical luminosity vs supercritical luminosity regime).
In this case, we did not find clear theoretical predictions to be compared and this is certainly an interesting avenue for future work.
\subsection{System geometry}
We need to address the question on what makes \vzero\ so 
peculiar among all the other X-ray pulsars. Based on our knowledge, 
there is no other known source with the same set of 
characteristics as this one: very low PF values below the cyclotron 
line energy, strong PF increase and drastic pulse profile change 
at the cyclotron energy, extreme depth of the 
cyclotron line in the energy spectrum. This set of characteristics does not depend  
on the accretion rate, as we found these characteristics ubiquitous in 
all examined dataset. The NS magnetic field value, spin period, 
system physical orbital parameters, and the general 
spectral shape are standard when compared to the whole HMXB sample \citep{Kim2023}. 
We retain that the only key factor to possibly differentiate this 
source from all the others might be the geometrical factor: it is 
possible that our line-of-sight is well aligned with the spin 
axis, while the magnetic obliquity (i.e. the angle between the spin and 
the magnetic axis), has an \textit{extreme value}, either very low, or
close to be orthogonal.\\
Let us first consider the low spin axis angle hypothesis:
even if the posterior probability of such geometry is very low, 
when considering the whole sample of known HMXBs, 
such probability becomes no longer marginal. 
If we take the most recent mass function estimate 
\citep[$f_{M} = 0.44(1) M_{\odot}$][]{Doroshenko2016}, 
and we assume a canonical 1.4  $M_{\odot}$ NS mass and 
a possible mass range of 15--30 $M_{\odot}$ for the companion star 
\citep[an unevolved O8–9Ve star][]{Negueruela1999}, 
we obtain a very narrow range for the system inclination angle $i$\,=\,14--19 deg, 
thus validating this first assumption. 

\begin{figure} 
  \centering
  \includegraphics[width=\columnwidth]{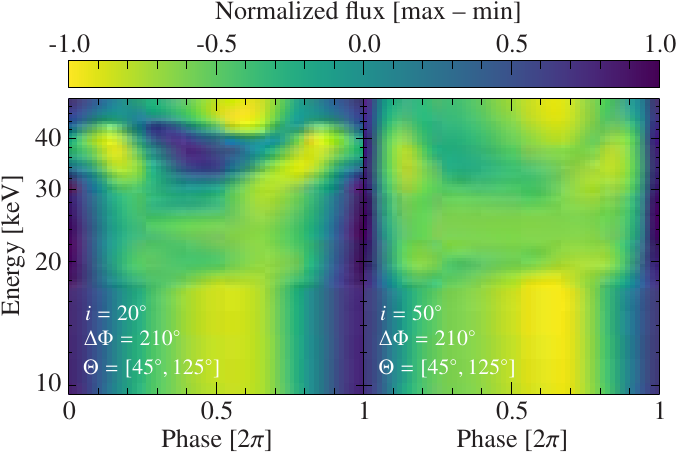}
  \caption{Phase-energy maps simulated from the solution of radiative transfer equation in a slab of highly magnetised plasma based on a model described by \citep[][]{Sokolova-Lapa2023} and combined with relativistic ray tracing by \citet{Falkner2018}. A colour map shows the variation of a normalised total observed flux. The two maps are calculated for observer's inclinations $i$ and the same location of two identically emitting poles, specified by their polar angle from the spin axis $\Theta$ and their azimuthal separation $\Delta\Phi$
  (the values are indicated on the panels). For the geometry displayed on the left, phase shifts occur over a broad range around the line, being caused by the redistribution of photons in the wings. The blue wing is particularly strongly affected by the process due to relatively high electron temperature $kT=7\,\mathrm{keV}$.} \label{fig:simu_map}
\end{figure}

Regarding the magnetic obliquity,
let us first consider the orthogonal rotator case. This geometry needs to assume
a fan beam pattern for all observations; this can explain at the same time
the overall low PF values (most likely driven by inhomogeneous 
reflection patterns on the NS by the two accretion columns as seen 
by the distant observer) and the large optical depth of the cyclotron 
line in the energy spectra (most of the observed photons are 
observed coming perpendicularly to the magnetic field, where 
the scattering/absorption cross-section for cyclotron absorption 
is the highest). 
On the contrary, in the case of a small obliquity, i.e. 
less than 30$^{\circ}$ \citep[a geometrical configuration 
that has been also independently advocated in a recent study, see][]{Mushtukov2024}, 
on the contrary a pencil beam pattern for all observations must be assumed.
To illustrate the feasibility of such geometry, we
provide an example of simulated phase-energy maps. For this, we utilise
a physical model for emission from a slab of highly magnetised plasma
on the surface of a NS, as presented in \citet{Sokolova-Lapa2023}.
The simulations are based on a radiative transfer code, which addresses
the formation of both the continuum and the cyclotron resonance.
We chose a model described there as ``inhomogeneous accretion mound'',
which includes a variation of the density profile with depth but
assumes the constant electron temperature and magnetic field strength
throughout the emission region. Here, we fix the electron temperature
to $7\,\mathrm{keV}$ and the magnetic field to 3.4$\times$10$^{12}$ G,  
corresponding to a cyclotron energy of $30\,\mathrm{keV}$. 
To study the variation of the observed signal with
geometry (i.e., the observer's inclination and the location of the
emission regions), we combine the obtained emission, determined for a
wide range of photon energies and angles to the magnetic field, with
a relativistic ray-tracing code by \citet{Falkner2018}. This code
calculates the projection of a rotating neutron star with defined
emitting regions onto the observer's sky, assuming the Schwarzschild
metric. We chose a simple setup of two emitting identical poles on
the neutron star surface. 
Figure~\ref{fig:simu_map} shows simulated
phase-energy maps for the observed emission for two distinct
geometries, chosen to illustrate two principally different
results. The left-hand side indicates a striking variation of emission
patterns in different energy bands, emphasised by a chosen geometry. The
(fundamental) cyclotron line appears in the observed phase-averaged
spectra at around $25\,\mathrm{keV}$, affected by gravitational
redshift and, to a smaller degree, by angular variability of the line
profile. From this example, it is clear that a wide region around the
cyclotron line energy -- red and blue wings -- is affected by redistribution much
stronger than the line core. While the energies below (${\sim}18$--$25\,\mathrm{keV}$),
corresponding to the red wing, show relatively weak, although noticeable 
phase shift, for a wide range of energies contributing to the blue wing,
the structure of the map changes dramatically. We attribute this difference
to more enhanced redistribution of photons into the blue wing during
the line formation due to the relatively high electron temperature, as well
as to the geometrical effect.

The right-hand panel presents the same geometry of the poles with the
larger observer inclination, which result in almost completely aligned
magnetic and spin axes. This configuration leads to a very low variability
of the flux in a phase-energy map, including the cyclotron line region.
Our simulations indicate a strong dependency of phase-energy maps
on the assumed geometry. In this setup, the offset larger than ${\sim}10^\circ$
is required to create complex variation of the emission with the phase.

However, a thorough exploration of the parameter space is required to
draw further conclusions, which will be presented in a future work. 
The presented maps serve exclusively to illustrate the effect of a complex angular 
behaviour of emission from a warm, strongly magnetised plasma, 
as expected close to the surface of the accreting NS. 
This affects not only the cyclotron line but also a broad range of surrounding energies, 
due to significant redistribution in the line wings. With these simulations, 
we aim to support the general interpretation of the emission wings presented in this work 
and highlight the primary influence of geometry on their phase variability. 
However, we note that incorporating additional physical processes into the model 
-- such as bulk, in addition to thermal Comptonization, and a more 
realistic vertical extension of the emission region -- 
is necessary to fully describe the complex accretion 
picture in the super-critical regime \citep[see also][]{Zhang2022}.\\
In conclusion, both scenarios require that the emission mode, either fan 
or pencil, remains largely the same from all the examined observations,
which likely encompass both super- and sub-critical regimes, thus 
also explaining the 
lack of significant changes in the observed PFS for observations 
at low and high accretion rates.\\
Interestingly, recent spectro-polarimetric 
observations of accreting pulsars seem to suggest that the rotational 
and magnetic angle are distributed according to such bimodality: 
either sources are orthogonal \citep[like X-Persei and 
GRO J1008--57][]{Mushtukov2023a, Tsygankov2023}, or show 
a very small magnetic obliquity \citep[like Cen X--3 and 
Her X--1][]{Doroshenko2022, Tsygankov2022a}.
Future polarimetric observations of \vzero\ will be, therefore,
key to definitively determine the source's geometry.

An open question remains the robustness of the correlation of the cyclotron line energy 
with the source luminosity \citep{Tsygankov2006, Cusumano2016a, Doroshenko2016, Ferrigno2016}. 
We retain that the complexity of the line profile, taking into account the presence 
of emission wings on both side of the absorption feature, might not significantly 
alter the determination of the cyclotron line energy, although a deeper spectral analysis investigation with 
ad-hoc physical modelling of the line \citep[e.g. using the most recent MC codes][]{Schwarm2017a} 
is required for a definitive proof. 

\section{Conclusions}
In this paper we have presented an in-depth study of 
the variations  in the pulse profile of \vzero\ with 
energy using all available \nustar\ observations to date. 
Remarkable profile variations are observed on both sides 
of the putative fundamental cyclotron 
line energy ($E_{\textrm{cycl}} \sim$ 28\,keV) that 
we interpret as features arising from the reprocessing of cyclotron line 
photons (either scattered or re-emitted from higher excited Landau levels) in the line-forming region. 
Consistent with previous theoretical predictions of 
how cyclotron lines should appear to the distant 
observer, we associate these profile changes 
to the emergence of the so-called cyclotron line 
\textit{wings}. We infer that the \vzero\ is a source
seen at low inclination angle and the magnetic axis direction is
either $\lesssim$ 30$^{\circ}$, or perpendicular to the 
rotational spin axis.
Preliminary theoretical work shown in this paper 
support our observational findings, though a detailed determination of 
all the physical ingredients in the line forming region is demanded to 
a future publication. A robust \textit{physical}  model of cyclotron line shapes
in \vzero\ is needed in order to firmly test  any correlation between  cyclotron 
energy position and other observables, such as the source luminosity, as 
this in turn has key physical implications to correctly posit the 
cyclotron line forming region \citep{Loudas2024}.
Finally, we hope that this work will encourage authors to model the energy-pulse shape 
dependence in their Monte Carlo radiative-transfer calculations, taking proper account of light propagation, 
as this work highlights the power of this method to identify and study the CRSF profile, 
which in principle encode important information about the plasma properties in the line-forming region.
\begin{acknowledgements} \smallskip
The authors would like to acknowledge and thank \mbox{Domitilla De Martino} and Jakob Stierhof for very useful comments and discussion. \\

This research was supported by the International Space Science Institute (ISSI) in Bern, through the ISSI Working Group project \href{https://collab.issibern.ch/neutron-stars/}{Disentangling Pulse Profiles of (Accreting) Neutron Stars}.\\

This research benefited from funding by the 
Space Science Faculty of the European Space Agency. The research has received funding
from the European Union’s Horizon 2020 Programme
under the AHEAD2020 project (grant agreement n. 871158). AD, EA, GC, VLP acknowledge funding from the Italian Space Agency, contract ASI/INAF n. I/004/11/4. CP, MDS, AD, FP, GR acknowledge support from SEAWIND grant funded by NextGenerationEU. ESL acknowledges support from Deutsche Forschungsgemeinschaft grant WI 1860/11-2. FP, AD, MDS, GR and CP aknowledge the INAF GO/GTO grant number 1.05.23.05.12 for the project OBIWAN (Observing high B-fIeld Whispers from Accreting Neutron stars). CM acknowledges funding from the Italian Ministry of University and Research (MUR), PRIN 2020 (prot. 2020BRP57Z) “\textit{Gravitational and Electro magnetic-wave Sources in the Universe with current and next generation detectors (GEMS)}” and the INAF Research Grant \textit{Uncovering the optical beat of the fastest magnetised neutron stars} (FANS).\\
We made use of Heasoft and NASA archives for the \nustar data.
We developed our own timing code for the epoch folding, orbital correction,
building of time-phase and energy-phase matrices.
This code is based partly on available Python packages such as:
\texttt{astropy} \citep{astropy:2022},
\texttt{lmfit} \citep{lmfit},
\texttt{matplotlib} \citep{matplotlib},
\texttt{emcee} \citep{emcee},
\texttt{stingray} \citep{stingray},
\texttt{corner} \citep{corner},
\texttt{scipy} \citep{scipy}.
An online service that reproduces our current results is available on the Renku-lab platform of the Swiss Science Data Centre at \href{https://renkulab.io/projects/carlo.ferrigno/ppanda-light/sessions/new?autostart=1}{this link}.\\
\end{acknowledgements}
\bibliographystyle{aa}
\bibliography{refs}
\newpage
\appendix
\section{Energy-phase matrices and energy-resolved 
pulse profiles of observations} \label{appendix:pprofiles}
Here we show the general evolution of \vzero\ pulse profiles for 
all the different \nustar\ observations examined in this paper.
Energy-phase colour matrices offer a quick-look view of the 
pulse shape in the full 3--70 keV energy band. 
The profiles have been aligned using as reference the phase 
of minima of the first harmonic.
\begin{figure*}[h]
\centering
\begin{tabular}{cc}
\includegraphics[width=0.8\columnwidth]{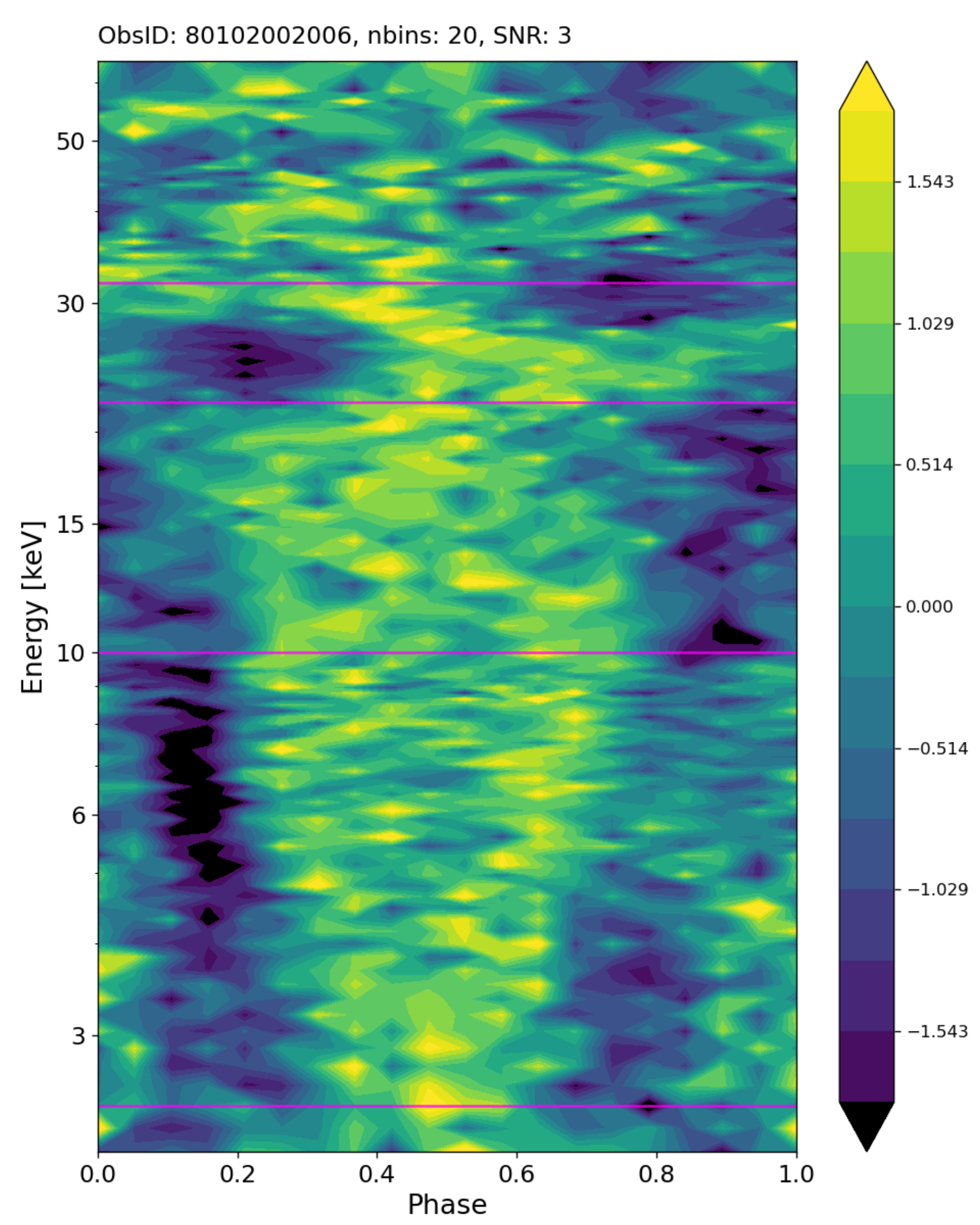}  & 
\includegraphics[width=0.8\columnwidth]{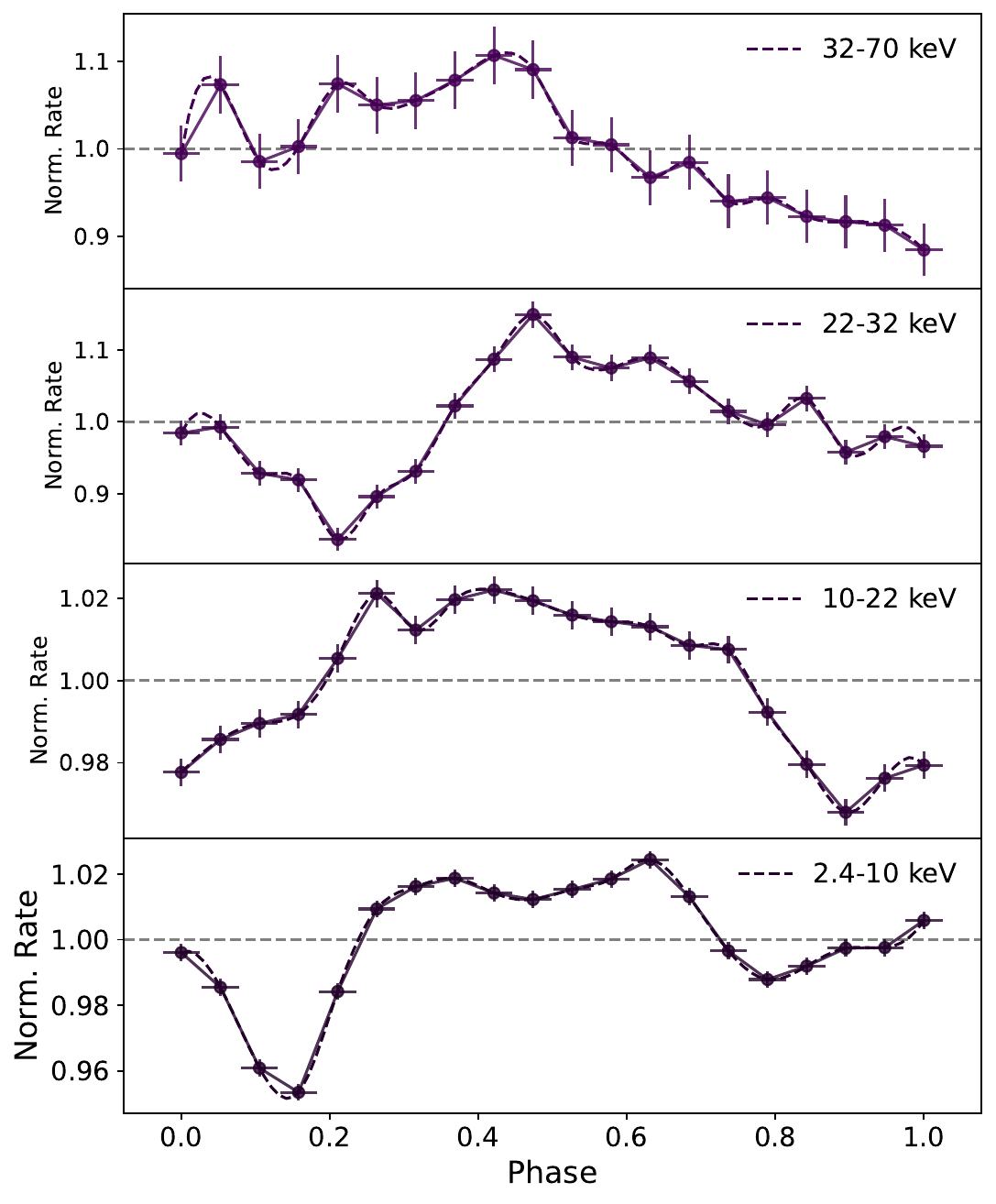} \\
\end{tabular}
\caption{Energy-phase colour maps and energy-selected pulse profiles 
for ObsIDs 806.}
\label{fig:pp806}
\end{figure*}

\begin{figure*}[h]
\centering
\begin{tabular}{cc}
\includegraphics[width=0.8\columnwidth]{figures2/colmap808.pdf}  & 
\includegraphics[width=0.8\columnwidth]{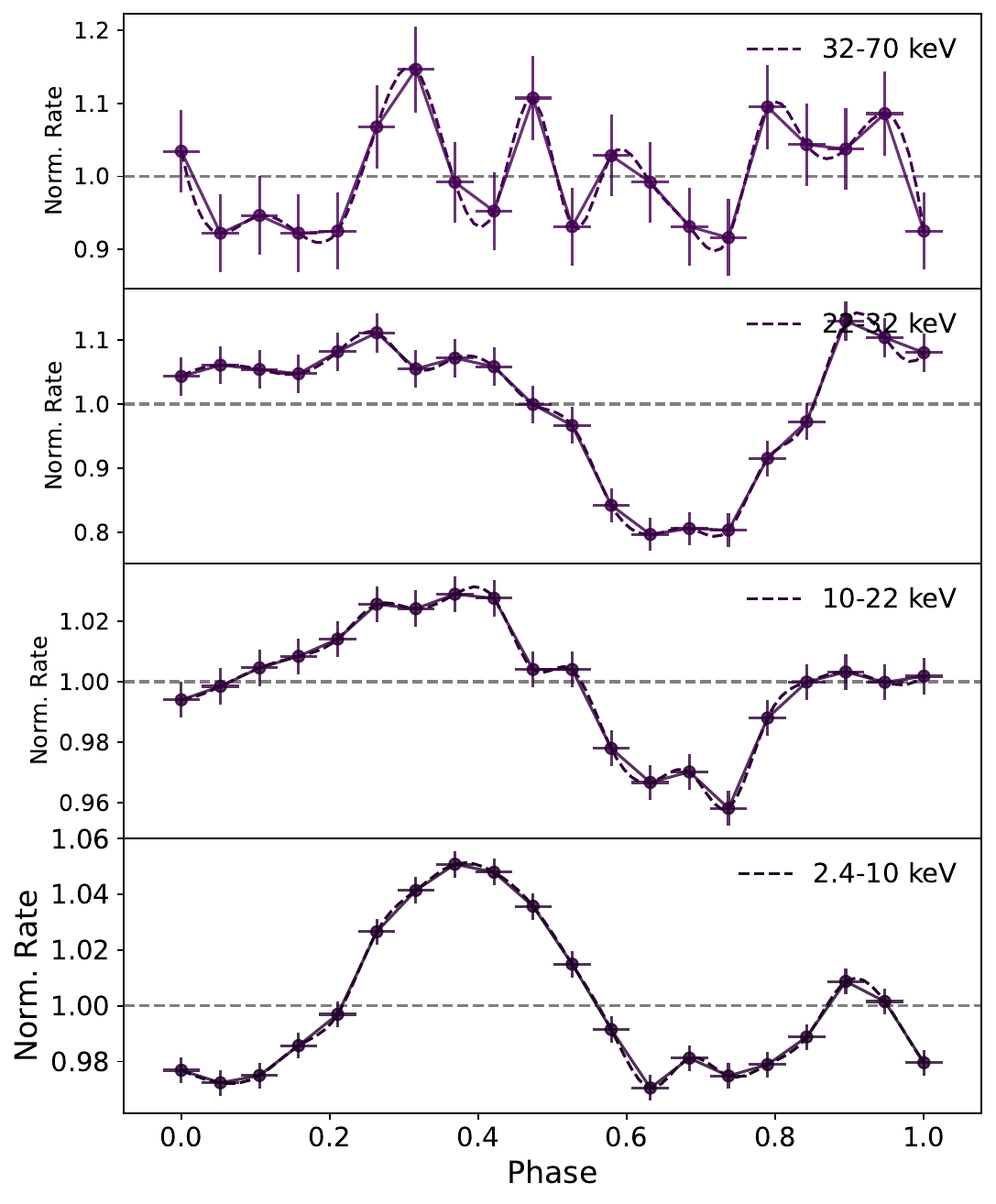} \\
\end{tabular}
\caption{Energy-phase colour maps and energy-selected pulse profiles 
for ObsIDs 808.}
\label{fig:pp808}
\end{figure*}

\begin{figure*}[h]
\centering
\begin{tabular}{cc}
\includegraphics[width=0.8\columnwidth]{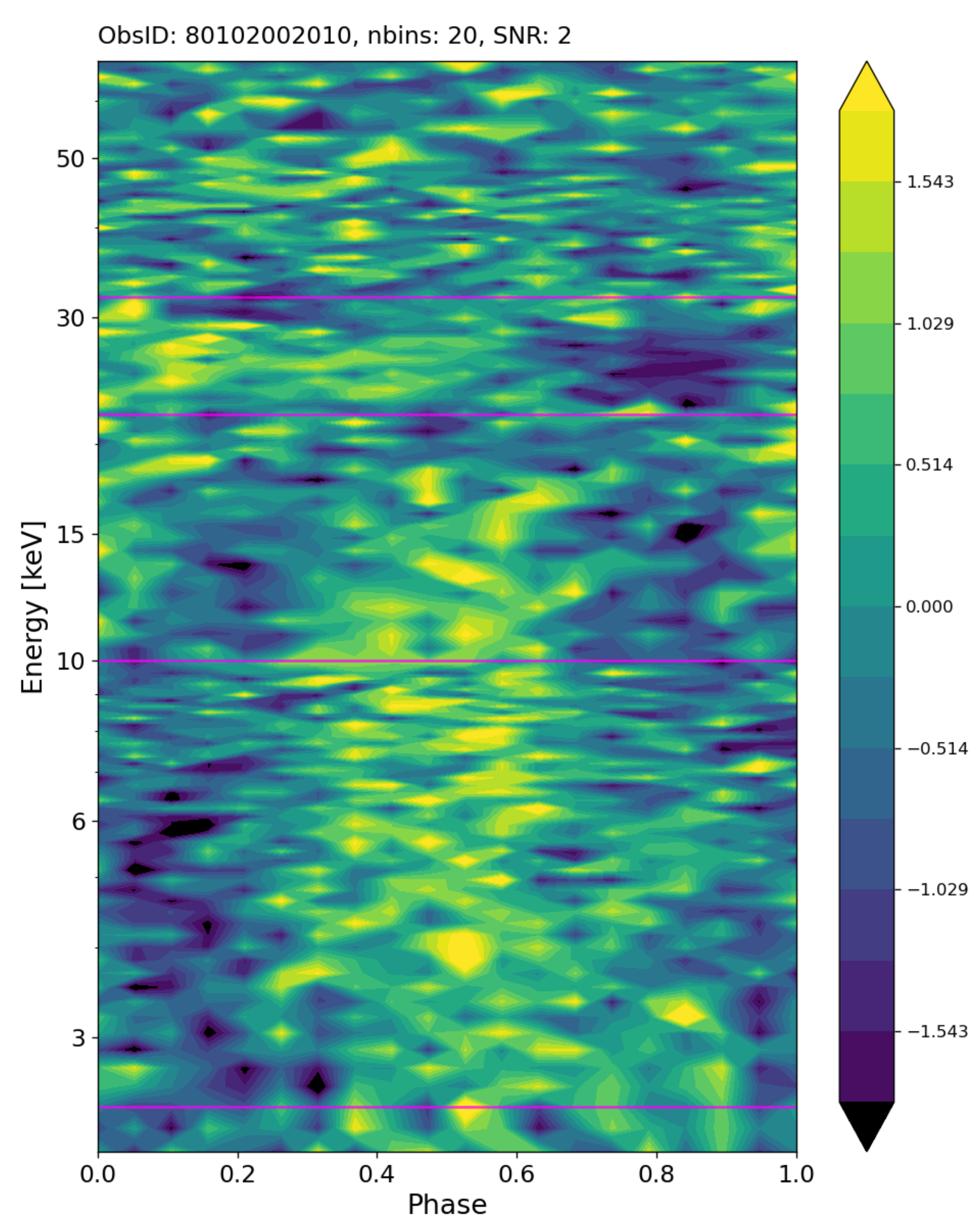}  & 
\includegraphics[width=0.8\columnwidth]{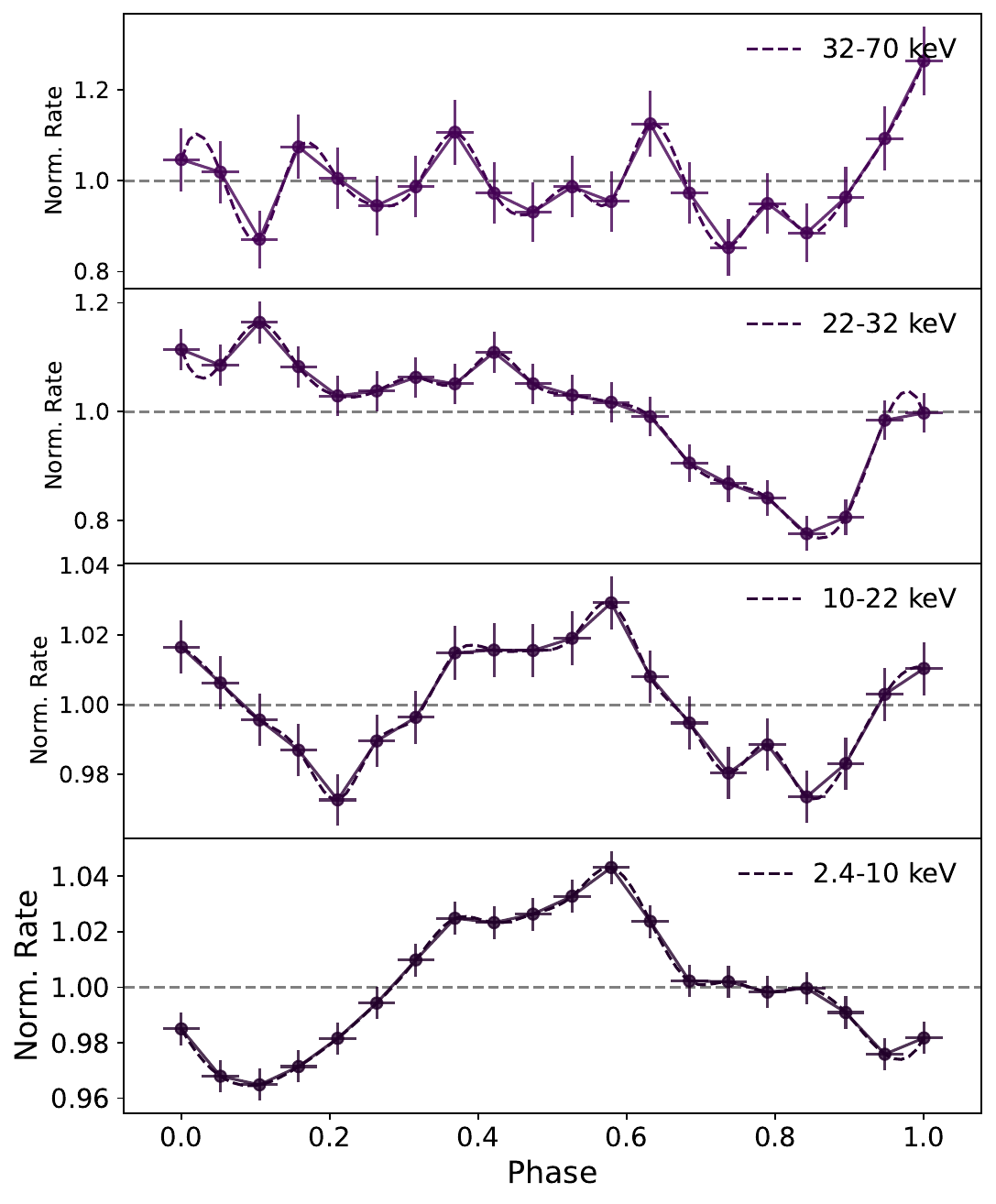} \\
\end{tabular}
\caption{Energy-phase colour maps and energy-selected pulse profiles 
for ObsIDs 810.}
\label{fig:pp806_pp808}
\end{figure*}

\begin{figure*}[h]
\centering
\begin{tabular}{cc}
\includegraphics[width=0.8\columnwidth]{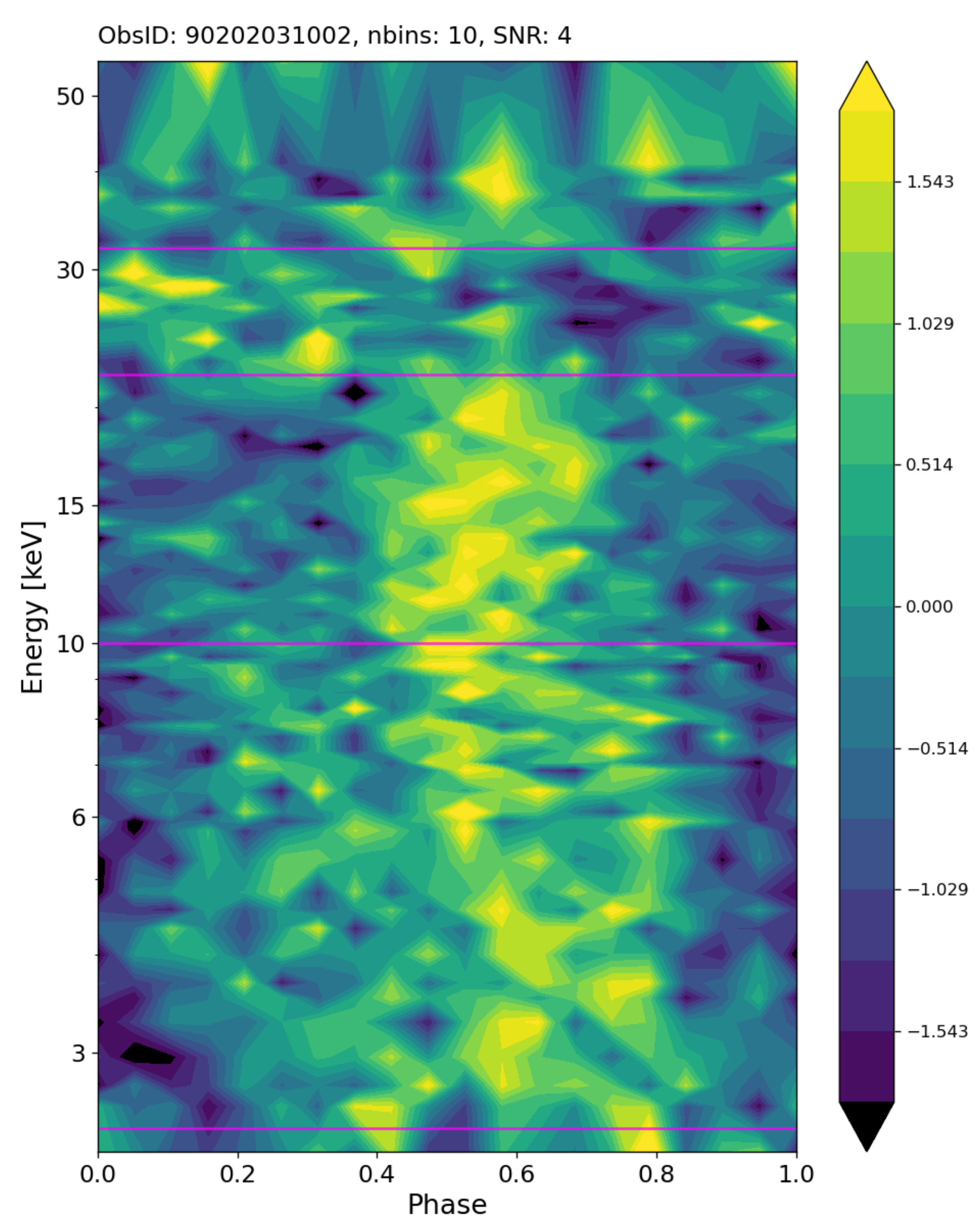}  & 
\includegraphics[width=0.8\columnwidth]{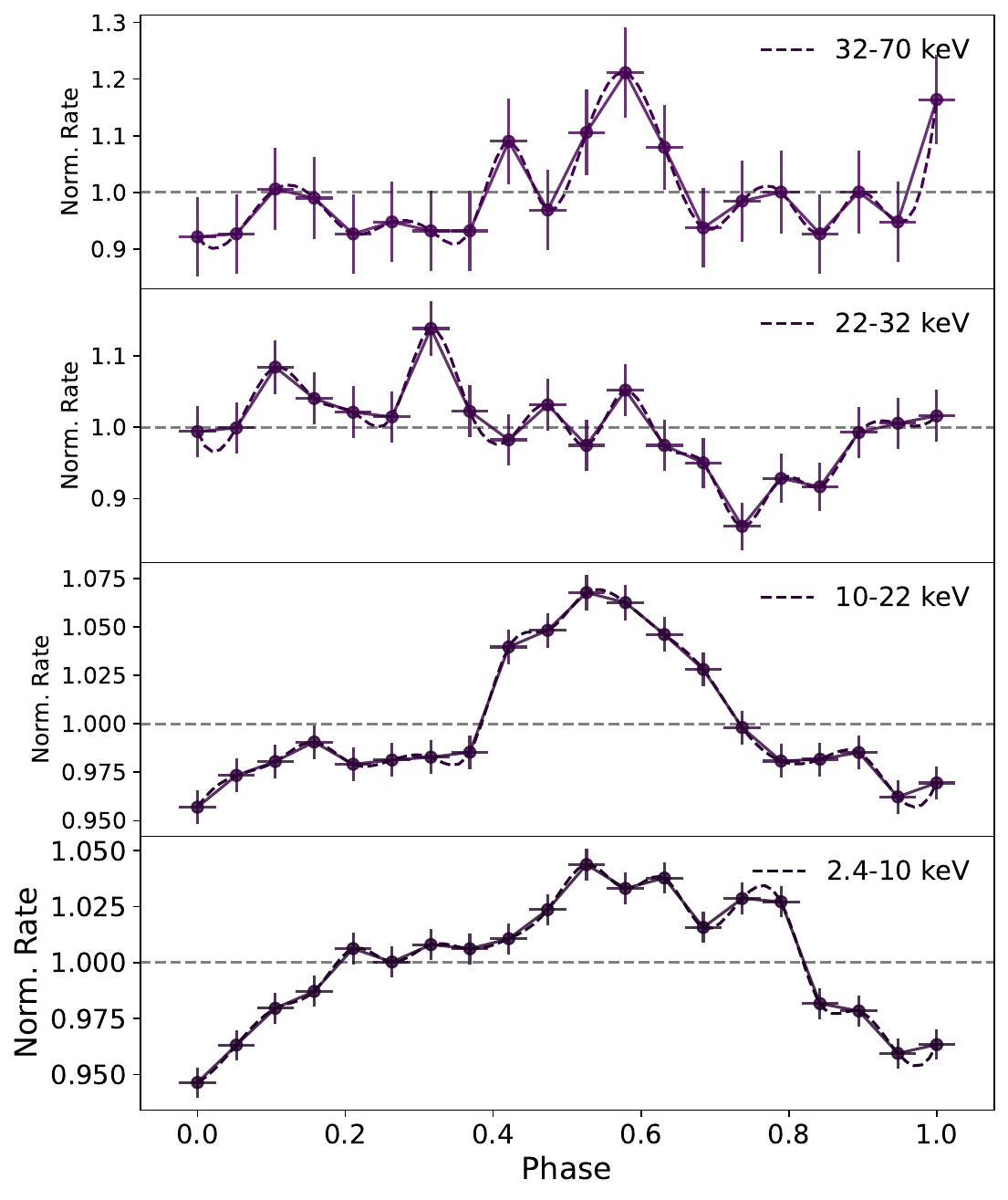} \\
\end{tabular}
\caption{Energy-phase colour maps and energy-selected pulse profiles 
for ObsIDs 902.}
\label{fig:pp902}
\end{figure*}

\begin{figure*}[h]
\centering
\begin{tabular}{cc}
\includegraphics[width=0.8\columnwidth]{figures2/colmap904.pdf}  & 
\includegraphics[width=0.8\columnwidth]{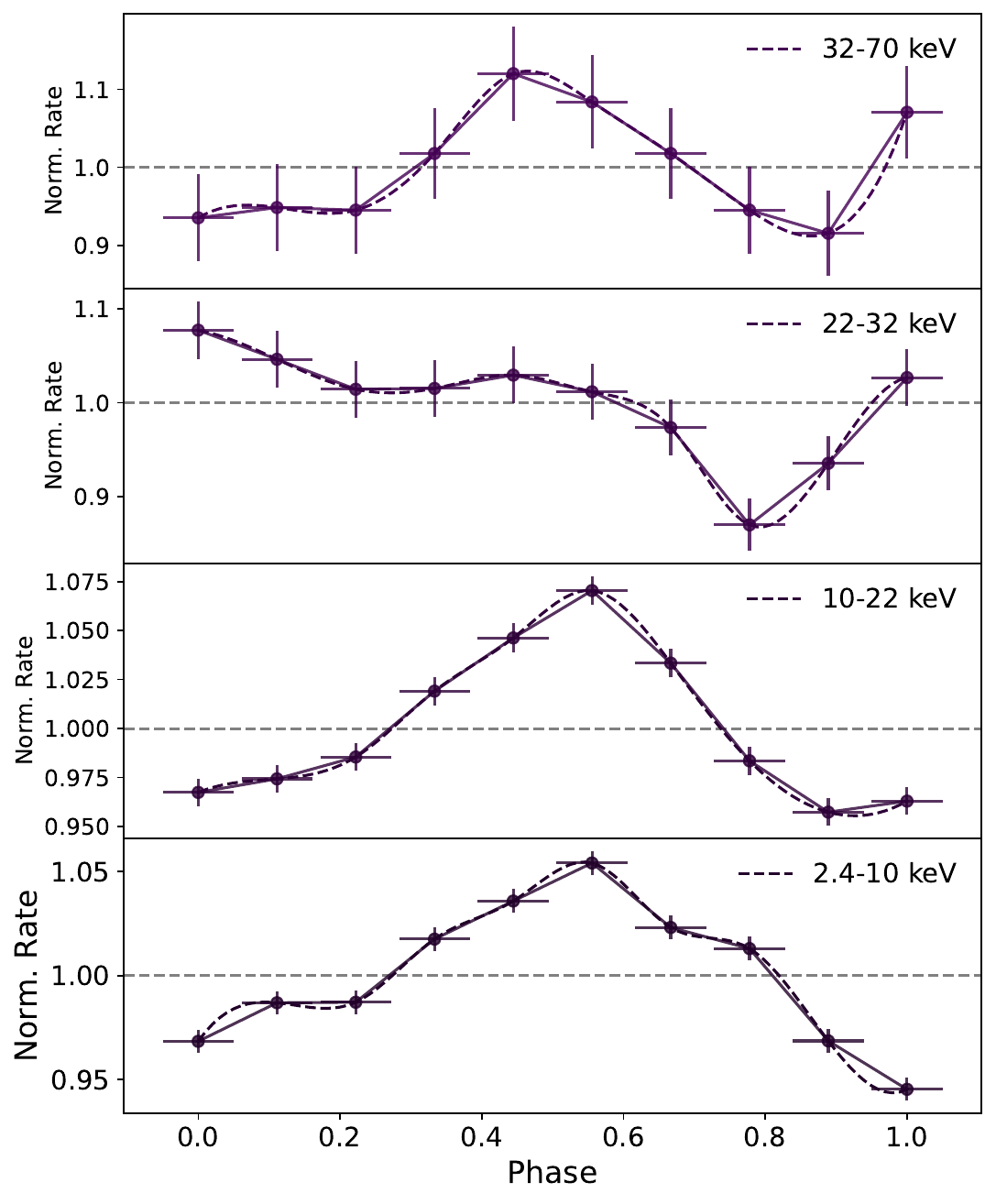} \\
\end{tabular}
\caption{Energy-phase colour maps and energy-selected pulse profiles 
for ObsIDs 904.}
\label{fig:pp904}
\end{figure*}
\newpage

\section{Harmonic decomposition: energy-dependence for the fundamental and the second harmonic amplitudes} \label{appendix:harmonics}
\begin{figure*}
\centering
\begin{tabular}{cc}
\includegraphics[width=\columnwidth]{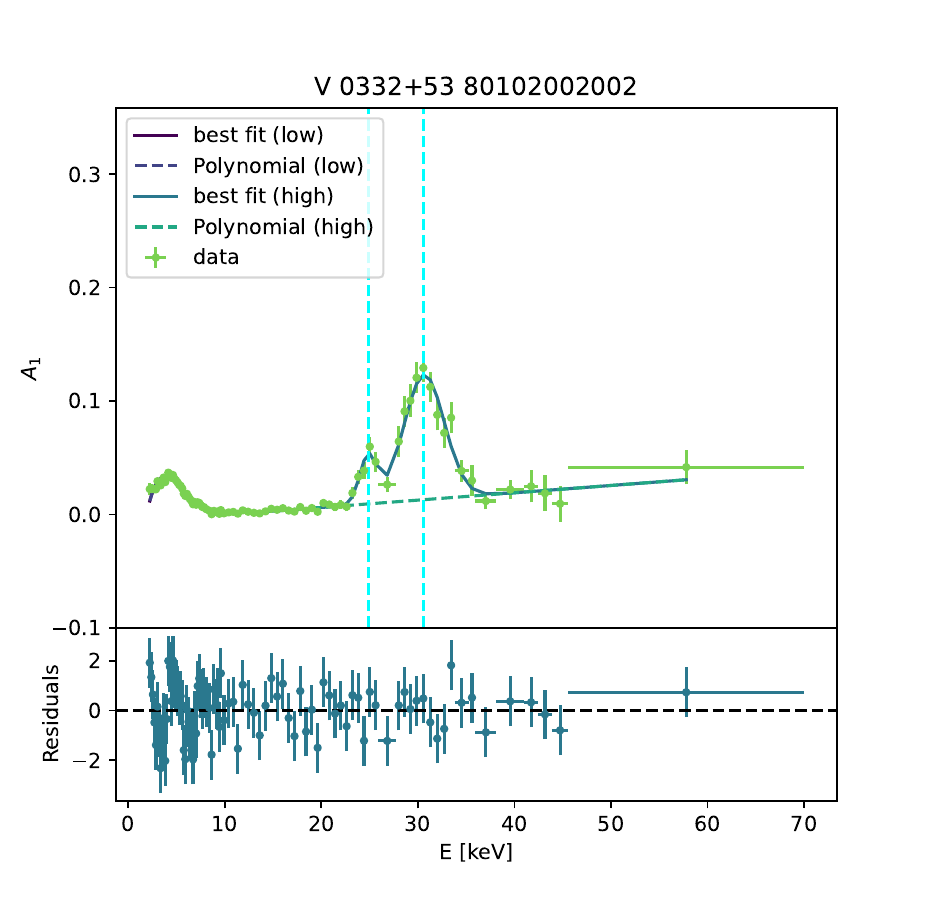}  & 
\includegraphics[width=\columnwidth]{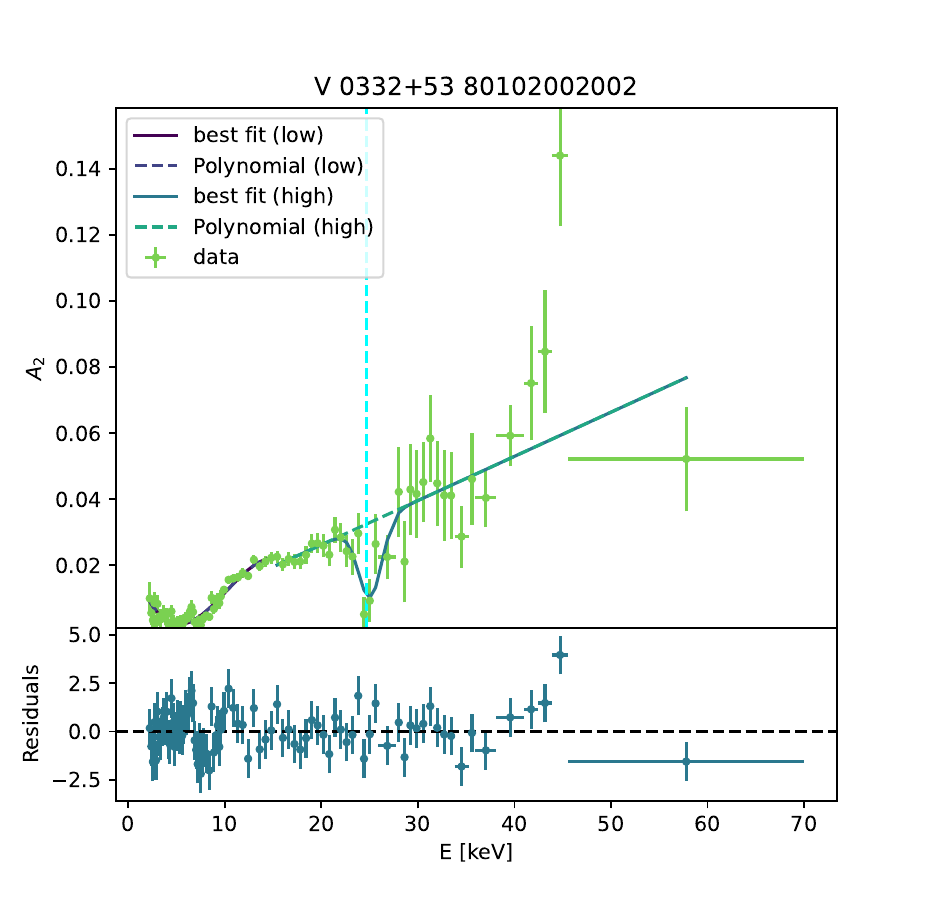} \\
\includegraphics[width=\columnwidth]{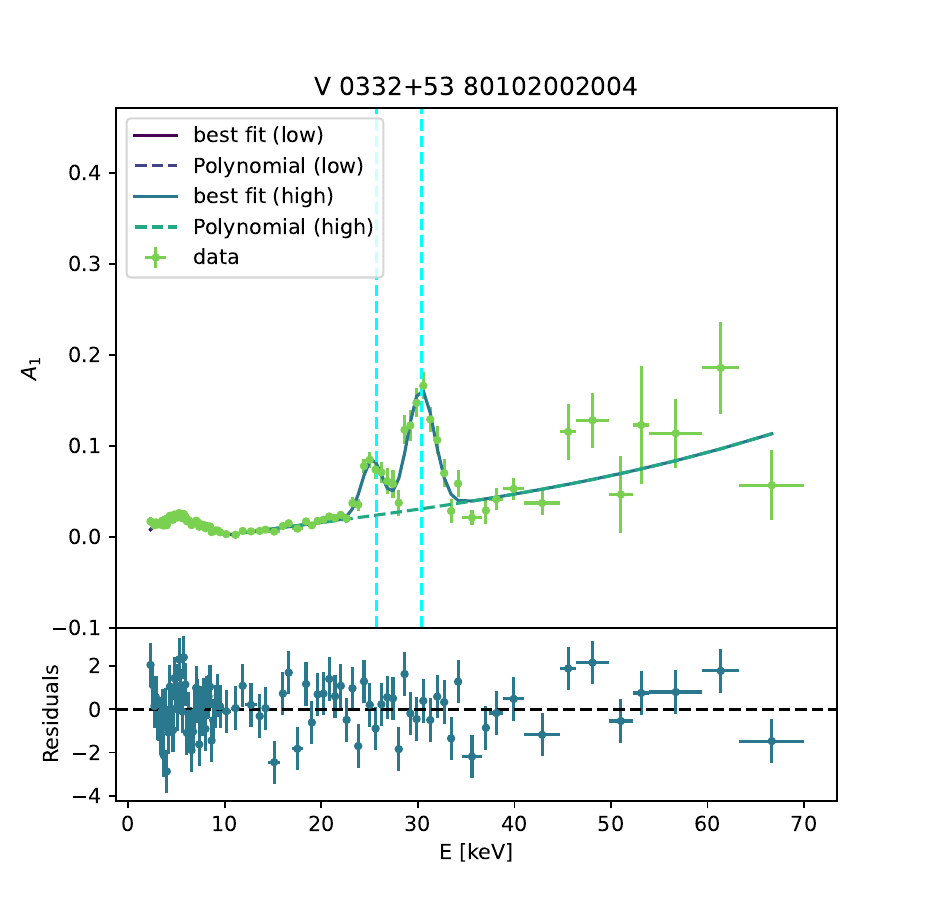}  & 
\includegraphics[width=\columnwidth]{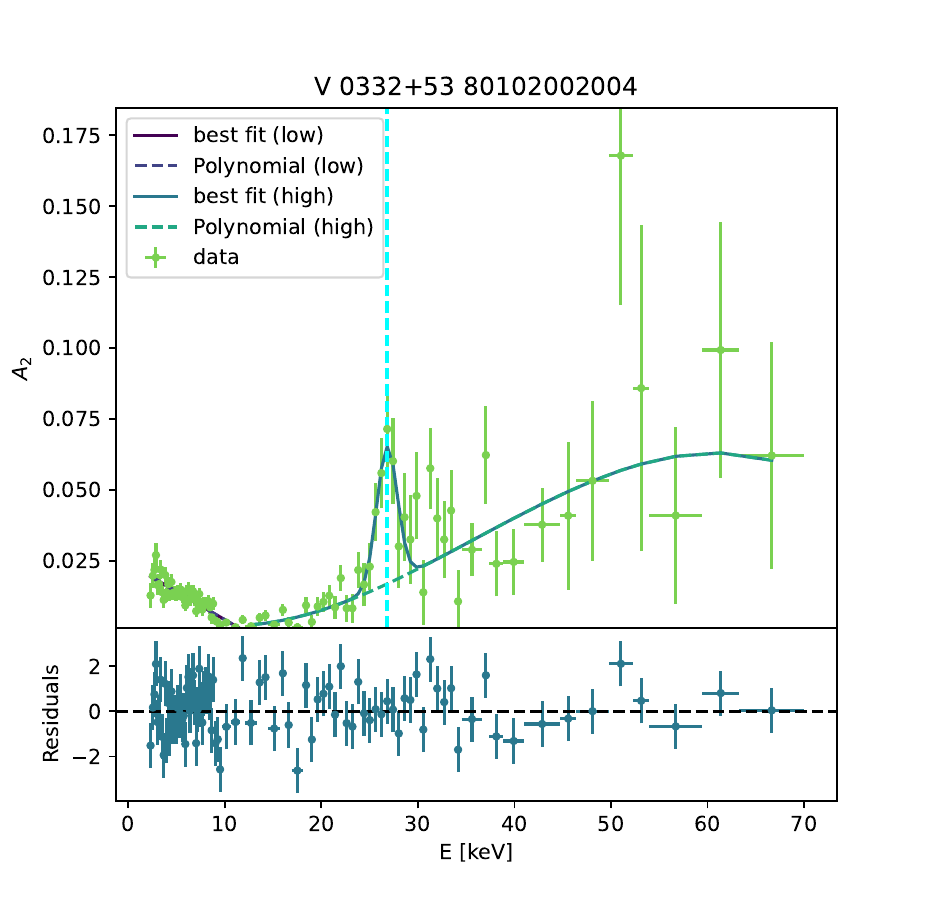} \\
\end{tabular}
\caption{Energy-dependent amplitudes (first harmonic: left panels; second harmonic: 
right panels), best-fitting models and residuals for ObsID 802 (upper panels) 
and 804 (lower panels).}
\label{fig:harmonics1}
\end{figure*}

\begin{figure*}
\centering
\begin{tabular}{cc}
\includegraphics[width=\columnwidth]{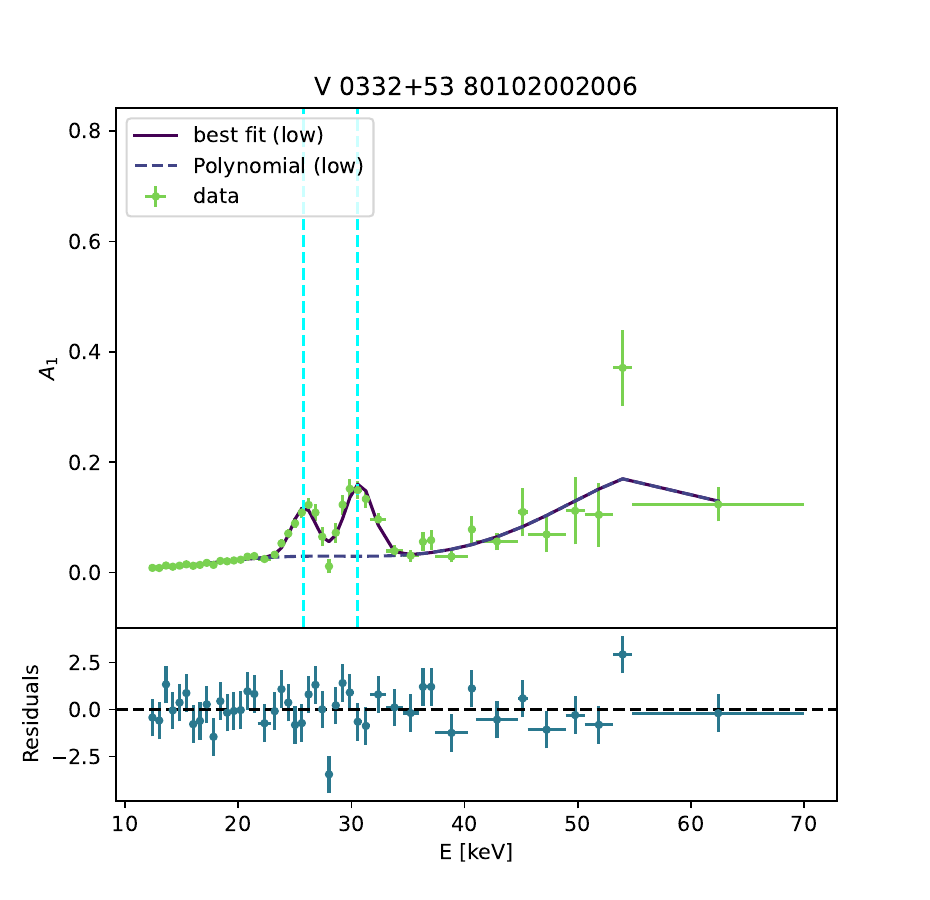}  & 
\includegraphics[width=\columnwidth]{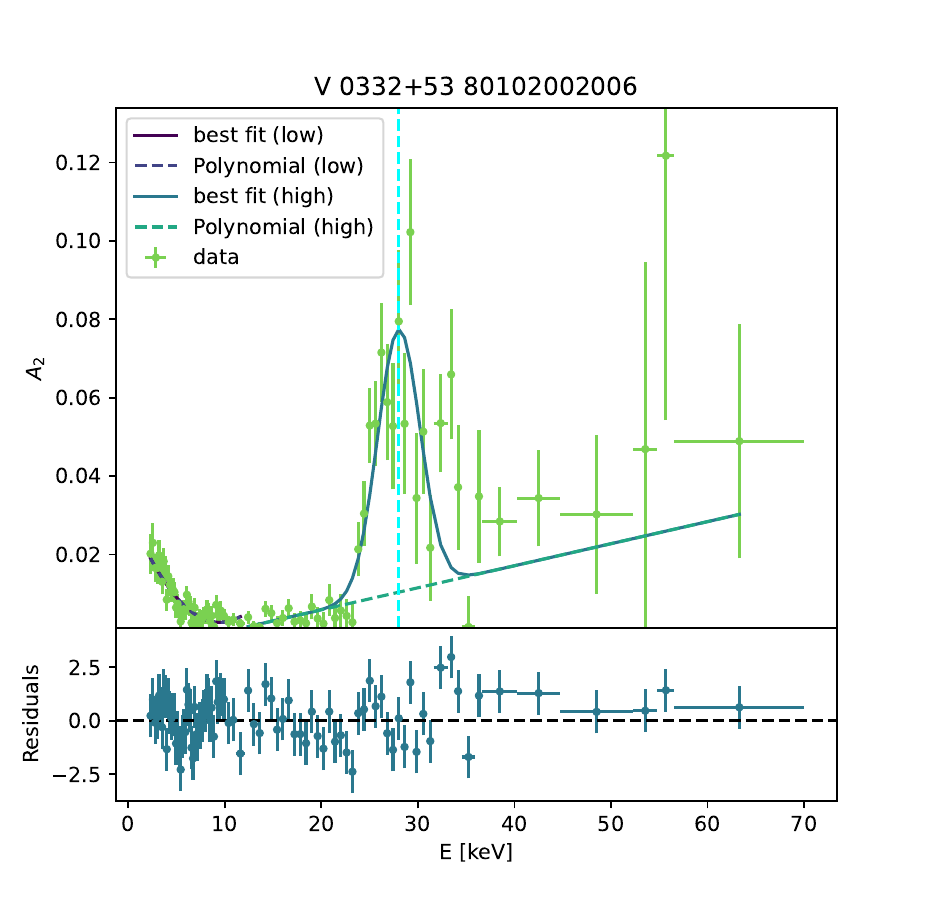} \\
\vspace{-0.5cm}
\includegraphics[width=\columnwidth]{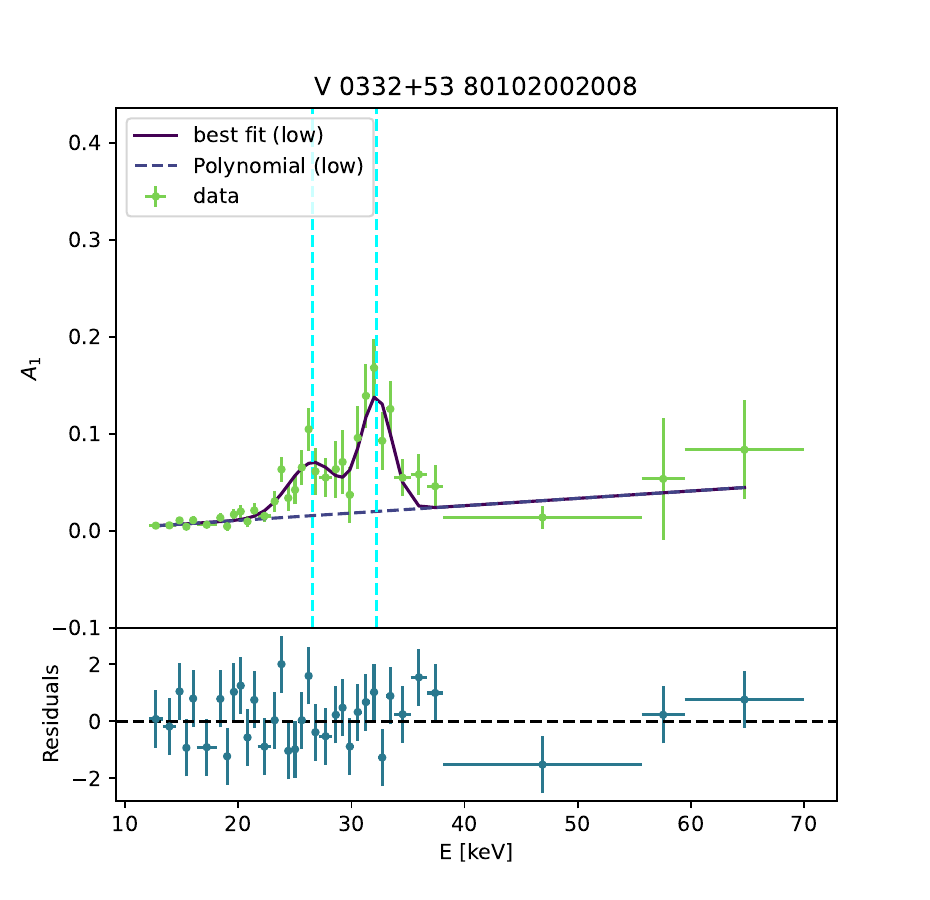}  & 
\includegraphics[width=\columnwidth]{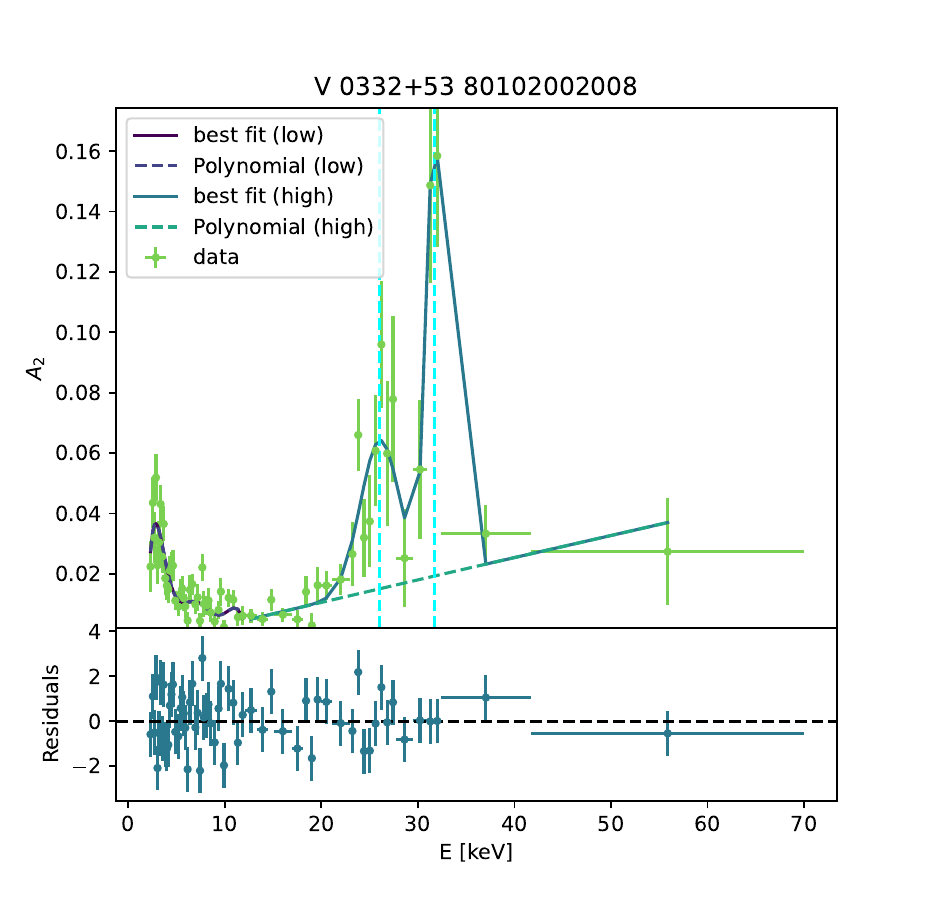} \\
\end{tabular}
\caption{Energy-dependent amplitudes (first harmonic: left panels; second harmonic: right panels), best-fitting models and residuals for ObsID 806 (upper panels) and 808 (lower panels).}
\label{fig:harmonics2}
\end{figure*}

As shown in \citet{ferrigno2023}, the PF is the simplest scalar used to  physically describe
the overall strength of the pulsed emission over the total emission. When the pulsed 
profile is decomposed in its Fourier terms, the single amplitudes might not have a direct
correspondence with physical quantities. Nonetheless, it is worth investigating
if the complex structure we observed in the PF spectrum might be determined only from 
the trend of the fundamental, and what is the specific contribution for the second harmonic 
at the same time. 
With the same methods we used to describe the PF trend in the general PF spectra, 
we repeated the same steps to describe the amplitudes of the fundamental and the 
second harmonic (there remains too little power and statistics to attempt any 
description of the higher harmonic components). We comparatively show these two amplitudes 
side by side in Fig.~\ref{fig:harmonics1} and \ref{fig:harmonics2}. 
We restrict this comparative analysis only for the first four observations, as the 
harmonic amplitudes of the low-statistics observations result too poorly determined.

As expected, the general PF trend is mostly carried out by the amplitude of the 
fundamental, which by itself, reproduces the double-peaked pattern and its 
essential variability for the different observations. 
The amplitude of the second harmonic does not follow the trend of the
fundamental only for the brightest observation (ObsID 802), where there is a drop. 
In all the remaining observation, the second harmonic amplitude displays a 
very similar pattern to the fundamental shape. 

A comprehensive summary of the best-fitting results, both for the PF spectrum 
and the energy-dependent amplitudes of the fundamental and second harmonic 
is shown in Tables \ref{tab:fit_parameters1} and \ref{tab:fit_parameters2}.

\newpage
\section{Spectral fit results} \label{sect:appendix2}
\begin{figure*}
\centering
\begin{tabular}{cc}
\vspace{-0.5cm}
\includegraphics[width=0.8\columnwidth]{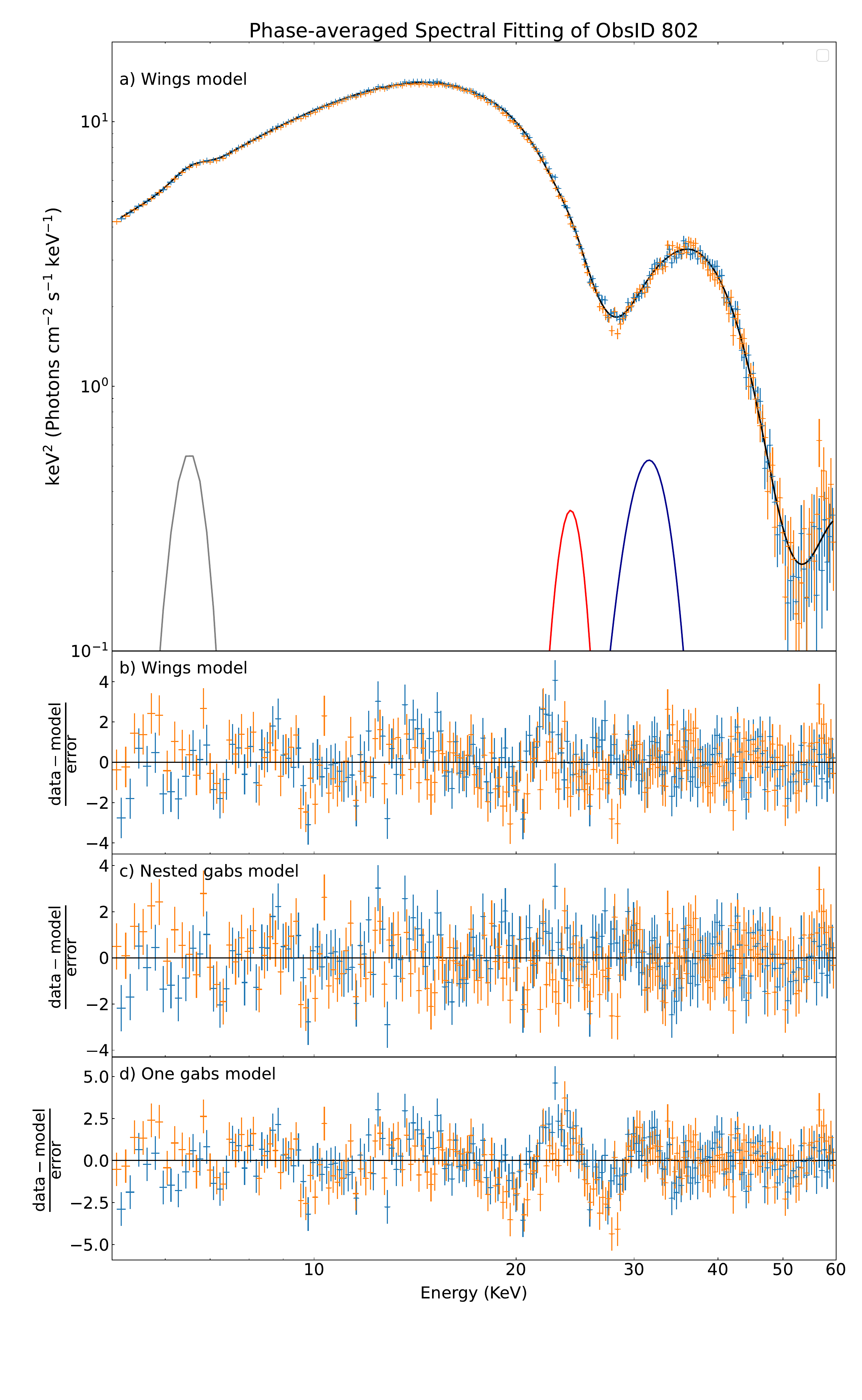}  & 
\includegraphics[width=0.8\columnwidth]{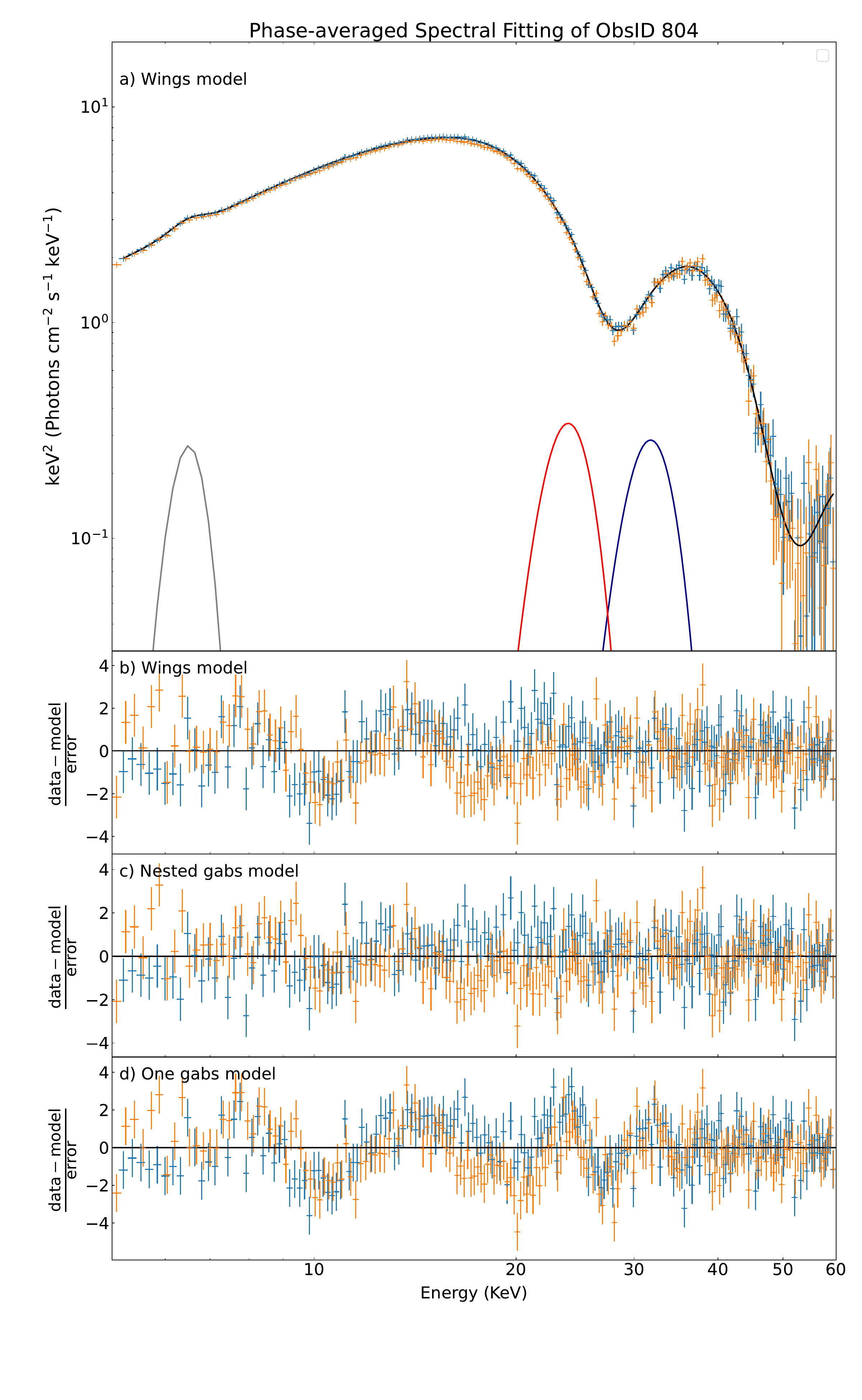} \\
\vspace{-0.5cm}
\includegraphics[width=0.8\columnwidth]{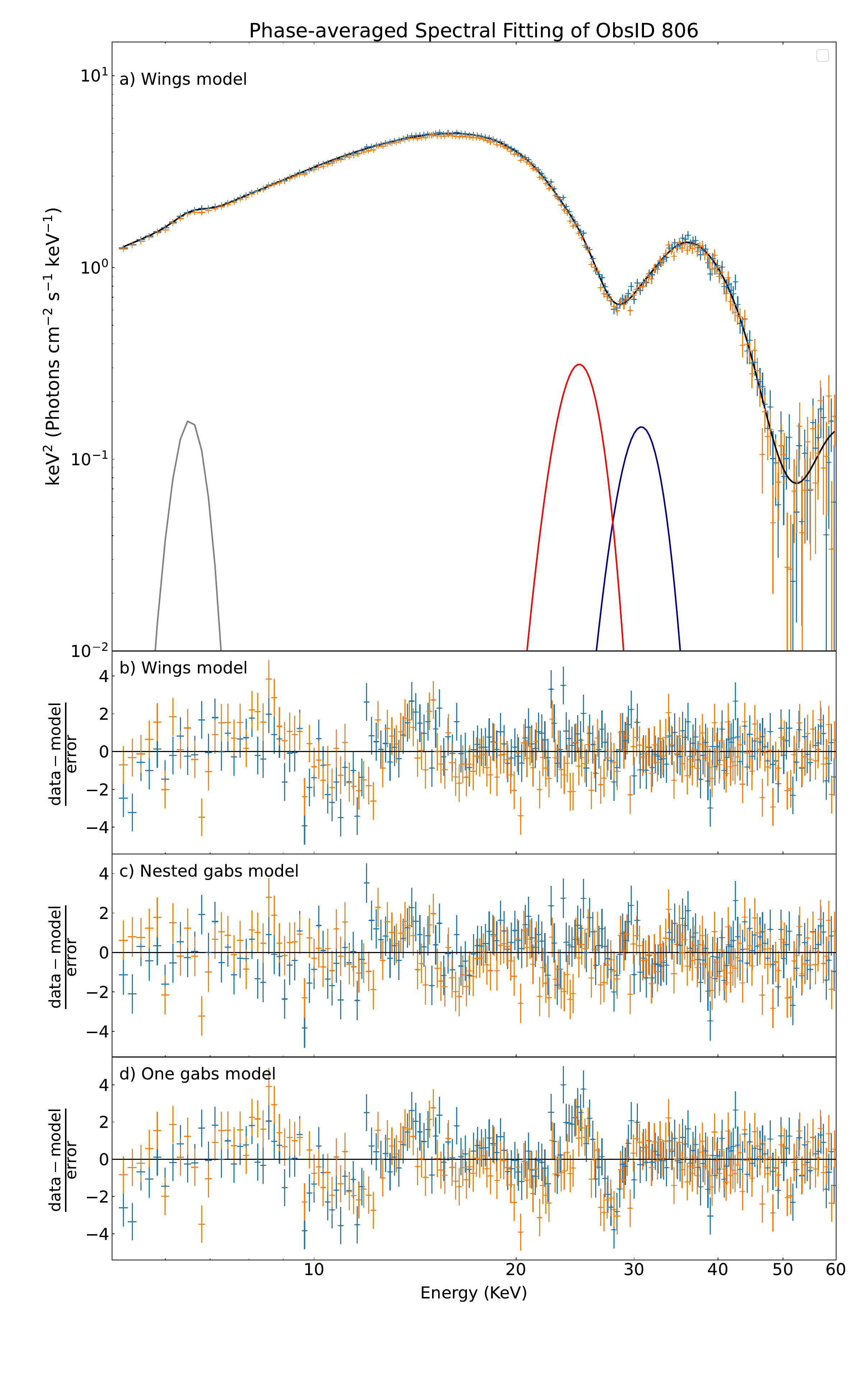} & 
\includegraphics[width=0.8\columnwidth]{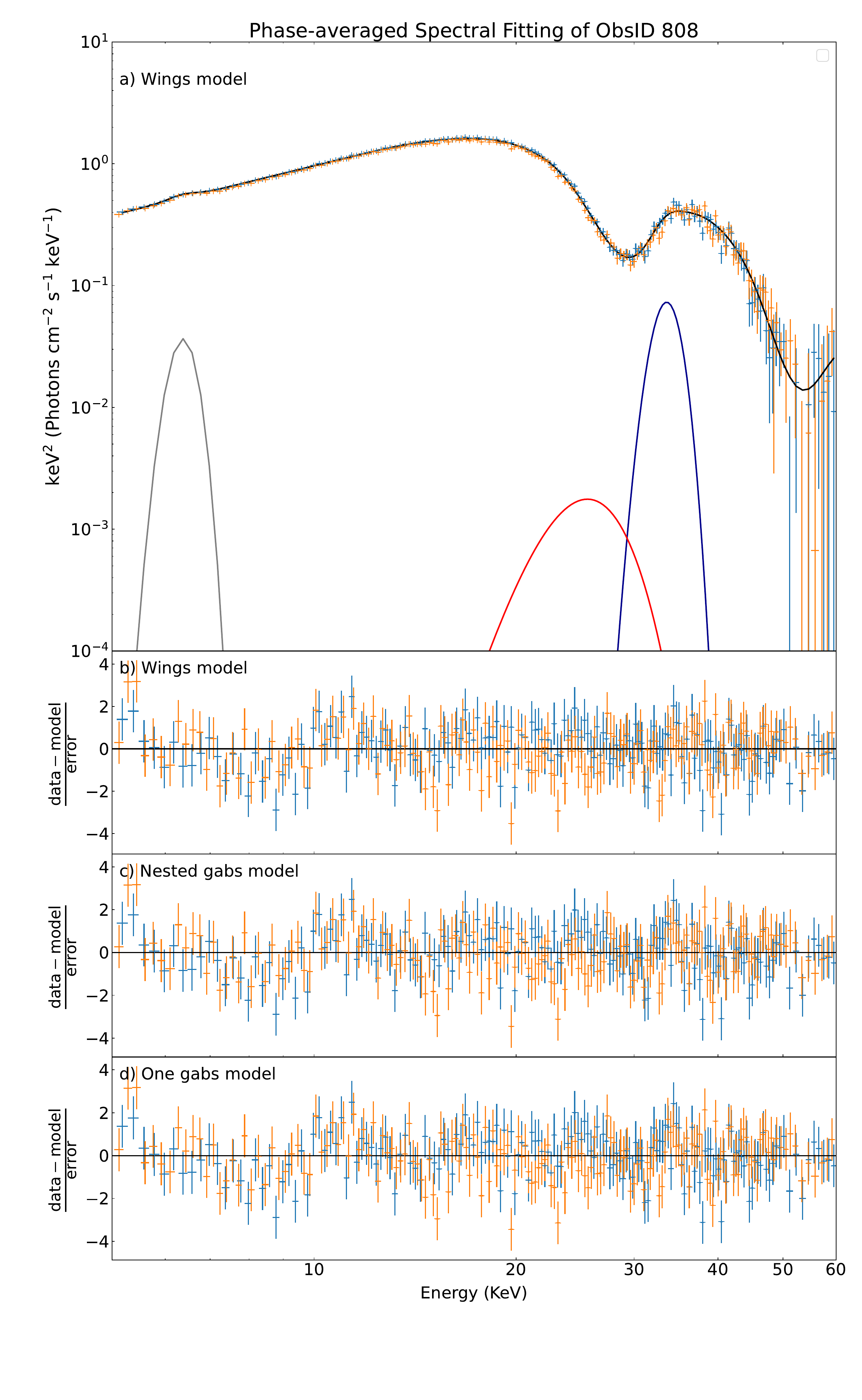} \\
\end{tabular}
\caption{Phase-averaged spectral fitting of ObsIDs 802, 804, 806 and 808. 
    a) Modelling the fundamental cyclotron line with one Gaussian feature in absorption 
    and two Gaussian features in emission representing the blue and red peaks or \textit{wings} 
    with the values derived from the phase-resolved analysis were used as Gaussian priors, 
    described in Sect.~\ref{sect:phaseresolved}. 
    b) Residuals of (a), 
    c) residuals of modelling the line with two Gaussian features in absorption, 
    d) residuals of modelling the line with just one Gaussian feature in absorption.}
\label{fig:energyspectra}
\end{figure*}

We present here the best-fitting results from fitting the phase-averaged
spectra of observations 802, 804, 806, and 808 with the models described
in Sect.\ref{sect:spectralanalysis} in Tables \ref{tab:spectralfits802}, 
\ref{tab:spectralfits804}, \ref{tab:spectralfits806}, and \ref{tab:spectralfits808},
respectively. 
Data, residuals (for all models) and the best-fitting \textit{wings} model 
superimposed on the data are shown in Fig.\ref{fig:energyspectra}. 

\newpage
\begin{table*}[h]
\centering
\caption{Best-fitting spectral results for ObsID 802.}\label{tab:spectralfits802}
\begin{tabular}{llr@{}lr@{}lr@{}lr@{}ll}
\hline
\hline
 & &\multicolumn{2}{c}{\textit{wings}} & \multicolumn{2}{c}{\textit{one gabs}} & \multicolumn{2}{c}{\textit{nested gabs}} \\
\hline 
\\ [-0.5em]
Parameter & Units & \multicolumn{6}{c}{Best-fitting Values} \\ [0.5em]
$E_\mathrm{RW}$ &keV             &24.02 &$\pm$0.08 & -- & -- &  -- & -- \\
$\sigma_\mathrm{RW}$ &keV        &1.08 &$\pm$0.07 & -- & -- &  -- & --\\
$N_\mathrm{RW}$ &$\times10^{-3}$ &1.6 &$\pm$0.3  & -- & -- &  -- & -- \\ [0.5em]

$E_\mathrm{BW}$& keV             &31.28 &$_{-0.10}^{+0.12}$ & -- & -- &  -- & -- \\
$\sigma_\mathrm{BW}$& keV      &2.19 &$_{-0.10}^{+0.14}$ & -- & -- &  -- & --\\
$N_\mathrm{BW}$ &$\times10^{-3}$ &3.0 &$_{-0.4}^{+0.6}$  & -- & -- &  -- & --\\ [0.5em]

$E_\mathrm{Fe}$& keV           &6.485 &$\pm$0.010 & 6.488 &$\pm$0.011 & 6.492 &$\pm$0.011\\
$\sigma_\mathrm{Fe}$& keV       &0.338 &$\pm$0.012 & 0.344 &$\pm$0.014 & 0.308 &$\pm$0.014 &\\ 
$N_\mathrm{Fe}$ &$\times10^{-3}$ &11.4 &$\pm$0.4 & 11.7 &$\pm$0.5 & 9.8&$\pm$0.4  \\ [0.5em]

$E_\mathrm{Cyc1}$&  keV    &28.20 &$_{-0.07}^{+0.09}$ & 27.84 &$\pm$0.03 & 27.79 &$\pm$0.02 \\
$\sigma_\mathrm{Cyc1}$& keV  &3.99 &$\pm$0.04 & 3.83 &$\pm$0.03 & 4.50 &$_{-0.08}^{+0.14}$ \\
$\tau_\mathrm{Cyc1}$ & keV &14.8 &$_{-0.3}^{+0.4}$ & 12.65 &$\pm$0.12 & 10.9 &$_{-0.4}^{+0.3}$ \\ 
$\sigma_\mathrm{Cyc1N}$& keV  &-- & -- & -- & -- & 2.35 &$_{-0.09}^{+0.13}$ \\ 
$\tau_\mathrm{Cyc1N}$ & keV   &-- & -- & -- & -- &  2.7 &$_{-0.3}^{+0.6}$\\ [0.5em]

$E_\mathrm{Cyc2}$ &keV      &51.68 &$\pm$0.16 & 51.89 &$\pm$0.16 & 51.72 &$\pm$0.17\\
$\tau_\mathrm{Cyc2}$ &keV   &18.7  &$\pm$0.4  & 17.8 &$\pm$0.5 & 18.0 &$\pm$0.4 \\ [0.5em]

$E_\mathrm{C}$&keV          &11.51 &$\pm$0.06 & 11.60 &$\pm$0.06 & 11.05 &$\pm$0.14 \\
$E_\mathrm{F}$&keV        &8.15 &$\pm$0.04 & 8.06 &$\pm$0.04 & 7.80 &$\pm$0.05 \\
$E_\mathrm{W}$ &keV         &9.62 &$\pm$0.15 & 9.46 &$\pm$0.15 & 12.2 &$_{-0.3}^{+0.5}$  \\
$\Gamma$        &               &0.497 &$\pm$0.008 & 0.504 &$\pm$0.008 &  0.396 &$_{-0.022}^{+0.018}$  \\
$N_\mathrm{PL}$  &              &0.375 &$\pm$0.005 & 0.380 &$\pm$0.006 & 0.317 &$\pm$0.010 \\ 
$C_{\textrm{cal}}$&             &0.9884 &$\pm$0.0008 & 0.9884 &$\pm$0.0007 & 0.9884 &$\pm$0.0008 \\ [0.5em]
Cstat/dof              &            & 530&/385 &  687&/391 &  493&/389 \\ 
\hline \hline 
\end{tabular}
\tablefoot{$E_\mathrm{RW}$, $\sigma_\mathrm{RW}$ and $N_\mathrm{RW}$ are line position, sigma and 
normalization (in units of photons/cm$^{2}$/s) for the red wing line. $E_\mathrm{BW}$, $\sigma_\mathrm{BW}$ and $N_\mathrm{BW}$ are line position, sigma and  normalisation (in units of photons/cm$^{2}$/s) for the blue wing line. $E_\mathrm{Fe}$, $\sigma_\mathrm{Fe}$ and $N_\mathrm{Fe}$ are line position, sigma and 
normalization (in units of photons/cm$^{2}$/s) for the iron emission line. $E_\mathrm{Cyc1}$, $\sigma_\mathrm{Cyc1}$, and $\tau_\mathrm{Cyc1}$  are line position, width and strength of the \texttt{gabs} component of the fundamental cyclotron line. $\sigma_\mathrm{Cyc1N}$ and $\tau_\mathrm{Cyc1N}$ are width and strength of the nested \texttt{gabs} at the fundamental cyclotron line energy. $E_\mathrm{Cyc2}$, $\sigma_\mathrm{Cyc2}$, and $\tau_\mathrm{Cyc2}$  are line position, width and strength of the \texttt{gabs} component of the second harmonic cyclotron line. $E_\mathrm{C}$, $E_\mathrm{F}$, $E_\mathrm{W}$, $\Gamma$, and $N_\mathrm{PL}$ are the cut-off energy, the folding energy, the smoothing width, the photon-index and component normalisation (in units of photons/keV/cm${2}$/s at 1 keV) of the \texttt{newhcut} continuum component \citep[see][]{Burderi2000}. $C_{\textrm{cal}}$ is a multiplicative constant for the FPMB dataset (this constant is fixed to 1 for the FPMA dataset).}
\end{table*}

\begin{table*}[h]
\centering
\caption{Best-fitting spectral results for ObsID 804.}\label{tab:spectralfits804}
\begin{tabular}{llr@{}lr@{}lr@{}ll}
\hline
\hline
 & &\multicolumn{2}{c}{\textit{wings}} & \multicolumn{2}{c}{\textit{one gabs}} & \multicolumn{2}{c}{\textit{nested gabs}} \\
\hline
\\ [-0.5em]
Parameter & Units & \multicolumn{6}{c}{Best-fitting Values} \\ [0.5em]
$E_\mathrm{RW}$ & keV     & 23.677 &$_{-0.016}^{+0.012}$ & -- & -- & -- & -- \\
$\sigma_\mathrm{RW}$& keV& 1.725  &$_{-0.008}^{+0.010}$ & -- & -- & -- & -- \\
$N_\mathrm{RW}$  &$\times10^{-3}$   & 2.61   &$\pm$0.11 &-- & -- & -- & --  \\ [0.5em]

$E_\mathrm{BW}$ &keV     & 31.42 &$\pm$0.04 &-- & -- & -- & -- \\
$\sigma_\mathrm{BW}$ &keV& 2.271 &$\pm$0.007& -- & -- & -- & -- \\
$N_\mathrm{BW}$  &$\times10^{-3}$   & 1.63 &$\pm$0.10& -- & -- & -- & --\\ [0.5em]

$E_\mathrm{Fe}$ &keV & 6.463 &$\pm$0.009 & 6.470 &$\pm$0.012 & 6.456 &$\pm$0.013 \\
$\sigma_\mathrm{Fe}$ &keV& 0.3578 &$\pm$0.0013& 0.363 &$_{-0.017}^{+0.021}$ & 0.295 &$_{-0.016}^{+0.019}$ \\
$N_\mathrm{Fe}$ &$\times10^{-3}$      & 5.8 &$\pm$0.2& 5.9 &$_{-0.2}^{+0.3}$ & 4.6 &$\pm$0.2 \\ [0.5em]

$\sigma_\mathrm{Cyc1N}$ &keV  & -- & -- & 2.39 &$\pm$0.06 \\
$\tau_\mathrm{Cyc1N}$ &keV   & -- & -- & 4.06 &$\pm$0.20 \\
$E_\mathrm{Cyc1}$  &keV      & 28.31 &$\pm$0.03 &28.15 &$\pm$0.03 & 27.89 &$_{-0.02}^{+0.03}$ \\
$\sigma_\mathrm{Cyc1}$ &keV  & 3.828 &$\pm$0.010 & 3.64 &$\pm$0.03 & 5.71 &$\pm$0.07 \\
$\tau_\mathrm{Cyc1}$ &keV   & 15.09 &$\pm$0.02  & 12.39 &$\pm$0.13 & 17.3 &$\pm$0.5 \\ [0.5em]

$E_\mathrm{Cyc2}$ &keV   & 51.64 &$\pm$0.08& 51.83 &$_{-0.19}^{+0.22}$ & 50.97 &$_{-0.19}^{+0.22}$ \\
$\tau_\mathrm{Cyc2}$ &keV    & 21.92 &$\pm$0.02& 21.4 &$_{-0.7}^{+0.8}$ & 19.7 &$_{-0.6}^{+0.8}$ \\ [0.5em]
 
$E_\mathrm{C}$  &keV         & 13.28 &$\pm$0.03& 13.21 &$\pm$0.05 & 17.4 &$_{-0.5}^{+0.4}$ \\
$E_\mathrm{F}$  &keV& 8.255 &$\pm$0.018 & 8.22 &$\pm$0.04 & 6.80 &$_{-0.10}^{+0.13}$ \\
$E_\mathrm{W}$  &keV & 10.484 &$\pm$0.010  & 9.81 &$\pm$0.14 & 28 &$_{-2}^{+1}$  \\
$\Gamma$  && 0.530 &$\pm$0.002& 0.538 &$\pm$0.006 & 0.415 &$\pm$0.019 \\
$N_\mathrm{PL}$ && 0.1788 &$\pm$0.0008 & 0.181 &$\pm$0.002 & 0.149 &$_{-0.003}^{+0.004}$  \\ 
$C_{\textrm{cal}}$ && 0.9787 &$\pm$0.0014& 0.9787 &$\pm$0.0009 & 0.9787 &$\pm$0.0009  \\ [0.5em]
Cstat/dof              &&  583&/375 & 721&/381 &  488&/379 \\
\hline \hline
\end{tabular}
\tablefoot{See notes in Table \ref{tab:spectralfits802} for parameter definitions.}
\end{table*}

\begin{table*}[h]
\centering
\caption{Best-fitting spectral results for ObsID 806.}\label{tab:spectralfits806}
\begin{tabular}{llr@{}lr@{}lr@{}ll}
\hline
\hline
 & &\multicolumn{2}{c}{\textit{wings}} & \multicolumn{2}{c}{\textit{one gabs}} & \multicolumn{2}{c}{\textit{nested gabs}} \\
\hline
\\ [-0.5em]
Parameter & Units & \multicolumn{6}{c}{Best-fitting Values} \\ [0.5em]

$E_\mathrm{RW}$ & keV     &24.65 &$\pm$0.08 & --& -- & -- & --\\
$\sigma_\mathrm{RW}$ &keV &1.56 &$\pm$0.03 & --& -- & -- & --\\
$N_\mathrm{RW}$ &$\times10^{-3}$     & 2.0 &$\pm$0.3 & --& -- & --& --\\ [0.5em]

$E_\mathrm{BW}$ &keV   &30.51 &$\pm$0.07& -- & -- & -- & --\\
$\sigma_\mathrm{BW}$ &keV&1.90 &$\pm$0.03& -- & -- & -- & --\\
$N_\mathrm{BW}$ &$\times10^{-4}$    & 7.5 &$\pm$1.0 & -- & -- & --  & --\\ [0.5em]

$E_\mathrm{Fe}$  &keV    &6.501&$\pm$0.016&6.504 &$\pm$0.014 & 6.474 &$\pm$0.015 \\
$\sigma_\mathrm{Fe}$ &keV&0.310 &$\pm$0.008&0.308 &$_{-0.018}^{+0.024}$ & 0.32 &$\pm$0.02 \\
$N_\mathrm{Fe}$ &$\times10^{-3}$    & 2.98 &$\pm$0.14 &2.96 &$\pm$0.17& 3.05 &$\pm$0.18\\ [0.5em]

$E_\mathrm{Cyc1}$  &keV     &28.26 &$\pm$0.06&28.38 &$\pm$0.03 & 28.22 &$_{-0.04}^{+0.05}$ \\
$\sigma_\mathrm{Cyc1}$ &keV &3.58 &$\pm$0.03&3.70 &$\pm$0.03 & 5.52 &$\pm$0.08 \\
$\tau_\mathrm{Cyc1}$  &keV  &14.52 &$\pm$0.07&12.69 &$\pm$0.17 & 19.6 &$\pm$0.7 \\
$\sigma_\mathrm{Cyc1N}$ &keV& -- & -- & -- & -- & 2.17 &$\pm$0.08 \\
$\tau_\mathrm{Cyc1N}$  &keV &-- & -- & -- & -- & 3.2 &$\pm$0.2 \\ [0.5em]
$E_\mathrm{Cyc2}$ &keV  &51.00 &$\pm$0.08&50.98 &$\pm$0.17 & 50.02 &$\pm$0.20 \\
$\tau_\mathrm{Cyc2}$ &keV&21.93 &$_{-0.18}^{+0.15}$&22.2 &$\pm$0.7 & 20.5 &$\pm$0.9 \\ [0.5em]

$E_\mathrm{C}$   &keV      &14.00 &$\pm$0.05&13.95 &$\pm$0.04 & 20.3 &$_{-0.7}^{+0.6}$ \\
$E_\mathrm{F}$   &keV       &8.50 &$\pm$0.05&8.55 &$\pm$0.05  & 6.82 &$_{-0.19}^{+0.22}$ \\
$E_\mathrm{W}$   &keV      &8.70 &$\pm$0.07&8.42 &$_{-0.15}^{+0.19}$ & 25 &$_{-2}^{+1}$  \\
$\Gamma$         &       &0.552 &$\pm$0.004&0.555 &$\pm$0.006 & 0.515 &$\pm$0.008 \\
$N_\mathrm{PL}$   &      &0.1197 &$\pm$0.0009&0.1203 &$\pm$0.0015 & 0.1116 &$\pm$0.0018  \\
$C_{\textrm{cal}}$ &                 &0.9777 &$\pm$0.0014&0.9777 &$\pm$0.0010 & 0.9777 &$\pm$0.0011  \\ [0.5em]
Cstat/dof               &    & 543&/369& 670&/375 &  523&/373  \\
\hline
\end{tabular}
\tablefoot{See notes in Table \ref{tab:spectralfits802} for parameter definitions.}
\end{table*}

\begin{table*}[h]
\centering
\caption{Best-fitting spectral results for ObsID 808.}\label{tab:spectralfits808}
\begin{tabular}{llr@{}lr@{}lr@{}ll}
\hline
\hline
 & &\multicolumn{2}{c}{\textit{wings}} & \multicolumn{2}{c}{\textit{one gabs}} & \multicolumn{2}{c}{\textit{nested gabs}} \\
\hline
\\ [-0.5em]
Parameter & Units & \multicolumn{6}{c}{Best-fitting Values} \\ [0.5em]

$E_\mathrm{RW}$ &keV       & 24.802 &$_{-0.006}^{+0.011}$ &-- & -- & -- & -- \\
$\sigma_\mathrm{RW}$ &keV  &3.102 &$_{-0.007}^{+0.005}$   & -- & -- & -- & -- \\
$N_\mathrm{RW}$  &$\times10^{-4}$     &$<$ & 2.3& -- & -- & -- & --  \\ [0.5em]

$E_\mathrm{BW}$ &keV      &33.422 &$_{-0.012}^{+0.005}$& -- & -- & -- & -- \\
$\sigma_\mathrm{BW}$&keV   &1.42 &$\pm$0.03& -- & -- & -- & -- \\
$N_\mathrm{BW}$ & $\times10^{-4}$  &2.3 &$_{-0.6}^{+0.8}$& -- & -- & -- & -- \\ [0.5em]

$E_\mathrm{Fe}$ & keV &6.357 &$_{-0.006}^{+0.011}$ & 6.35 &$_{-0.04}^{+0.03}$ & 6.357 &$\pm$0.008 \\
$\sigma_\mathrm{Fe}$ & keV  &0.2677 &$_{-0.0014}^{+0.0029}$  & 0.28 &$_{-0.04}^{+0.08}$ & 0.266 &$\pm$0.003 \\
$N_\mathrm{Fe}$ &$\times10^{-4}$  &6.3 &$\pm$0.9  & 6.4 &$_{-0.9}^{+1.2}$  & 6.2 &$\pm$0.4 \\ [0.5em]

$E_\mathrm{Cyc1}$ &keV  &29.01 &$\pm$0.03 & 28.72 &$\pm$0.05 & 28.72 &$\pm$0.03 \\
$\sigma_\mathrm{Cyc1}$ &keV  &3.374 &$_{-0.009}^{+0.019}$& 3.19 &$\pm$0.05 & 3.200 &$\pm$0.010 \\
$\tau_\mathrm{Cyc1}$ &keV  &13.171 &$_{-0.015}^{+0.073}$  & 12.1 &$\pm$0.2 & 12.070 &$\pm$0.015 \\ 
$\sigma_\mathrm{Cyc1N}$ &keV  &-- & --& -- & -- & 1.357 &$\pm$0.002 \\
$\tau_\mathrm{Cyc1N}$ &keV  &-- & --& -- & -- & 0.0755 &$\pm$0.0006 \\ [0.5em]

$E_\mathrm{Cyc2}$ &keV &52.47 &$\pm$0.03& 52.4 &$_{-0.4}^{+0.5}$ & 52.42 &$\pm$0.08 \\
$\tau_\mathrm{Cyc2}$ &keV &24.65 &$_{-0.04}^{+0.05}$  & 24 &$_{-2}^{+2}$ & 24.51 &$\pm$0.07 \\ [0.5em]

$E_\mathrm{C}$ &keV  &15.937 &$_{-0.017}^{+0.022}$ & 15.85 &$\pm$0.07 & 15.86 &$\pm$0.02 \\
$E_\mathrm{F}$ &keV  &8.24 &$\pm$0.03 & 8.26 &$\pm$0.08 & 8.265 &$\pm$0.009 \\
$E_\mathrm{W}$ &keV  &7.61 &$_{-0.03}^{+0.03}$   & 7.3 &$\pm$0.3 & 7.353 &$\pm$0.011  \\
$\Gamma$&0.669 & &$\pm$0.004  & 0.668 &$_{-0.008}^{+0.006}$ & 0.6695 &$\pm$0.0004  \\
$N_\mathrm{PL}$& & 0.0451 &$\pm$0.0005 & 0.0451 &$_{-0.0008}^{+0.0006}$ & 0.04519 &$\pm$0.00018  \\
$C_{\textrm{cal}}$& & 0.9766 &$\pm$0.0020  & 0.9766 &$\pm$0.0018 & 0.977 &$\pm$0.005  \\ [0.5em]
Cstat/dof  & & 330&/322 &  380&/328 &  374&/326  \\
\hline \hline
\end{tabular}
\tablefoot{See notes in Table \ref{tab:spectralfits802} for parameter definitions.}
\end{table*}
\newpage
\section{Corner plots of the Wings model}  \label{appendix:cornerplots}

\begin{figure*}[htbp]
\centering
\includegraphics[width=\textwidth]{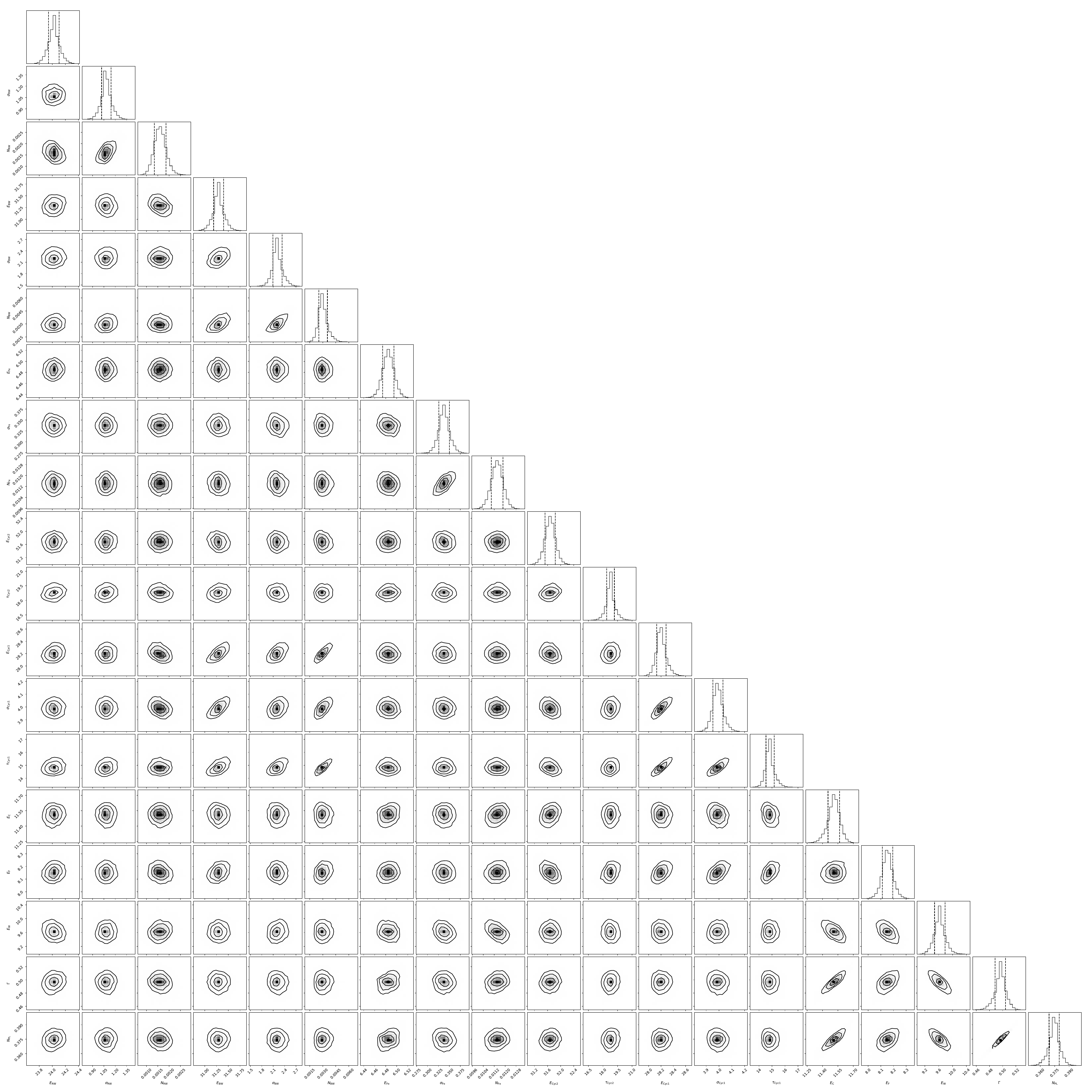}
\vspace{0.5cm}
\caption{Corner plots of the Wings model for ObsID 802}
\label{fig:relation_lines_802}
\end{figure*}

\begin{figure*}[htbp]
\centering
\includegraphics[width=\textwidth]{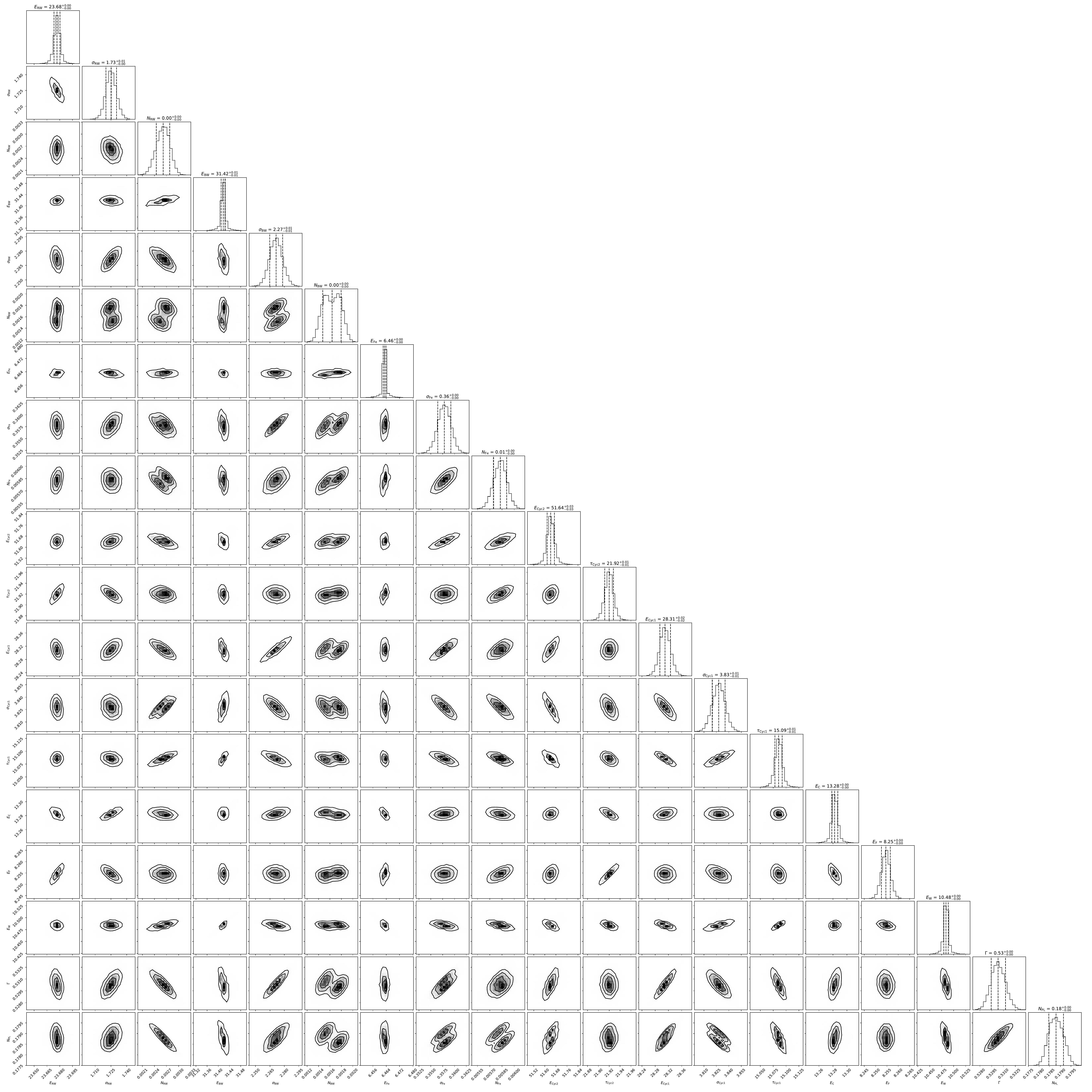}
\vspace{0.5cm}
\caption{Corner plots of the Wings model for ObsID 804}
\label{fig:relation_lines_804}
\end{figure*}

\begin{figure*}[htbp]
\centering
\includegraphics[width=\textwidth]{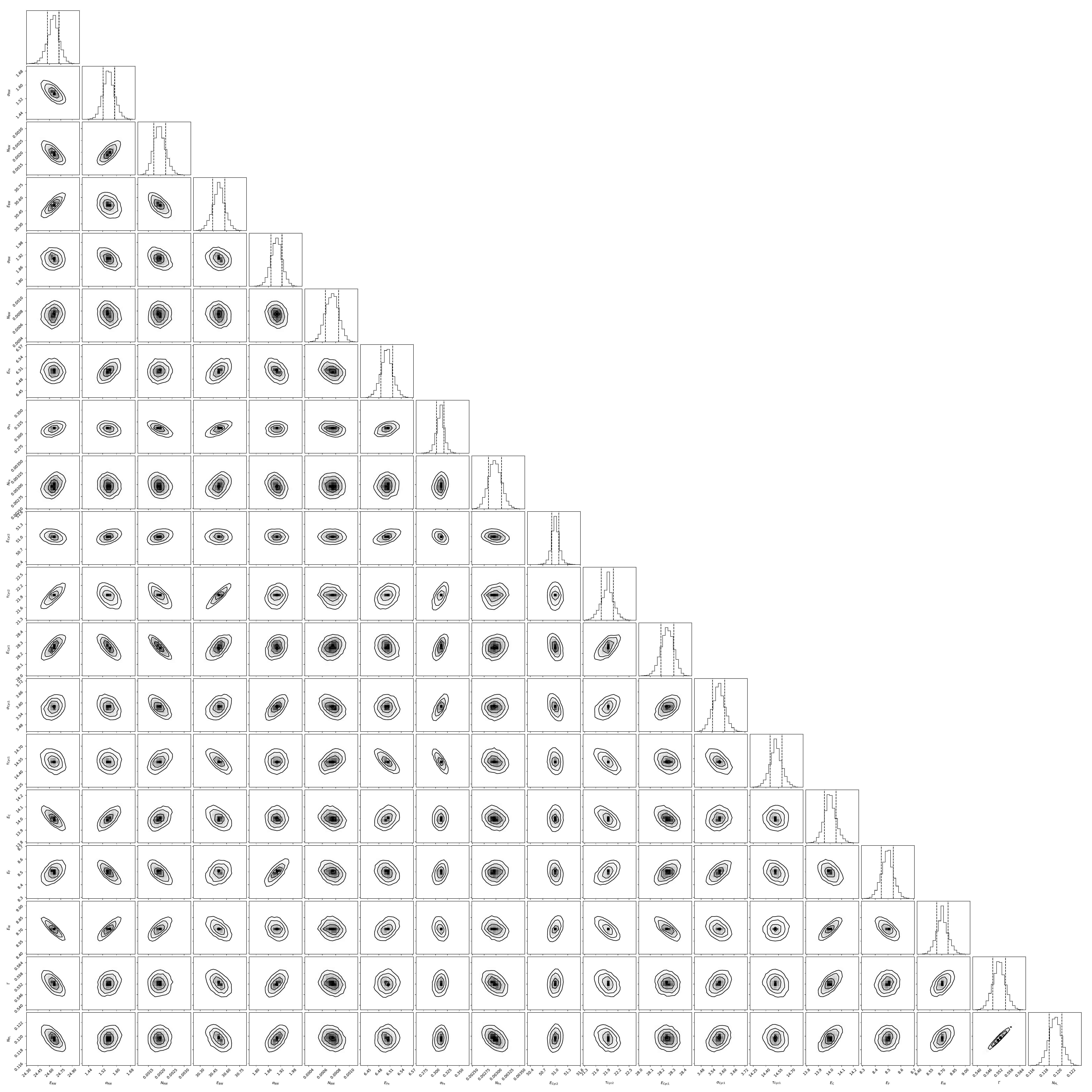}
\vspace{0.5cm}
\caption{Corner plots of the Wings model for ObsID 806}
\label{fig:relation_lines_806}
\end{figure*}

\begin{figure*}[htbp]
\centering
\includegraphics[width=\textwidth]{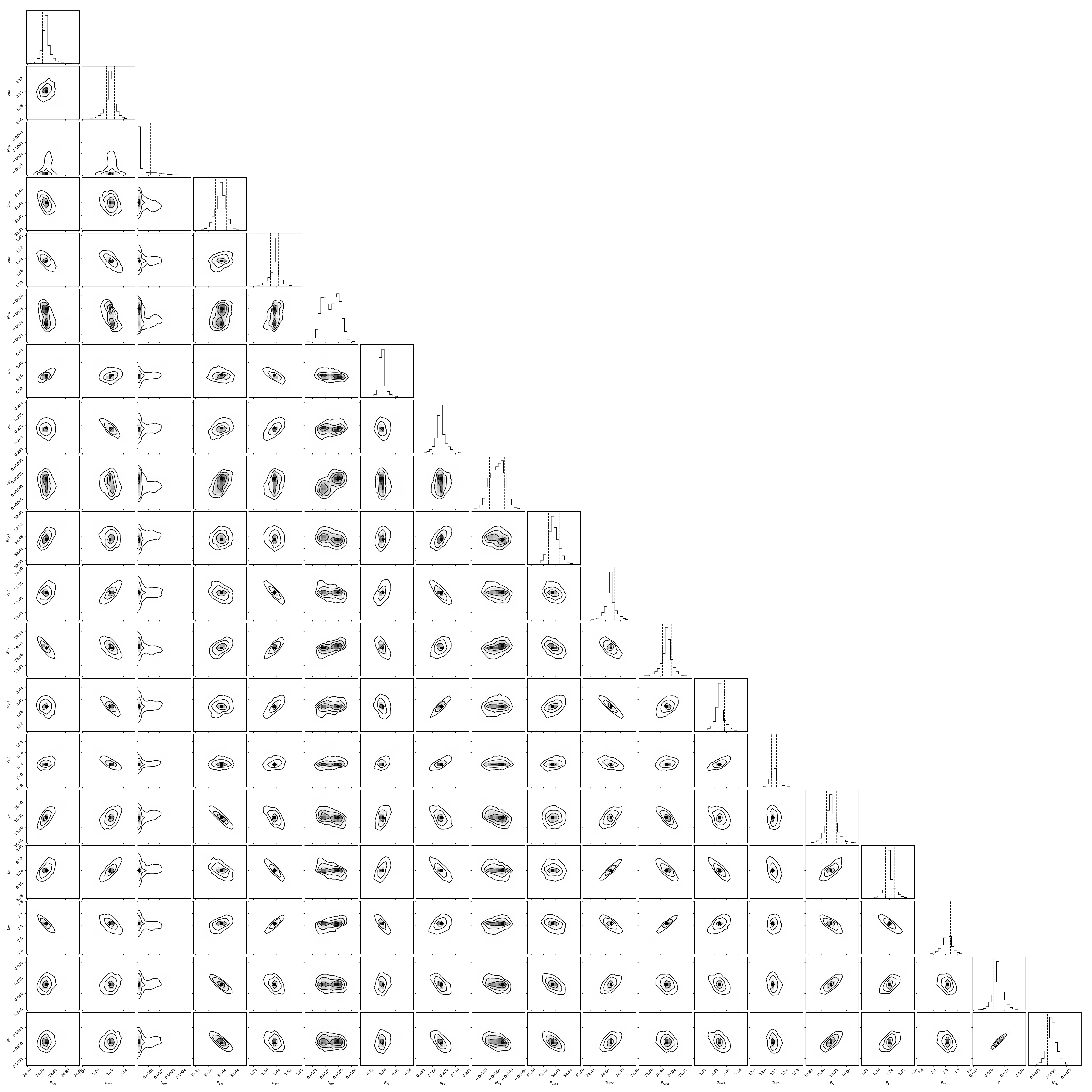}
\vspace{0.5cm}
\caption{Corner plots of the Wings model for ObsID 808}
\label{fig:relation_lines_808}
\end{figure*}
\end{document}